[ english version / português abaixo ]

# UFO, universe, reptilians and creatures communities on Brazilian Telegram: when the sky is not the limit and conspiracy theories seek answers beyond humanity

*Ergon Cugler de Moraes Silva*

Brazilian Institute of Information in
Science and Technology (IBICT)
Brasília, Federal District, Brazil

contato@ergoncugler.com
www.ergoncugler.com

## Abstract

Interest in extraterrestrial phenomena and conspiracy theories involving UFOs and reptilians has been growing on Brazilian Telegram, especially in times of global uncertainty, such as during the COVID-19 pandemic. Therefore, this study aims to address the research question: **how are Brazilian conspiracy theory communities on UFO, universe, reptilians and creatures topics characterized and articulated on Telegram?** It is worth noting that this study is part of a series of seven studies whose main objective is to understand and characterize Brazilian conspiracy theory communities on Telegram. This series of seven studies is openly and originally available on arXiv at Cornell University, applying a mirrored method across the seven studies, changing only the thematic object of analysis and providing investigation replicability, including with proprietary and authored codes, adding to the culture of free and open-source software. Regarding the main findings of this study, the following were observed: UFO communities act as gateways for theories about reptilians, connecting narratives of global control with extraterrestrial beings; Discussions about UFOs and the universe grew significantly during the Pandemic, reflecting a renewed interest in extraterrestrial phenomena; Reptilians remain a significant subculture within conspiracy theories, with a notable growth during the Pandemic; The thematic overlap between UFOs, reptilians and esotericism reveals a cohesive ecosystem of disinformation, making factual correction a challenge; UFO communities function as amplifiers of other conspiracy theories, connecting different themes and strengthening the disinformation network.

### Key findings

➔ UFO and universe communities serve as gateways to theories about reptilians and creatures, connecting global control and extraterrestrial narratives: With 1,268,407 posts, these communities directed 690 invites to reptilian groups, suggesting that interest in extraterrestrial life often leads to an exploration of theories that combine control with non-human beings;

➔ Theories about reptilians and creatures are strongly linked to the New World Order, reinforcing narratives of global domination: With 1,159 invites between these communities, belief in reptilian entities merges with narratives about global control, creating a unified discourse that suggests a worldwide conspiracy by non-human forces;

➔ Discussions about UFOs and the universe experienced a 4,400% growth during the pandemic, indicating renewed interest in extraterrestrial phenomena: Between 2020 and 2021, mentions of UFOs increased from 1,000 to 45,000, driven by government reports on sightings, reflecting growing interest accompanied by distrust in official explanations;

➔ Reptilians and creatures remain a significant subculture within conspiracy theories, with a 1,000% growth during the pandemic: Mentions of reptilians rose from 1,000 to 10,000 between 2020 and 2021, highlighting continued interest in extreme theories involving non-human entities and control narratives;

➔ UFO and universe communities are strongly interconnected with occultism and esotericism, forming a network of alternative beliefs: With 6,081 invites to occult communities, discussions about UFOs frequently intersect with esoteric beliefs, suggesting that the search for alternative explanations for unexplained phenomena extends to mystical practices, challenging conventional science;

➔ The narrative that natural tragedies are planned and controlled by extraterrestrial forces has gained traction in UFO communities: Discussions linking natural disasters to extraterrestrial intervention are common, reflecting distrust in scientific explanations and suggesting that natural events are part of a larger plan of global manipulation;

➔ The thematic overlap between UFOs, reptilians, and esotericism reveals a cohesive ecosystem of disinformation that is difficult to dismantle: These communities share members and content, creating a cycle of disinformation where different narratives support each other, making factual correction a challenge;

➔ The figure of reptilians as part of a global control plan resurfaces constantly, reinforcing the persistence of this narrative: Despite a lack of evidence, the idea of reptilians involved in conspiracies persists, fueling theories of domination and reinforcing distrust in authorities;

➔ UFO communities function as amplifiers of other conspiracy theories, connecting different themes and strengthening the disinformation network: Besides focusing on extraterrestrial phenomena, these communities act as hubs that connect their members to other conspiracy theories, such as occultism, NWO, and apocalypse, centralizing the spread of disinformation;

➔ Discussions about UFOs and reptilians are persistently intertwined, suggesting that these narratives are rooted in a broader conspiratorial worldview: Analyses show that these themes frequently overlap, indicating that they are part of a larger set of beliefs that challenge the conventional worldview.

## 1. Introduction

After analyzing thousands of Brazilian conspiracy theory communities on Telegram and extracting tens of millions of content pieces from these communities, created and/or shared by millions of users, this study aims to compose a series of seven studies that address the phenomenon of conspiracy theories on Telegram, focusing on Brazil as a case study. Through the identification approaches implemented, it was possible to reach a total of 66 Brazilian conspiracy theory communities on Telegram on UFO, universe, reptilians and creatures topics, summing up 1,427,011 content pieces published between May 2016 (initial publications) and August 2024 (date of this study), with 141,202 users aggregated from within these communities. Thus, this study aims to understand and characterize the communities focused on UFO, universe, reptilians and creatures present in this Brazilian network of conspiracy theories identified on Telegram.

To this end, a mirrored method will be applied across all seven studies, changing only the thematic object of analysis and providing investigation replicability. In this way, we will adopt technical approaches to observe the connections, temporal series, content, and overlaps of themes within the conspiracy theory communities. In addition to this study, the other six are openly and originally available on arXiv at Cornell University. This series paid particular attention to ensuring data integrity and respecting user privacy, as provided by Brazilian legislation (Law No. 13,709/2018 / Brazilian law from 2018).

Therefore, the question arises: **how are Brazilian conspiracy theory communities on UFO, universe, reptilians and creatures topics characterized and articulated on Telegram?**

## 2. Materials and methods

The methodology of this study is organized into three subsections: **2.1. Data extraction**, which describes the process and tools used to collect information from Telegram communities; **2.2. Data processing**, which discusses the criteria and methods applied to classify and anonymize the collected data; and **2.3. Approaches to data analysis**, which details the techniques used to investigate the connections, temporal series, content, and thematic overlaps within conspiracy theory communities.

### 2.1. Data extraction

This project began in February 2023 with the publication of the first version of TelegramScrap (Silva, 2023), a proprietary, free, and open-source tool that utilizes Telegram's Application Programming Interface (API) by Telethon library and organizes data extraction cycles from groups and open channels on Telegram. Over the months, the database was expanded and refined using four approaches to identifying conspiracy theory communities:

**(i) Use of keywords:** at the project's outset, keywords were listed for direct identification in the search engine of Brazilian groups and channels on Telegram, such as "apocalypse", "survivalism", "climate change", "flat earth", "conspiracy theory", "globalism", "new world order", "occultism", "esotericism", "alternative cures", "qAnon" "reptilians", "revisionism", "aliens", among others. This initial approach provided some communities whose titles and/or descriptions of groups and channels explicitly contained terms related to conspiracy theories. However, over time, it was possible to identify many other communities that the listed keywords did not encompass, some of which deliberately used altered characters to make it difficult for those searching for them on the network.

**(ii) Telegram channel recommendation mechanism:** over time, it was identified that Telegram channels (except groups) have a recommendation tab called "similar channels", where Telegram itself suggests ten channels that have some similarity with the channel being observed. Through this recommendation mechanism, it was possible to find more Brazilian conspiracy theory communities, with these being recommended by the platform itself.

**(iii) Snowball approach for invitation identification:** after some initial communities were accumulated for extraction, a proprietary algorithm was developed to identify URLs containing "t.me/", the prefix for any invitation to Telegram groups and channels. Accumulating a frequency of hundreds of thousands of links that met this criterion, the unique addresses were listed, and their repetitions counted. In this way, it was possible to investigate new Brazilian groups and channels mentioned in the messages of those already investigated, expanding the network. This process was repeated periodically to identify new communities aligned with conspiracy theory themes on Telegram.

**(iv) Expansion to tweets published on X mentioning invitations:** to further diversify the sources of Brazilian conspiracy theory communities on Telegram, a proprietary search query was developed to identify conspiracy theory-themed keywords using tweets published on X (formerly Twitter) that, in addition to containing one of the keywords, also included "t.me/", the prefix for any invitation to Telegram groups and channels, "https://x.com/search?q=lang%3Apt%20%22t.me%2F%22%20SEARCH-TERM".

With the implementation of community identification approaches for conspiracy theories developed over months of investigation and method refinement, it was possible to build a project database encompassing a total of 855 Brazilian conspiracy theory communities on Telegram (including other themes not covered in this study). These communities have collectively published 27,227,525 pieces of content from May 2016 (the first publications) to August 2024 (the period of this study), with a combined total of 2,290,621 users across the Brazilian communities. It is important to note that this volume of users includes two elements: first, it is a variable figure, as users can join and leave communities daily, so this value represents what was recorded on the publication extraction date; second, it is possible that the same user is a member of more than one group and, therefore, is counted more than once. In this context, while the volume remains significant, it may be slightly lower when considering the deduplicated number of citizens within these Brazilian conspiracy theory communities.

## 2.2. Data processing

With all the Brazilian conspiracy theory groups and channels on Telegram extracted, a manual classification was conducted considering the title and description of the community. If there was an explicit mention in the title or description of the community related to a specific theme, it was classified into one of the following categories: (i) "Anti-Science"; (ii) "Anti-Woke and Gender"; (iii) "Antivax"; (iv) "Apocalypse and Survivalism"; (v) "Climate Changes"; (vi) "Flat Earth"; (vii) "Globalism"; (viii) "New World Order"; (ix) "Occultism and Esotericism"; (x) "Off Label and Quackery"; (xi) "QAnon"; (xii) "Reptilians and Creatures"; (xiii) "Revisionism and Hate Speech"; (xiv) "UFO and Universe". If there was no explicit mention related to the themes in the title or description of the community, it was classified as (xv) "General Conspiracy". In the following table, we can observe the metrics related to the classification of these conspiracy theory communities in Brazil.

**Table 01.** Conspiracy theory communities in Brazil (metrics up to August 2024)

| Categories | Groups | Users | Contents | Comments | Total |
|---|---|---|---|---|---|
| **Anti-Science** | 22 | 58,138 | 187,585 | 784,331 | 971,916 |
| **Anti-Woke and Gender** | 43 | 154,391 | 276,018 | 1,017,412 | 1,293,430 |
| **Antivax** | 111 | 239,309 | 1,778,587 | 1,965,381 | 3,743,968 |
| **Apocalypse and Survivalism** | 33 | 109,266 | 915,584 | 429,476 | 1,345,060 |
| **Climate Changes** | 14 | 20,114 | 269,203 | 46,819 | 316,022 |
| **Flat Earth** | 33 | 38,563 | 354,200 | 1,025,039 | 1,379,239 |
| **General Conspiracy** | 127 | 498,190 | 2,671,440 | 3,498,492 | 6,169,932 |
| **Globalism** | 41 | 326,596 | 768,176 | 537,087 | 1,305,263 |
| **NWO** | 148 | 329,304 | 2,411,003 | 1,077,683 | 3,488,686 |
| **Occultism and Esotericism** | 39 | 82,872 | 927,708 | 2,098,357 | 3,026,065 |
| **Off Label and Quackery** | 84 | 201,342 | 929,156 | 733,638 | 1,662,794 |
| **QAnon** | 28 | 62,346 | 531,678 | 219,742 | 751,420 |
| **Reptilians and Creatures** | 19 | 82,290 | 96,262 | 62,342 | 158,604 |
| **Revisionism and Hate Speech** | 66 | 34,380 | 204,453 | 142,266 | 346,719 |
| **UFO and Universe** | 47 | 58,912 | 862,358 | 406,049 | 1,268,407 |
| **Total** | **855** | **2,296,013** | **13,183,411** | **14,044,114** | **27,227,525** |

Source: Own elaboration (2024).

With this volume of extracted data, it was possible to segment and present in this study only communities and content related to UFO, universe, reptilians and creatures themes. In parallel, other themes of Brazilian conspiracy theory communities were also addressed with studies aimed at characterizing the extent and dynamics of the network, which are openly and originally available on arXiv at Cornell University.

Additionally, it should be noted that only open communities were extracted, meaning those that are not only publicly identifiable but also do not require any request to access the content, being available to any user with a Telegram account who needs to join the group or channel. Furthermore, in compliance with Brazilian legislation, particularly the General Data Protection Law (LGPD), or Law No. 13,709/2018 (Brazilian law from 2018), which deals with privacy control and the use/treatment of personal data, all extracted data were anonymized for the purposes of analysis and investigation. Therefore, not even the identification of the communities is possible through this study, thus extending the user's privacy by anonymizing their data beyond the community itself to which they submitted by being in a public and open group or channel on Telegram.

## 2.3. Approaches to data analysis

A total of 66 selected communities focused on UFO, universe, reptilians and creatures themes, containing 1,427,011 publications and 141,202 combined users, will be analyzed. Four approaches will be used to investigate the conspiracy theory communities selected for the scope of this study. These metrics are detailed in the following table:

**Table 02.** Selected Communities for Analysis (metrics up to August 2024)

| **Categories** | **Groups** | **Users** | **Contents** | **Comments** | **Total** |
|---|---|---|---|---|---|
| **UFO and Universe** | 47 | 58.912 | 862.358 | 406.049 | 1.268.407 |
| **Reptilians and Creatures** | 19 | 82.290 | 96.262 | 62.342 | 158.604 |
| **Total** | **66** | **141.202** | **958.620** | **468.391** | **1.427.011** |

Source: Own elaboration (2024).

**(i) Network:** by developing a proprietary algorithm to identify messages containing the term "t.me/" (inviting users to join other communities), we propose to present volumes and connections observed on how **(a)** communities within the UFO, universe, reptilians and creatures theme circulate invitations for their users to explore more groups and channels within the same theme, reinforcing shared belief systems; and how **(b)** these same communities circulate invitations for their users to explore groups and channels dealing with other conspiracy theories, distinct from their explicit purpose. This approach is valuable for observing whether these communities use themselves as a source of legitimation and reference and/or rely on other conspiracy theory themes, even opening doors for their users to explore other conspiracies. Furthermore, it is worth mentioning the study by Rocha *et al.* (2024), where a network identification approach was also applied in Telegram communities, but by observing similar content based on an ID generated for each unique message and its similar ones;

**(ii) Time series:** the "Pandas" library (McKinney, 2010) is used to organize the investigation data frames, observing **(a)** the volume of publications over the months; and **(b)** the volume of engagement over the months, considering metadata of views, reactions, and comments collected during extraction. In addition to volumetry, the "Plotly" library (Plotly Technologies Inc., 2015) enabled the graphical representation of this variation;

**(iii) Content analysis:** in addition to the general word frequency analysis, time series are applied to the variation of the most frequent words over the semesters—observing from May 2016 (initial publications) to August 2024 (when this study was conducted). With the "Pandas" (McKinney, 2010) and "WordCloud" (Mueller, 2020) libraries, the results are presented both volumetrically and graphically;

**(iv) Thematic agenda overlap:** following the approach proposed by Silva & Sátiro (2024) for identifying thematic agenda overlap in Telegram communities, we used the "BERTopic" model (Grootendorst, 2020). BERTopic is a topic modeling algorithm that facilitates the generation of thematic representations from large amounts of text. First, the algorithm extracts document embeddings using sentence transformer models, such as “all-MiniLM-L6-v2”. These embeddings are then reduced in dimensionality using techniques like “UMAP”, facilitating the clustering process. Clustering is performed using “HDBSCAN”, a density-based technique that identifies clusters of different shapes and sizes, as well as outliers. Subsequently, the documents are tokenized and represented in a bag-of-words structure, which is normalized (L1) to account for size differences between clusters. The topic representation is refined using a modified version of “TF-IDF”, called “Class-TF-IDF”, which considers the importance of words within each cluster (Grootendorst, 2020). It is important to note that before applying BERTopic, we cleaned the dataset by removing Portuguese “stopwords” using “NLTK” (Loper & Bird, 2002). For topic modeling, we used the “loky” backend to optimize performance during data fitting and transformation.

In summary, the methodology applied ranged from data extraction using the own tool TelegramScrap (Silva, 2023) to the processing and analysis of the collected data, employing various approaches to identify and classify Brazilian conspiracy theory communities on Telegram. Each stage was carefully implemented to ensure data integrity and respect for user privacy, as mandated by Brazilian legislation. The results of this data will be presented below, aiming to reveal the dynamics and characteristics of the studied communities.

## 3. Results

The results are detailed below in the order outlined in the methodology, beginning with the characterization of the network and its sources of legitimation and reference, progressing to the time series, incorporating content analysis, and concluding with the identification of thematic agenda overlap among the conspiracy theory communities.

### 3.1. Network

The analysis of internal networks among communities discussing UFOs, the universe, reptilians, and other mysterious creatures reveals a complex web of interconnections that transcends the boundaries between different conspiracy narratives. These graphs expose how seemingly distinct themes, such as extraterrestrial life and the existence of reptilian beings, are intrinsically linked, forming a cohesive conspiratorial ecosystem. The communities that uphold these beliefs act not only as vehicles for the dissemination of ideas but also as gateways into and out of a vast universe of related theories, suggesting that initial engagement with one of these themes often leads to deeper immersion in other interrelated conspiracy theories. By analyzing the connections and invitation flows between these communities, we can observe how these narratives reinforce each other, creating an environment where beliefs in supernatural phenomena and global conspiracies are amplified and perpetuated. These networks not only sustain individual theories but also construct a worldview where the inexplicable and the occult gain coherence and meaning within a narrative structure that challenges scientific logic and institutional authority.

[ english version / português abaixo ]

**Figure 01.** Internal network between UFO, universe, reptilians and creatures communities

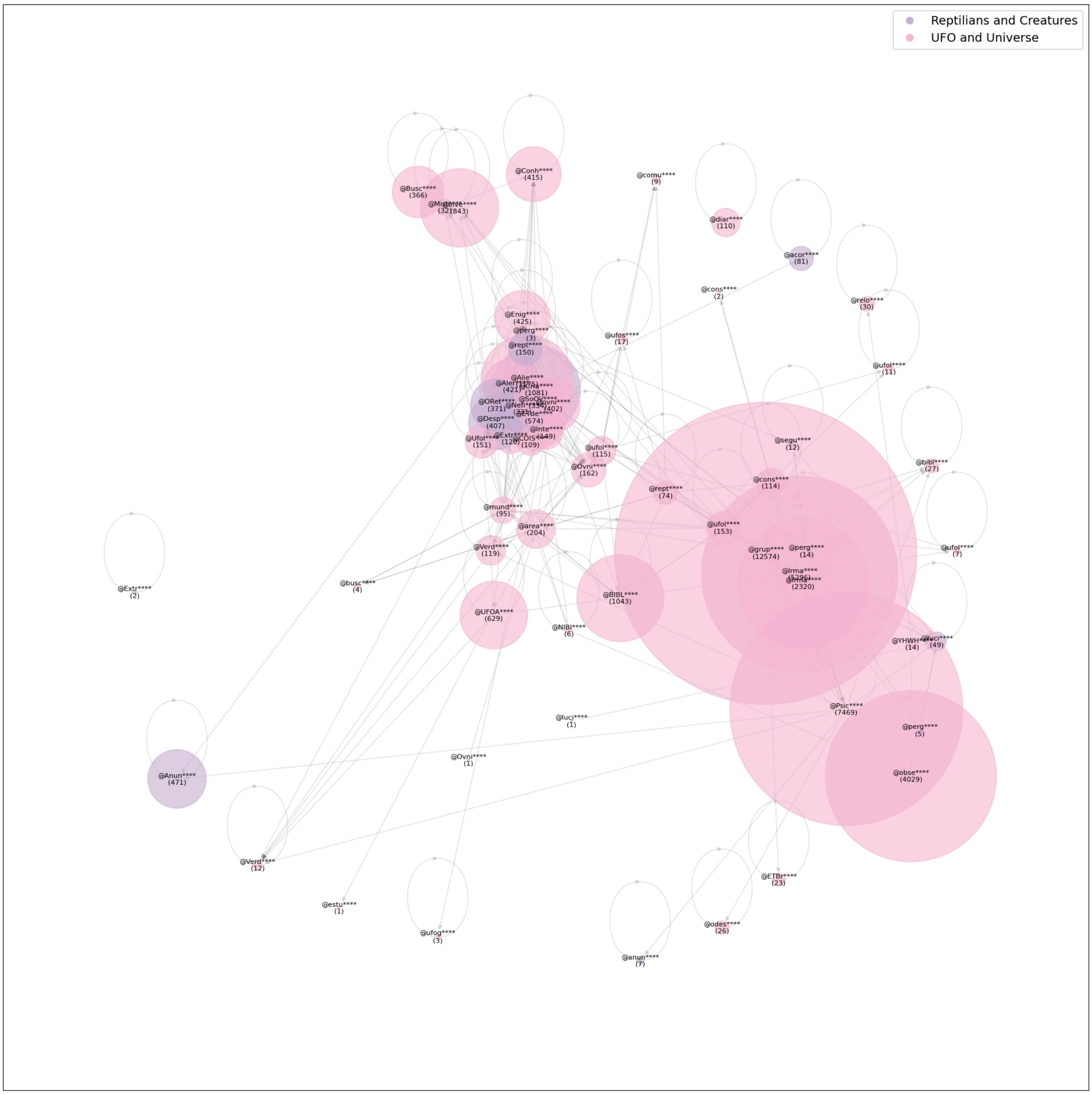

Source: Own elaboration (2024).

This graph depicts the internal network among communities discussing UFOs, the universe, reptilians, and mysterious creatures. The network, composed of two main hubs, reveals how theories related to extraterrestrial life and conspiracies about reptilian beings are deeply connected. The interactions suggest that these communities feed off each other, with interest in UFOs often leading to exploration of more radical theories about reptilians and unknown creatures. The large nodes indicate that certain groups have a disproportionate influence in propagating these narratives, serving as reference centers for smaller communities. The network, while less dense than some others, still demonstrates strong internal cohesion, suggesting that followers are constantly exposed to a variety of related theories that reinforce and expand the belief in the existence of extraterrestrial conspiracies.

[ english version / português abaixo ]

**Figure 02.** Network of communities that open doors to the theme (gateway)

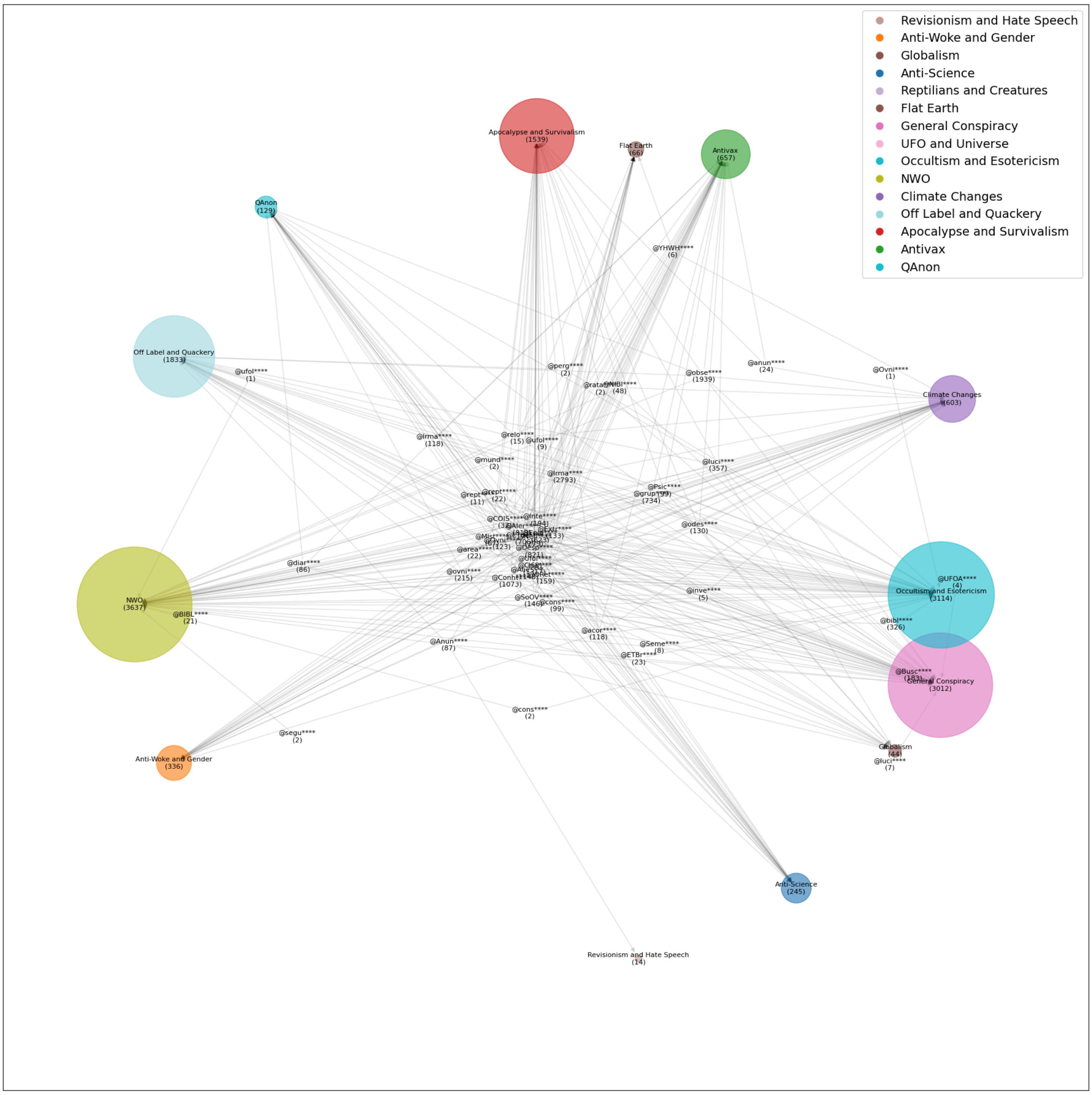

Source: Own elaboration (2024).

This figure presents the communities that act as gateways to discussions about UFOs, reptilians, and other mysterious creatures. The network is characterized by a strong concentration around a few central communities that attract those curious about extraterrestrial life and theories about reptilian beings. The graph highlights how interest in UFOs and the universe can easily lead to deeper explorations of conspiratorial theories about reptilians and creatures. These larger communities not only serve as hubs for disseminating these narratives but also connect followers to a vast network of related theories, suggesting that once within this network, individuals are exposed to a wide range of conspiratorial ideas, reinforcing their beliefs and expanding their engagement with other similar communities.

**Figure 03.** Network of communities whose theme opens doors (exit point)

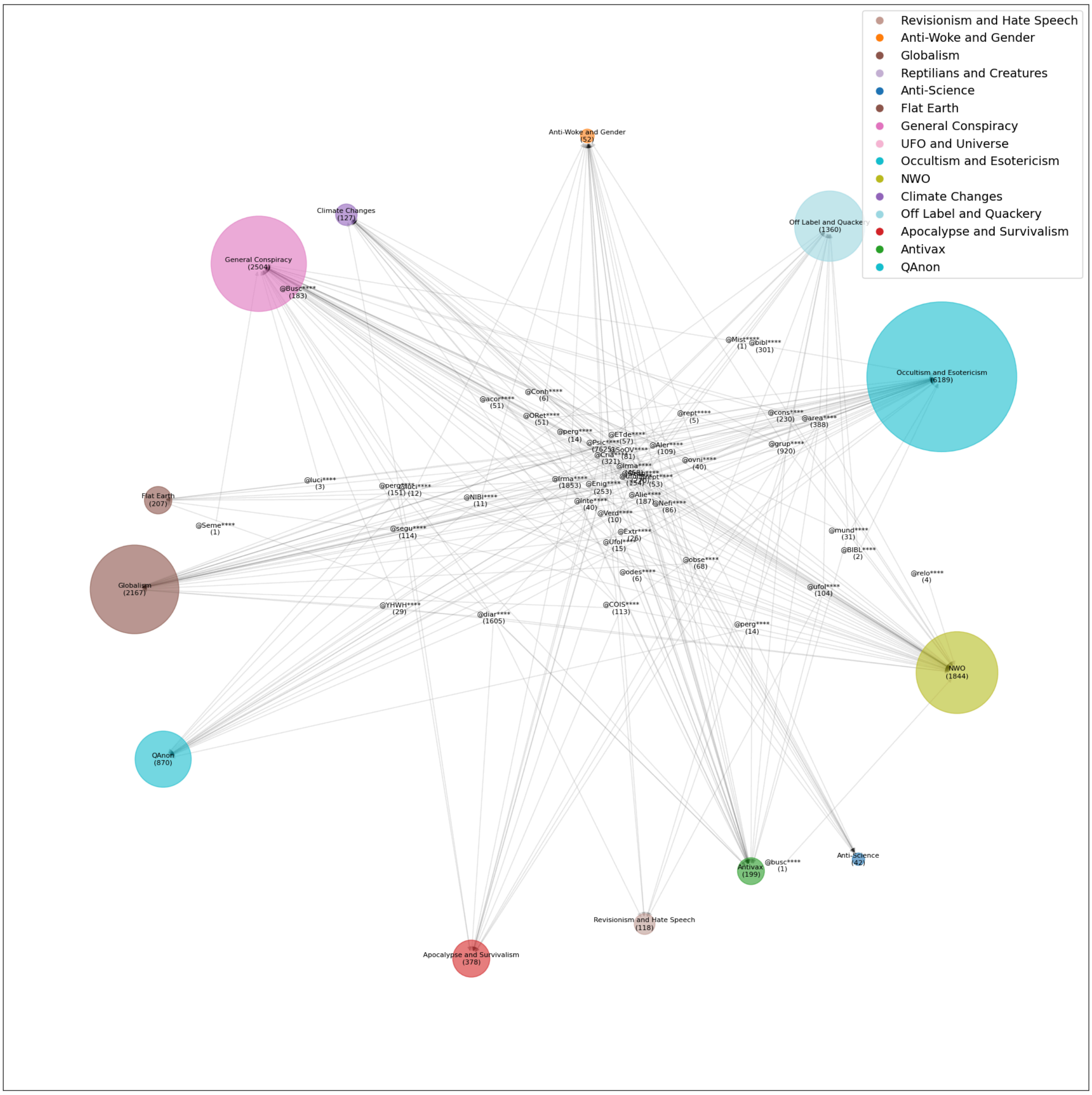

Source: Own elaboration (2024).

The network graph focused on UFOs, the universe, reptilians, and creatures highlights how these themes, which may seem isolated, are deeply interconnected with other conspiracy theories. The connections between these communities and other themes, such as Occultism, Flat Earth, and Globalism, suggest that discussions about UFOs and extraterrestrial life often pave the way for exploration of broader narratives of global conspiracy and secret government control. The figure suggests that initial interest in UFO theories can lead followers to immerse themselves in more complex and interconnected theories, even relating to occult and esoteric beliefs, where supernatural explanations and distrust of authorities intertwine, creating a belief ecosystem anchored in these convictions.

[ english version / português abaixo ]

**Figure 04.** Flow of invitation links between UFO and universe

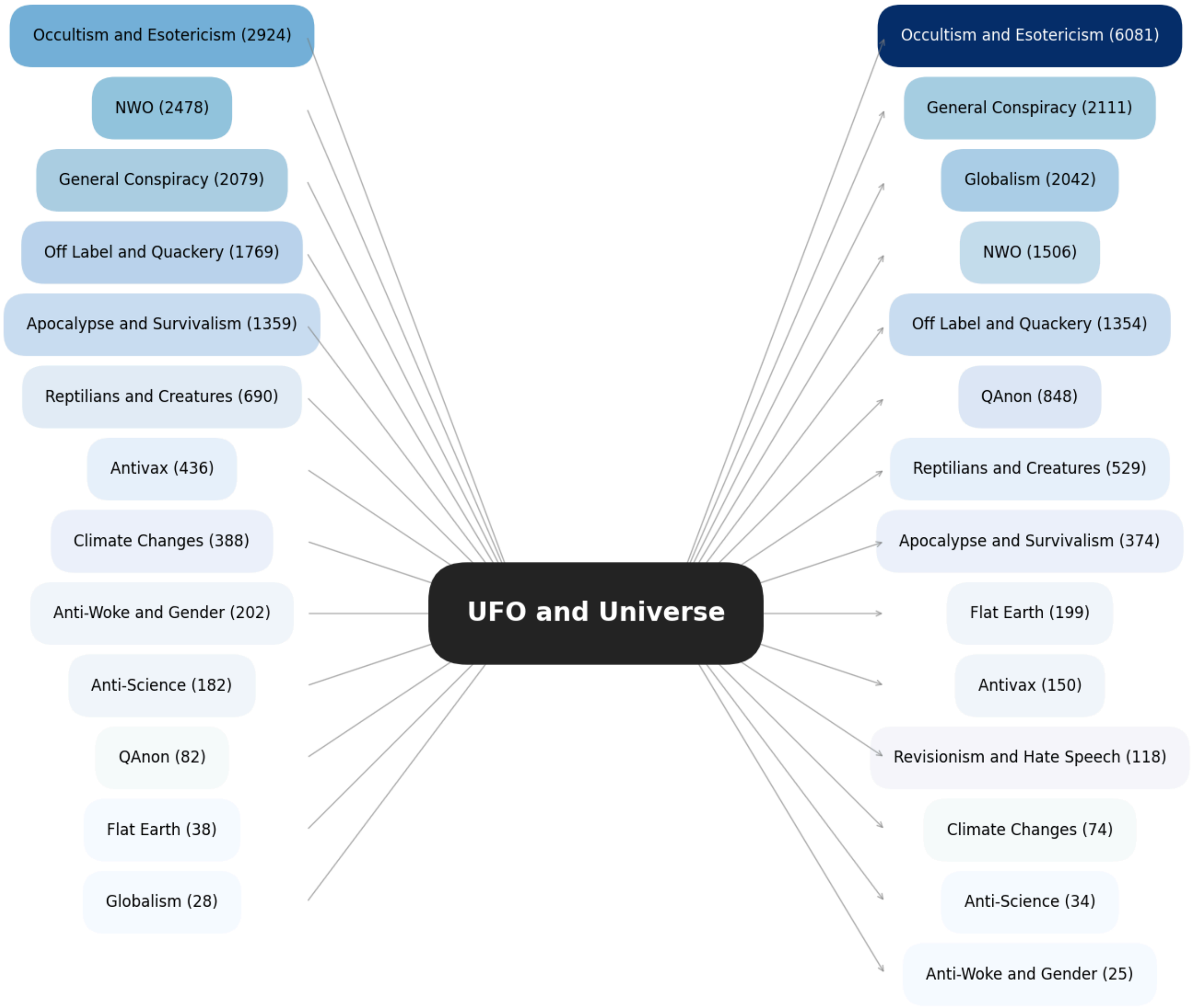

Source: Own elaboration (2024).

The UFO and Universe graph shows that this theme is strongly interconnected with Occultism and Esotericism (2,924 links) and General Conspiracies (2,079 links), suggesting that belief in extraterrestrial life and unexplained phenomena serves as an entry point into a larger network of theories that question perceived reality. The connection with Off-Label Medications (1,769 links) and Apocalypse and Survival (1,359 links) points to a trend where distrust of science and preparation for cataclysmic events intertwine with interest in phenomena unexplained by conventional science. When UFOs and the Universe issue invitations to Occultism and Esotericism (6,081 links) and General Conspiracies (2,111 links), the analysis indicates that these groups function as disseminators of a worldview where the inexplicable not only exists but is part of a larger plan of manipulation and concealment by powerful forces. This cycle of belief not only sustains the conspiracy narrative but also creates an environment where adherents are encouraged to constantly seek out new "truths" that validate their worldview. This shows how the UFO and Universe theme can function as an amplifier of theories that challenge the rational and scientific worldview, opening doors to widespread acceptance of conspiracy narratives.

[ english version / português abaixo ]

**Figure 05.** Flow of invitation links between reptilians and creatures

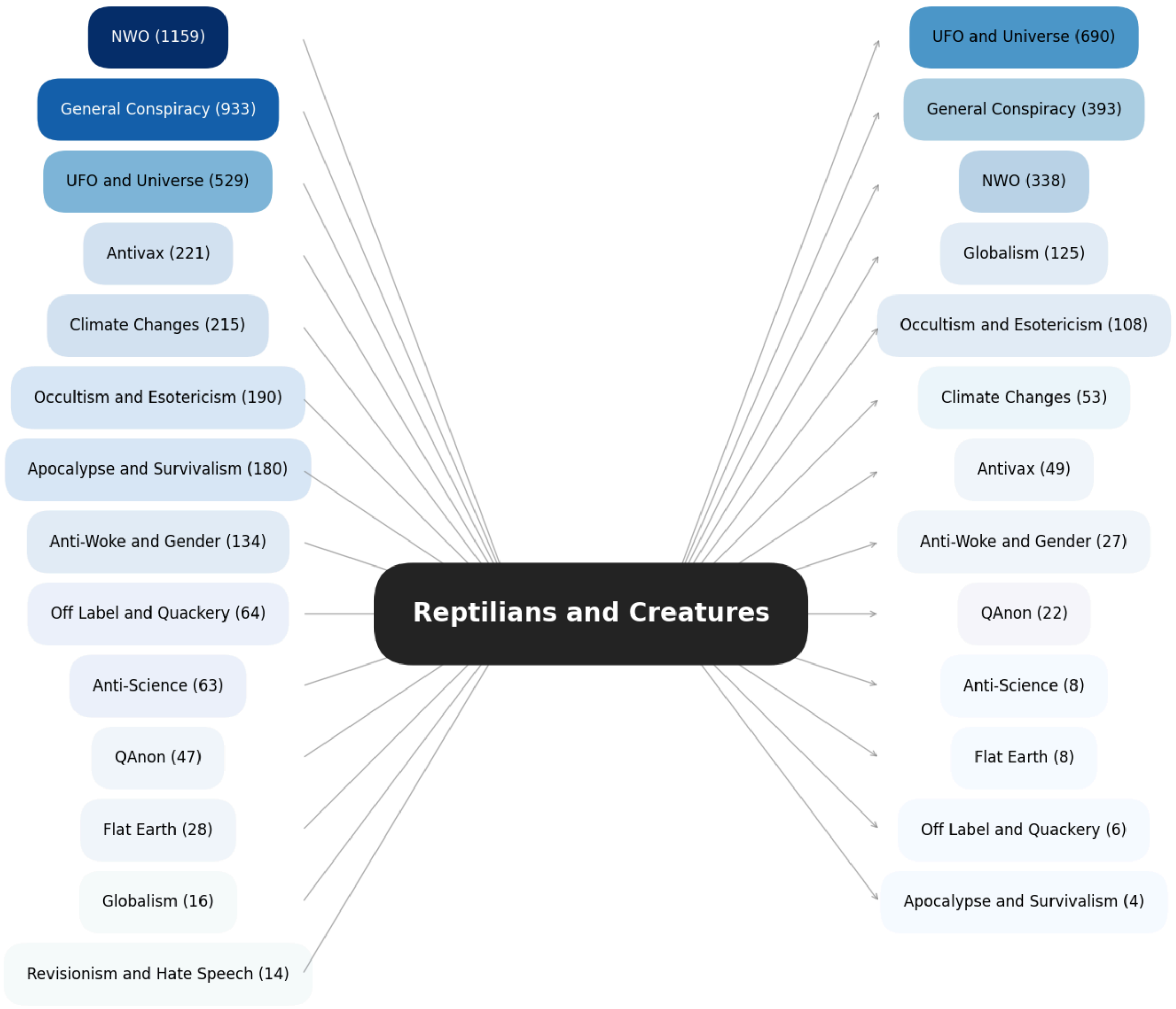

Source: Own elaboration (2024).

The Reptilian and Creature graph highlights the New World Order (NOM) as one of the main sources of invitations (1,159 links), suggesting that theories about reptilian creatures are deeply rooted in narratives of global control and domination. This link reflects a fusion of beliefs where the "other" is demonized and seen as part of a broader plan to subjugate humanity. The fact that Reptilians and Creatures primarily issue invitations to UFOs and the Universe (690 links) and General Conspiracies (393 links) indicates that this narrative acts as a bridge between belief in non-human entities and a broader conspiratorial worldview. This dynamic implies that the acceptance of the existence of reptilians is not just an isolated belief but part of a broader process of radicalization, where each narrative is used to reinforce the other. Thus, the Reptilians and Creatures theme not only introduces new adherents to the conspiratorial world but also serves as a means of perpetuating and expanding the influence of other theories, creating a cycle of belief and distrust that is difficult to break.

[ english version / português abaixo ]

### 3.2. Time series

In the analysis of the temporal series that follows, we will observe how discussions about UFOs and the Universe, as well as Reptilians and Creatures, intensified after 2020. The graph shows that mentions of UFOs saw an impressive increase, driven by new government reports and the growing public interest in extraterrestrial phenomena. Simultaneously, although on a smaller scale, discussions about Reptilians and other creatures also grew, reflecting a continued interest in more extreme theories. This data suggests that even in a context of increasing interest in extraterrestrial topics supported by governmental evidence, more radical theories continue to maintain their relevance within specific niches.

**Figure 06.** Line graph over the period

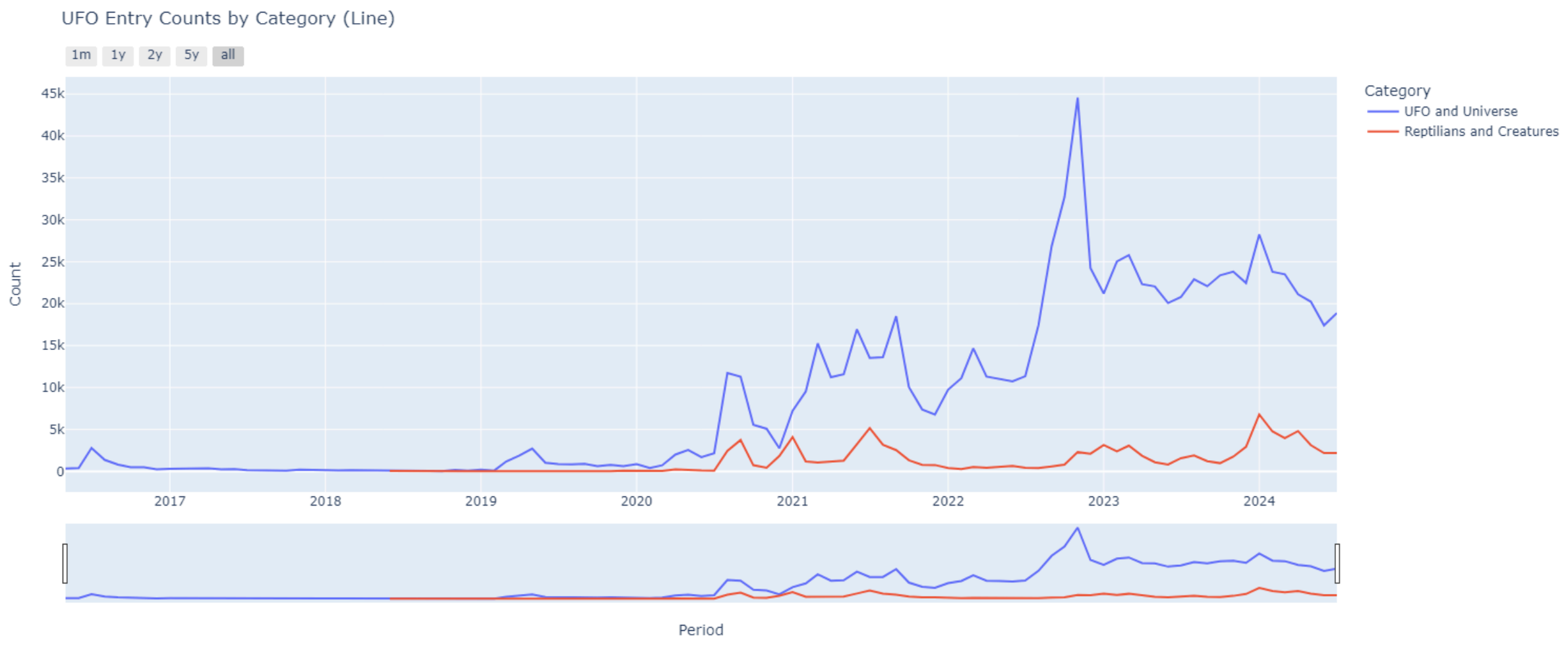

Source: Own elaboration (2024).

Publications about UFOs and the Universe experienced a 4,400% increase in mentions between 2020 and June 2021, rising from 1,000 to 45,000 mentions. This increase was driven by the release of government reports on UFO sightings, which revitalized public interest. Reptilians and Creatures saw a smaller increase in publications, about 1,000% between 2020 and 2021, rising from 1,000 to 10,000 mentions. Although smaller in absolute volume compared to UFOs, the percentage growth is significant, indicating renewed interest in more extreme theories regarding the presence of extraterrestrial creatures. After the peaks in 2021, both topics show stabilization at levels higher than pre-2020, with UFOs and the Universe still maintaining around 20,000 monthly mentions, and Reptilians and Creatures around 5,000, suggesting that these narratives continue to capture public interest.

[ english version / português abaixo ]

**Figure 07.** Absolute area chart over the period

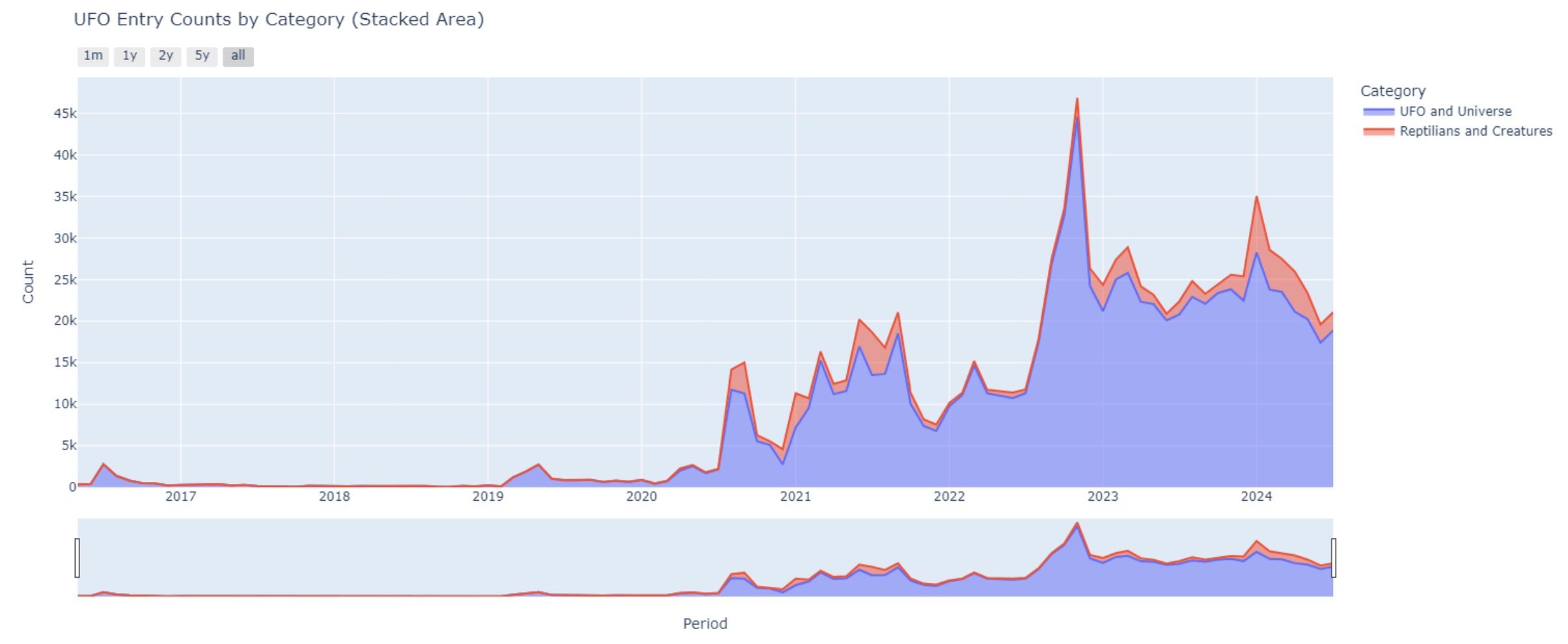

Source: Own elaboration (2024).

The absolute area graph reveals that discussions around UFOs and the Universe and Reptilians and Creatures grew considerably after 2020, with a sharp peak in 2022. UFOs and the Universe dominate the graph in absolute terms, especially after new government reports on unidentified aerial phenomena and the growing media interest in extraterrestrial life. The peak coincides with the release of reports on UFOs by the U.S. government, which increased public interest and revitalized discussions about alien life and related conspiracies. Although Reptilians and Creatures have a smaller presence, their growth during the same periods suggests that these narratives are intertwined, often linked to theories of global control by non-human entities.

**Figure 08.** Relative area chart over the period

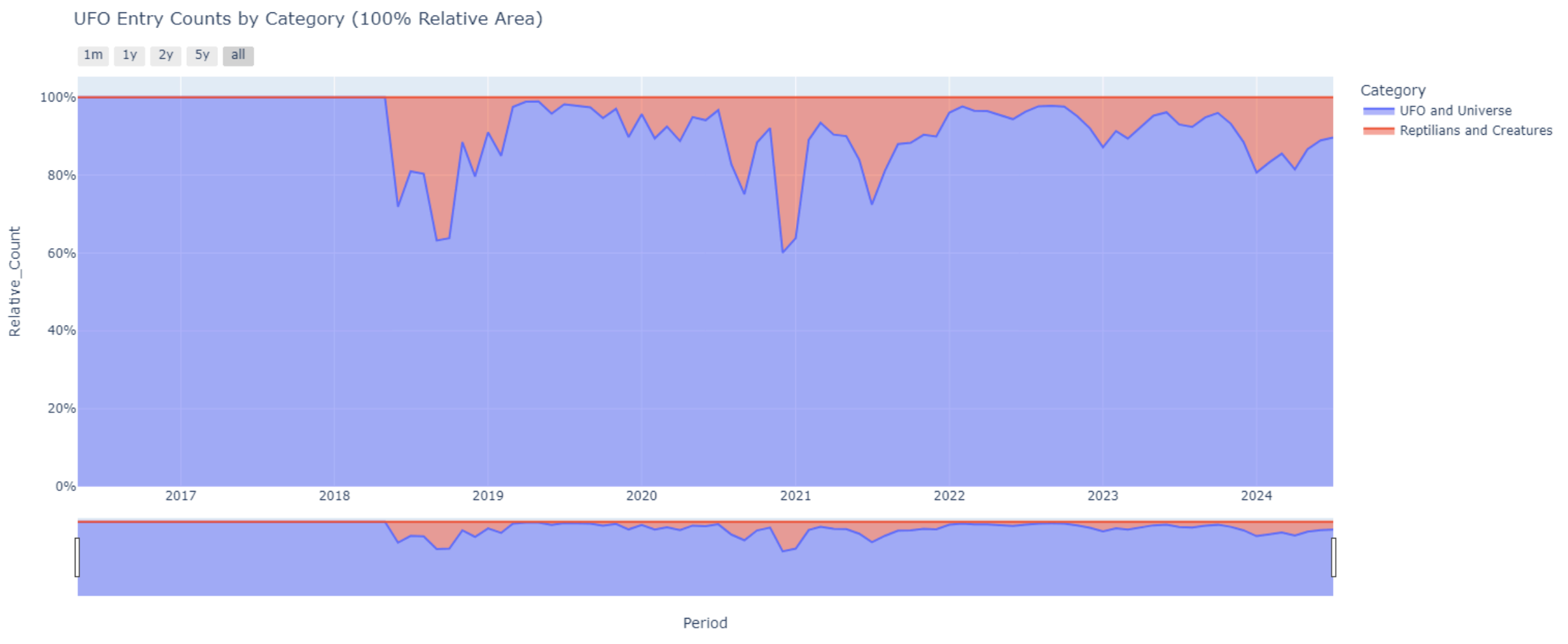

Source: Own elaboration (2024).

The relative area graph of the period highlights how publications from communities focused on UFOs and the Universe continue to dominate discussions over time, representing the majority of entries, especially after 2021. Reptilians and Creatures, while less predominant, still maintain a constant presence, with smaller peaks indicating the persistence of this narrative within conspiratorial discourse. This graph suggests that while discussions about UFOs gain more mainstream relevance, especially with the support of government reports, theories about reptilians remain more confined to specific niches but continue to be a significant component within the broader discussions of global conspiracies. The analysis shows how different facets of extraterrestrial conspiracy theories can coexist and reinforce each other within the same disinformation ecosystem.

### 3.3. Content analysis

The content analysis of the UFO, universe and reptilian communities through word clouds allows us to explore how these themes interact and consolidate in the collective imagination of their members over time. The words highlighted, such as “life”, “light”, “time”, and “God”, reveal a deep connection between these beliefs and a worldview that goes beyond the merely scientific, involving spiritual, esoteric, and existential dimensions. These narratives intertwine with a sense of urgency and transformation, reflected in the terms “now” and “energy”, which indicate both a search for understanding and preparation for cosmic events or global changes. The analysis of these keywords offers a deeper understanding of how these communities structure their discourses and maintain their beliefs, often in opposition to conventional science, and how they use platforms to perpetuate these ideas.

[ english version / português abaixo ]

**Figure 09.** Consolidated word cloud for UFO, universe, reptilians and creatures

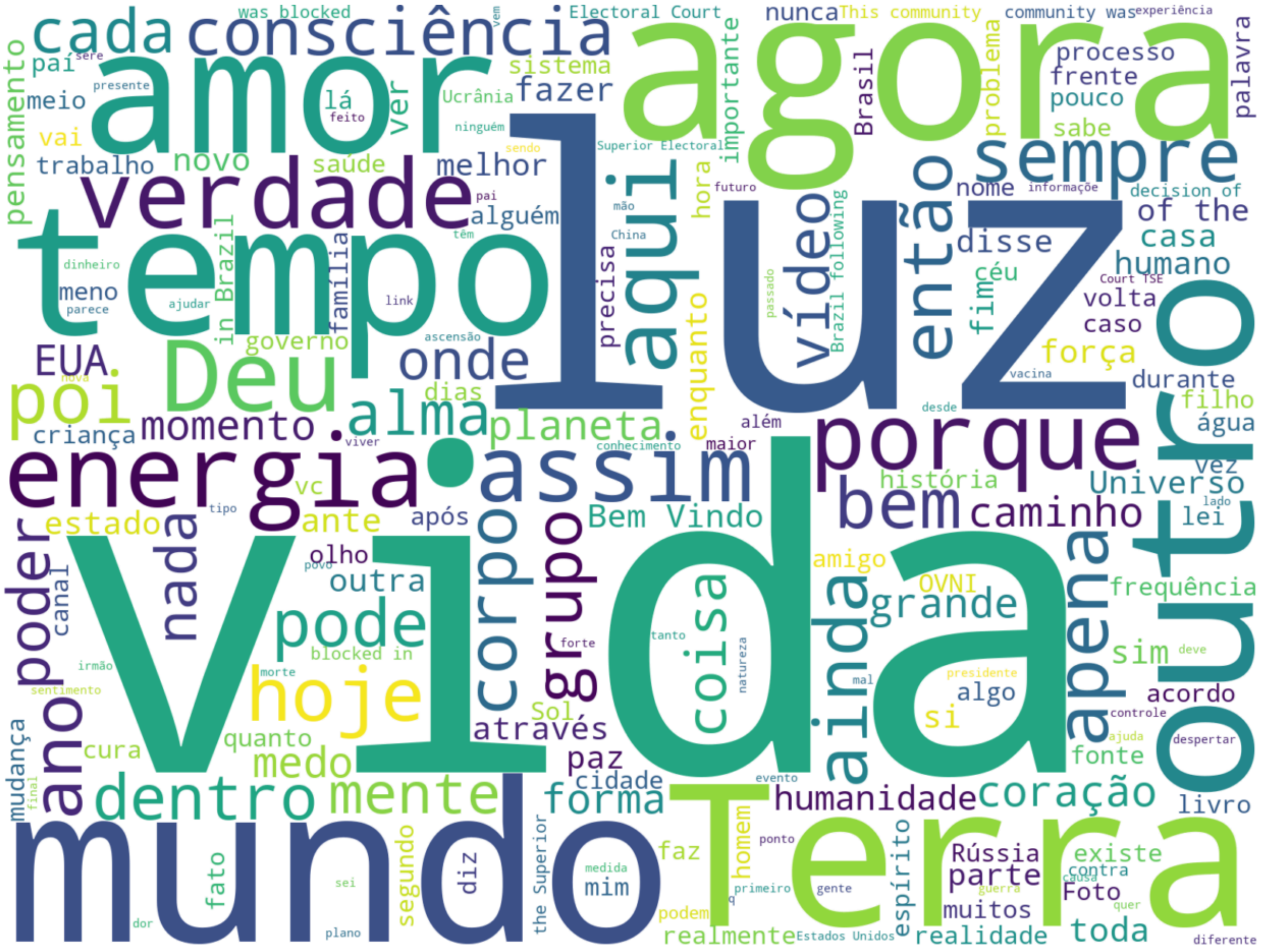

Source: Own elaboration (2024).

The consolidated word cloud of the themes of UFOs, the universe, reptilians, and creatures shows a concentration on terms such as “life”, “light”, “time”, “God”, and “energy”. The word “life” appears prominently, suggesting that the discussion revolves around concepts that transcend the physical and material, also involving spiritual and existential issues. The recurring presence of “light” and “energy” indicates a concern with invisible or cosmic forces, which are often interpreted as essential components of a reality greater than the perceived one. Terms like “God” and “truth” reinforce the idea that these discussions are shaped by a search for deep meanings and hidden truths, often associated with conspiracy theories or alternative views of the universe. The use of “time” and “now” suggests a sense of urgency, possibly related to predicted events or expected transformations in the global or cosmic scenario.

[ english version / português abaixo ]

**Chart 01.** Temporal word cloud series for UFO and universe

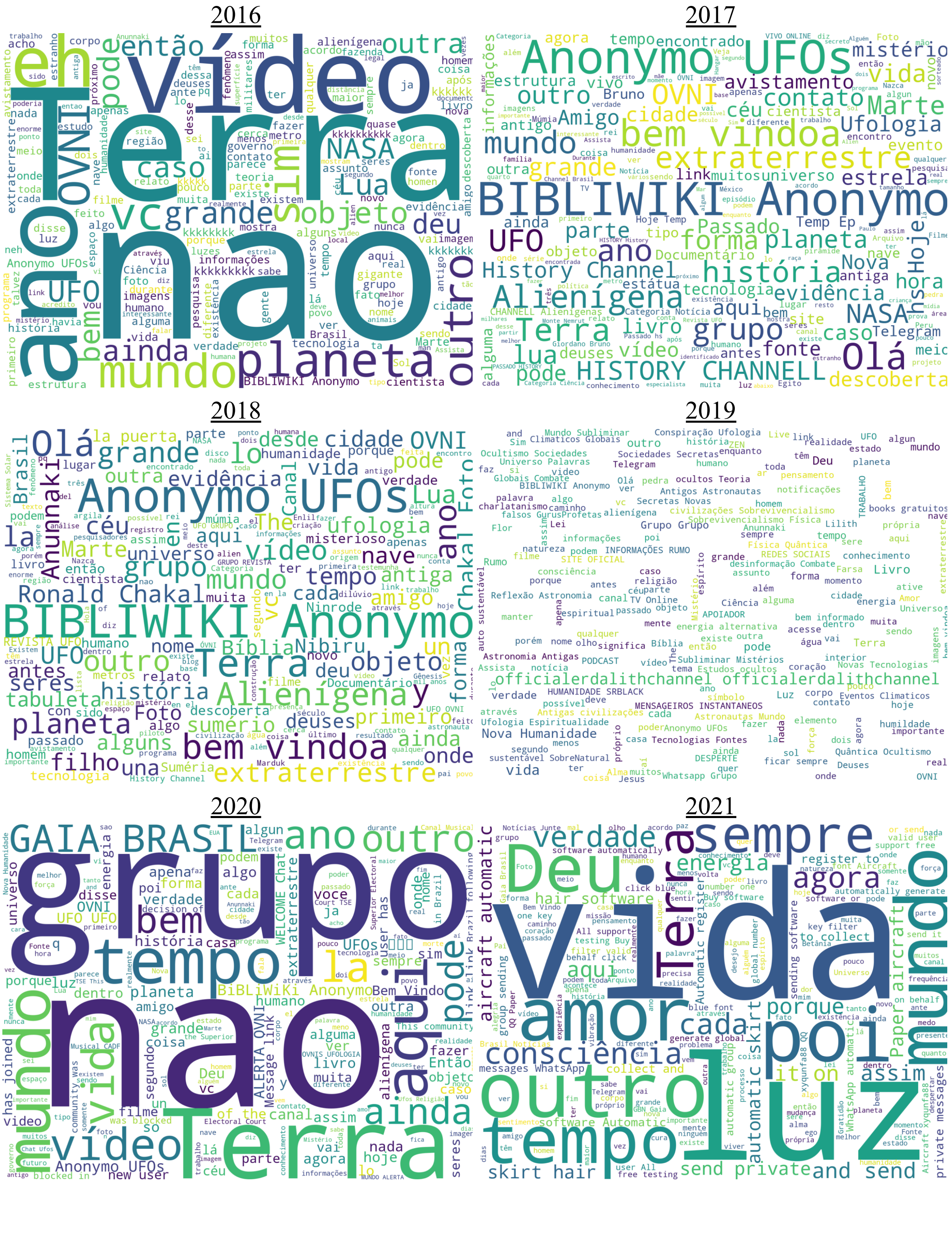

[ english version / português abaixo ]

2022

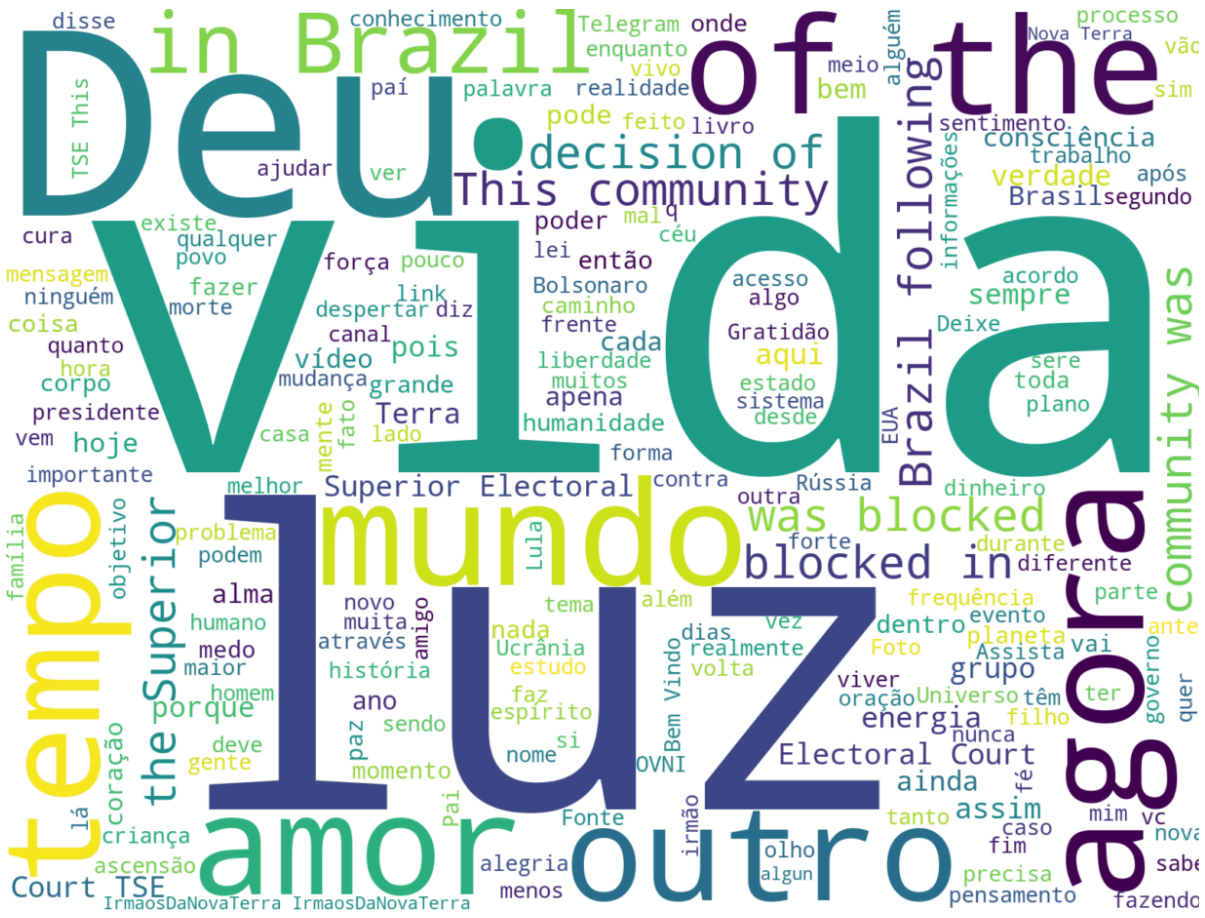

2023

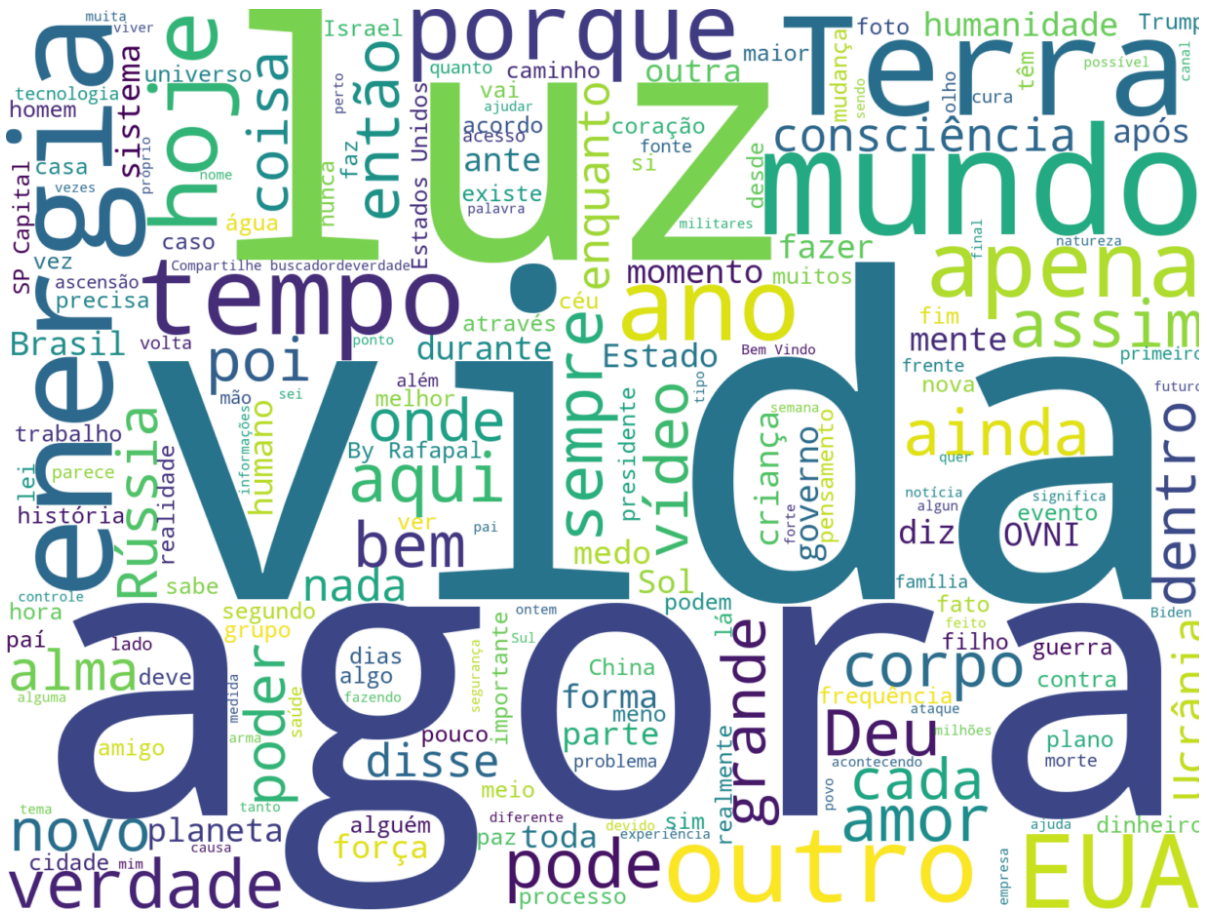

2024

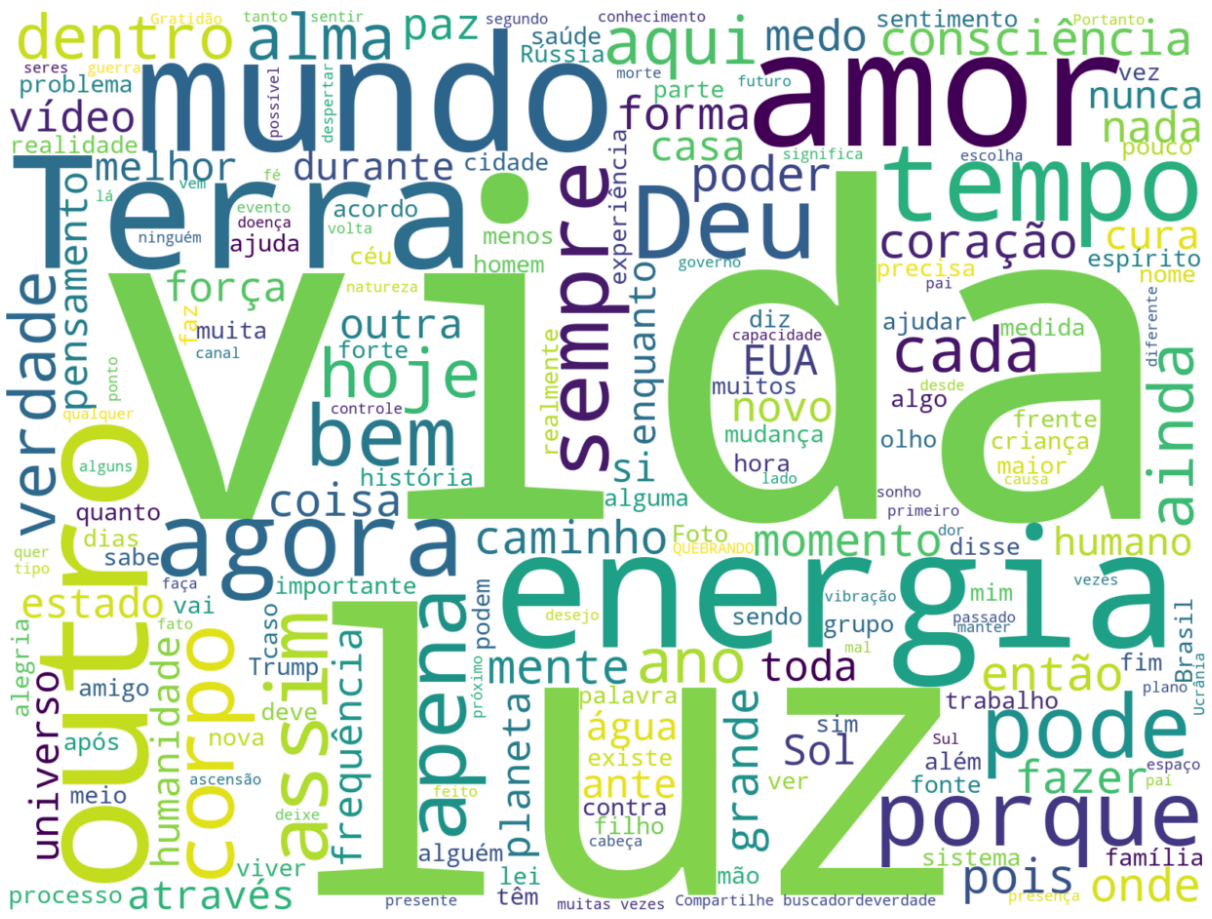

Source: Own elaboration (2024).

In Chart 01, the temporal analysis of the words reveals how discussions about UFOs and the universe evolved between 2016 and 2024. In 2016, terms like “video” and “guy” dominate the discussions, possibly indicating a focus on visual evidence and personal debates about the existence of extraterrestrial life. Over the years, a transition to more abstract and spiritual themes is observed, with the inclusion of words like “life”, “light”, and “God” in 2020, reflecting a shift from the concrete to the esoteric. In 2022 and 2023, discussions seem to consolidate around the idea of “life” and “time”, suggesting a growing concern with existential issues and the relationship of humanity with the universe. The year 2024 maintains this trend, with the inclusion of “energy” and “consciousness”, reinforcing the view that these communities seek an understanding of the cosmos.

[ english version / português abaixo ]

**Chart 02.** Temporal word cloud series for reptilians and creatures

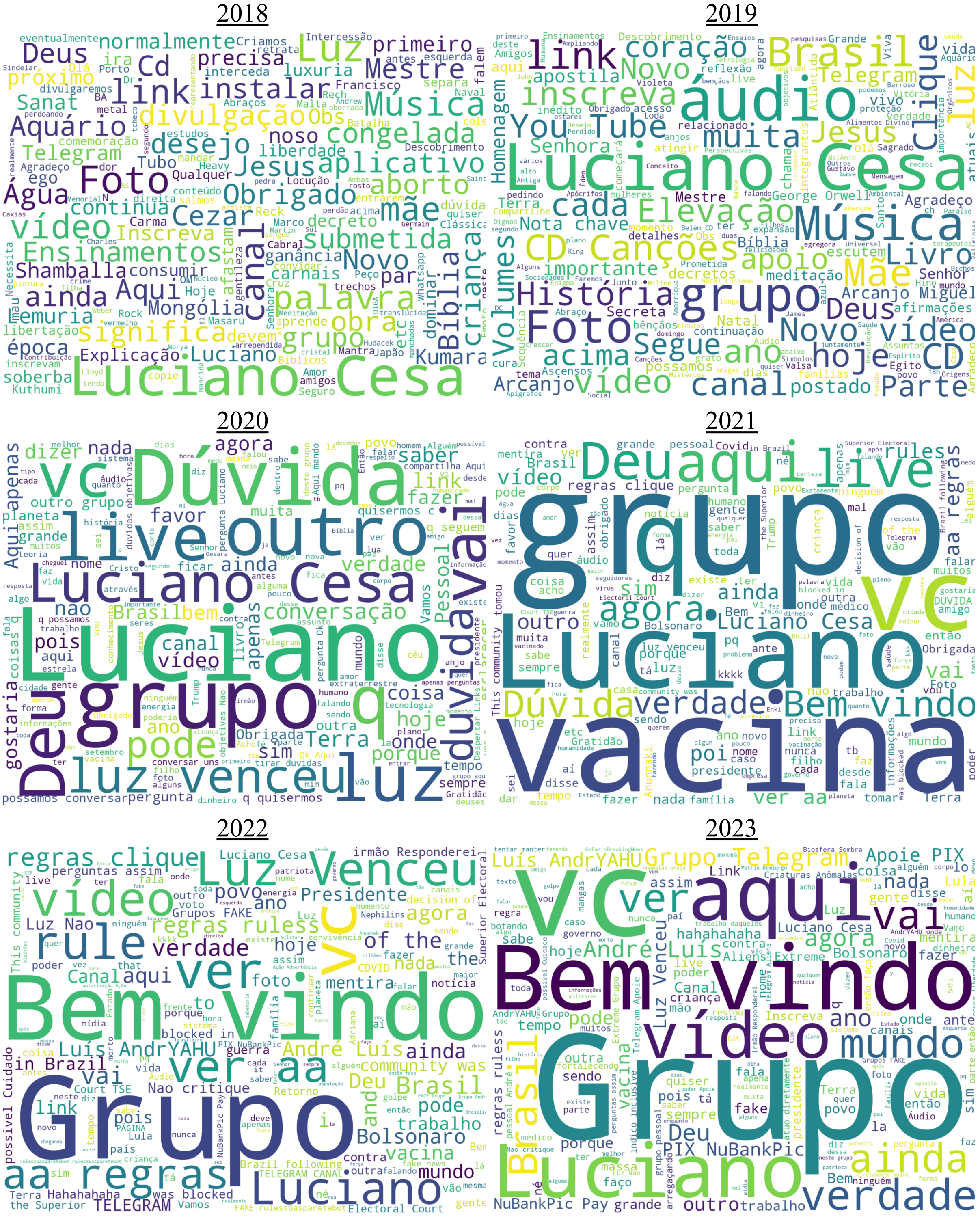

[ english version / português abaixo ]

2024

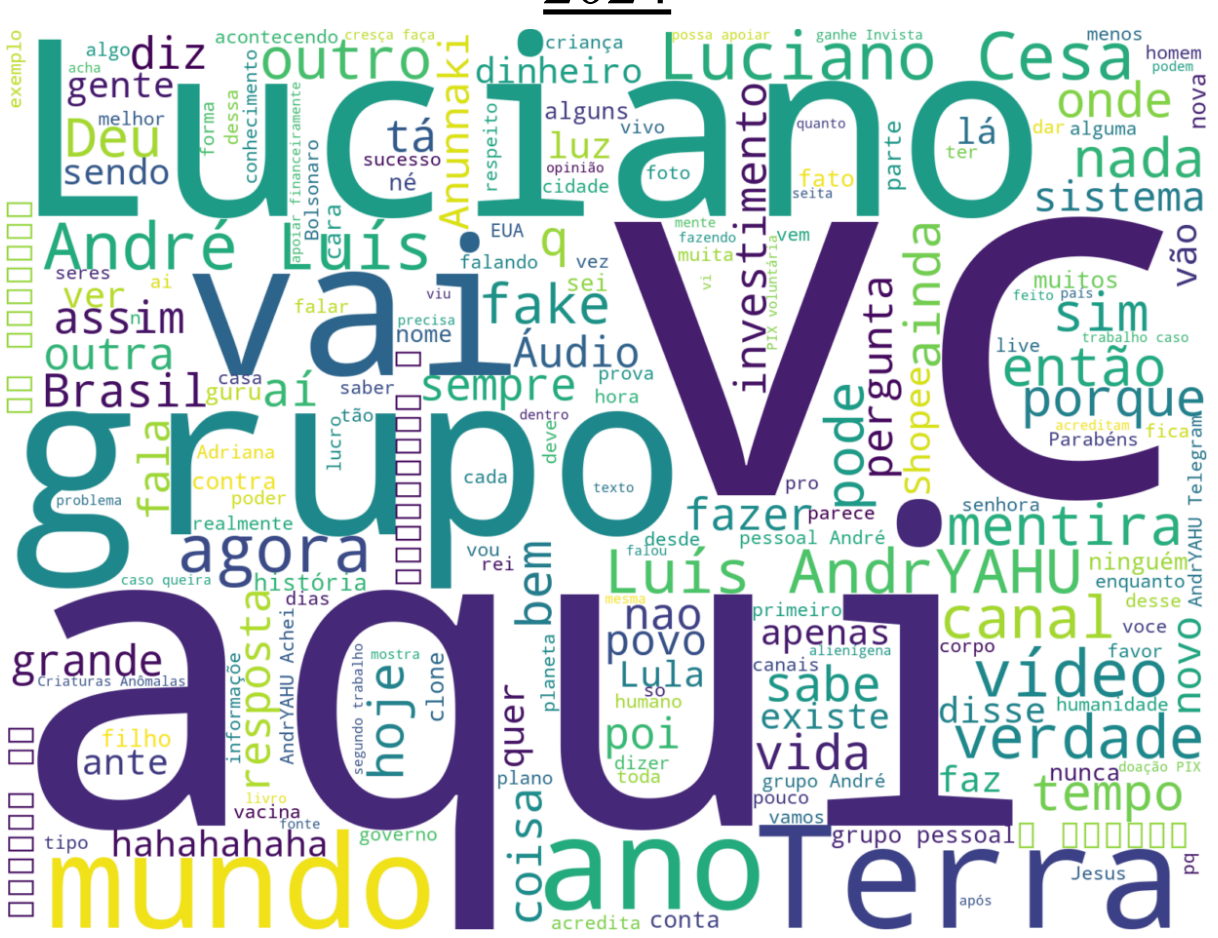

Source: Own elaboration (2024).

Chart 02 presents the evolution of discussions about reptilians and creatures over time, from 2018 to 2024. In 2018, the word “Luciano Cesa” (spiritual coach) stands out, indicating the importance of specific figures within these communities for spreading narratives. The word “group” also appears frequently, suggesting that these discussions are largely collaborative and occur in community environments, such as online groups. Over the years, we see a repetition and reaffirmation of these terms, especially “group” and “Luciano Cesa”, which may indicate the formation of subcultures or specific groups within the broader communities that follow particular leaders or influencers. The presence of “vaccine” in 2022 and 2023 suggests an intersection between reptilian narratives and discussions about public health and the pandemic, highlighting how these conspiracy theories can adapt to current events to maintain relevance and appeal. In 2024, the consistency of terms like “life”, “truth”, and “group” suggests a continuity of these narratives, where the search for an alternative understanding of the world continues to be mediated by figures and cohesive communities.

### 3.4. Thematic agenda overlap

The following figures show how UFOs, Universe, Reptilians, and Creatures themes overlap in discussions within conspiracy theory communities. These themes not only attract the interest of individuals curious about the unknown but are also used to sustain narratives of distrust toward institutions and global events. By integrating topics such as extraterrestrials, mysterious disappearances, and planned natural disasters, these communities construct a cohesive discourse that reinforces conspiratorial beliefs and complicates factual correction.

[ english version / português abaixo ]

**Figure 10.** UFO themes

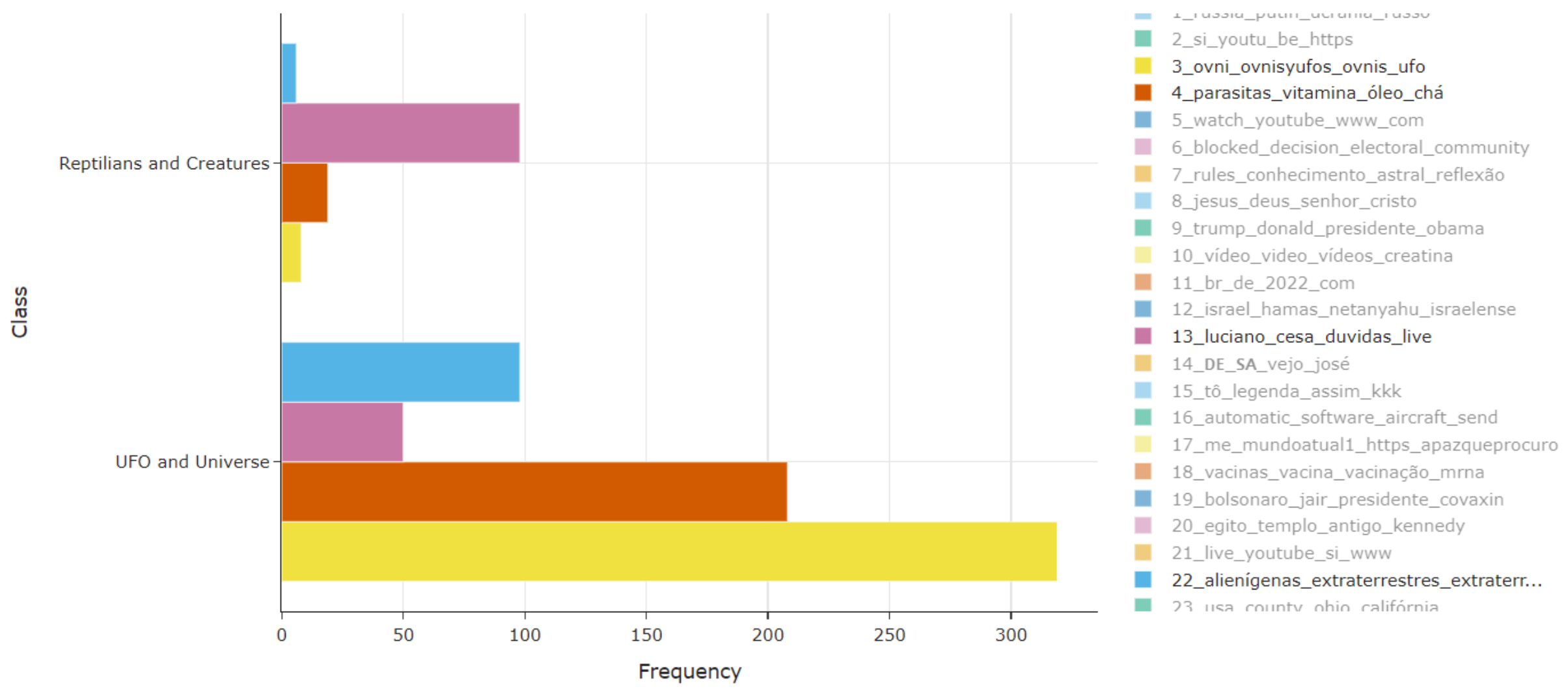

Source: Own elaboration (2024).

Figure 10 highlights how topics related to UFOs and the Universe dominate discussions in these communities, compared to themes about Reptilians and Creatures. Terms like “aliens” and “extraterrestrials” appear frequently, indicating that these communities are deeply involved in narratives suggesting interactions or conspiracies involving beings from other planets. These discussions often blend with theories about governments concealing information, creating a narrative that sustains widespread distrust toward authorities.

**Figure 11.** Geopolitical and electoral disputes themes

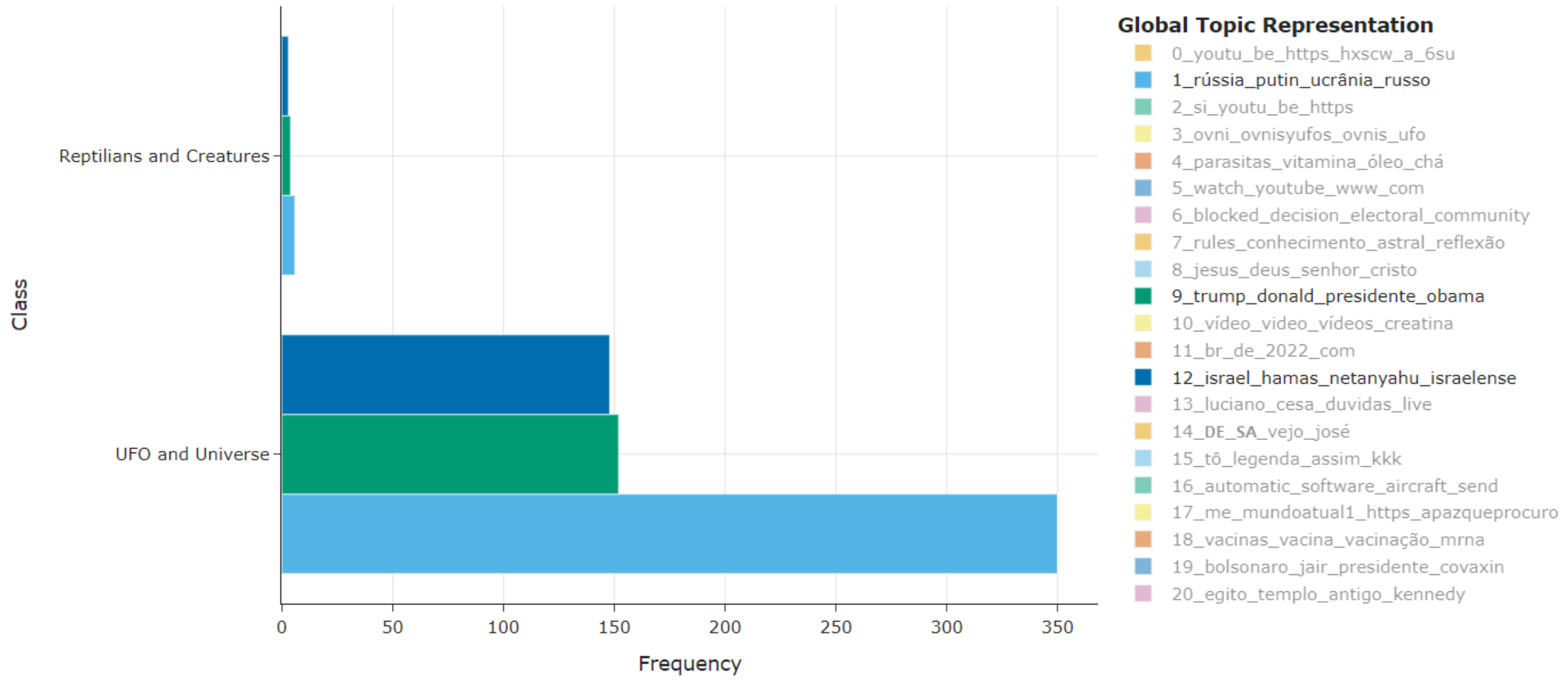

Source: Own elaboration (2024).

In Figure 11, the overlap of geopolitical themes with discussions about UFOs and the Universe is evident. Topics like “Russia” and “Putin” are frequently associated with global conspiracy narratives, suggesting that geopolitical events are interpreted as part of a larger plan involving extraterrestrial beings. The integration of these themes with electoral disputes reflects the politicization of conspiracy theories, where political leaders are seen as central figures in conspiracies involving forces beyond our planet.

**Figure 12.** Alleged disappearances or plane crashes themes

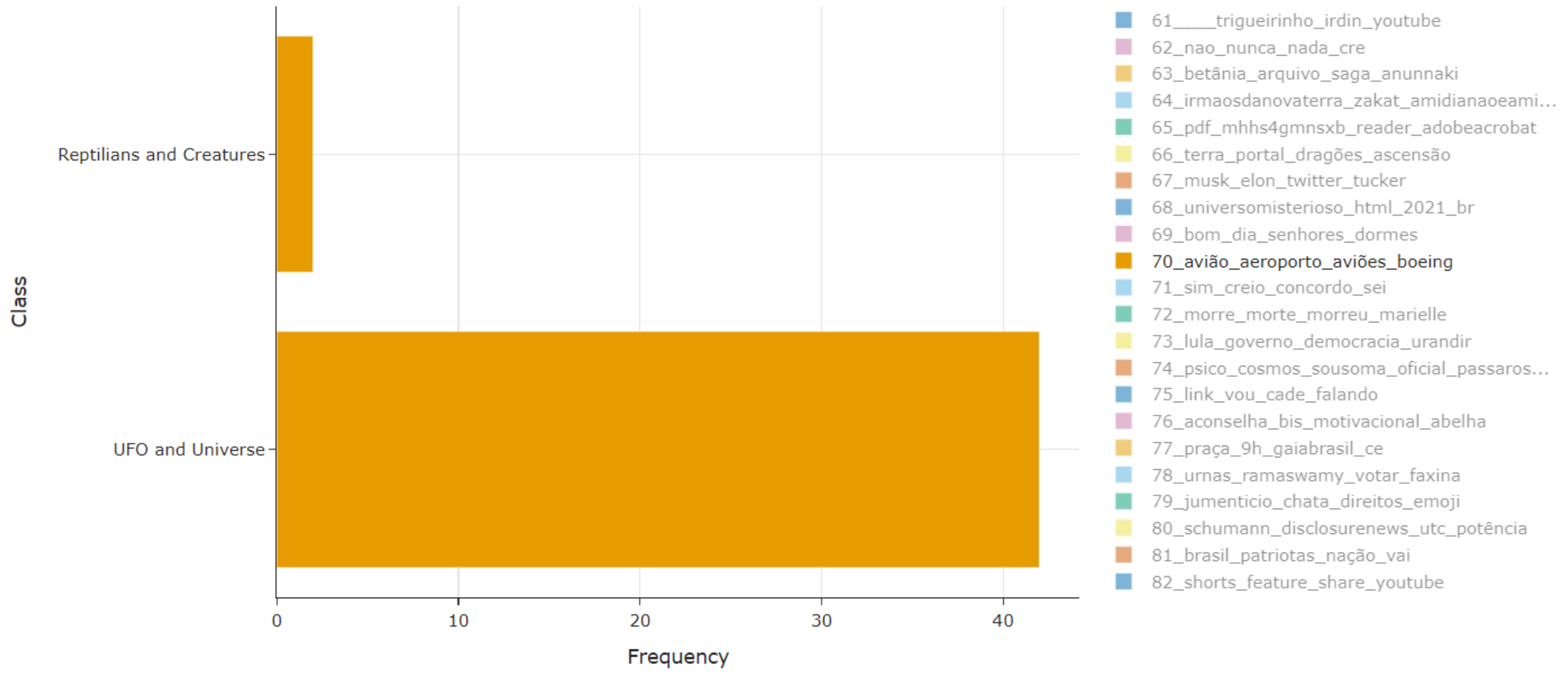

Source: Own elaboration (2024).

Figure 12 addresses how discussions about UFOs and the Universe intertwine with themes of mysterious disappearances and plane crashes. Topics like “disappearances” and “Boeing” are highlighted, suggesting that these communities associate incidents with extraterrestrial interventions or hidden conspiracies. The prevalence of these themes within UFO discussions reflects a tendency in these communities to interpret tragedies and unexplained events as evidence of alien activity or manipulation by secret powers.

**Figure 13.** Natural disasters treated as planned themes

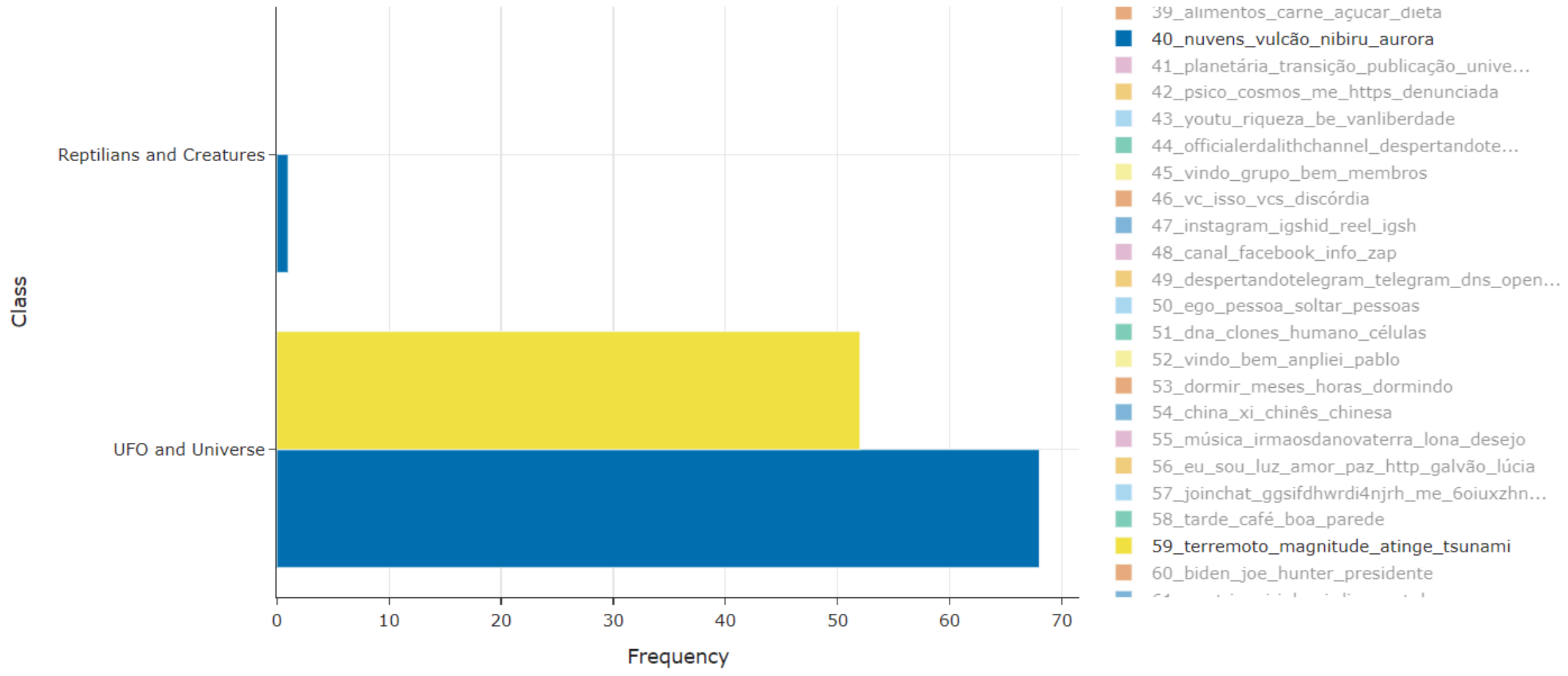

Source: Own elaboration (2024).

Figure 13 reveals how discussions about natural disasters, such as earthquakes and tsunamis, are integrated into narratives about UFOs and Reptilians. The idea that natural disasters are, in fact, planned or controlled by extraterrestrial forces or hidden elites is a recurring theme. These topics reinforce the view that the world is being manipulated in ways that are deliberately hidden from the general population, fueling widespread distrust of scientific explanations.

**Figure 14.** Religion and faith themes

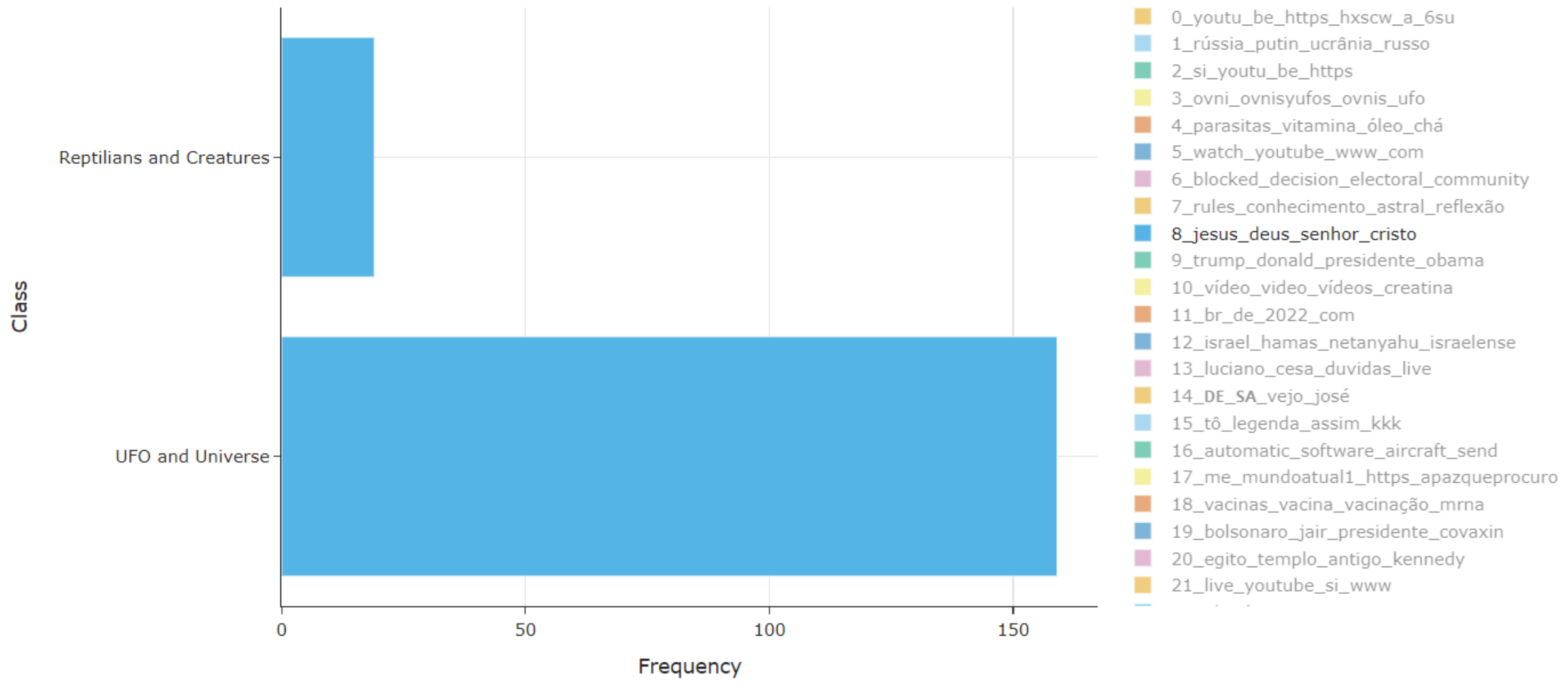

Source: Own elaboration (2024).

In Figure 14, the themes of UFOs and the Universe are combined with discussions about religion and faith. Topics like “Jesus” and “God” emerge alongside extraterrestrial narratives, suggesting that these communities integrate religious beliefs with theories about extraterrestrial life. This fusion reflects an attempt to explain religious or spiritual phenomena as interactions with beings from other worlds, reinterpreting religious dogmas in the light of UFO conspiracy theories.

**Figure 15.** Spirituality and so-called quantum DNA alterations themes

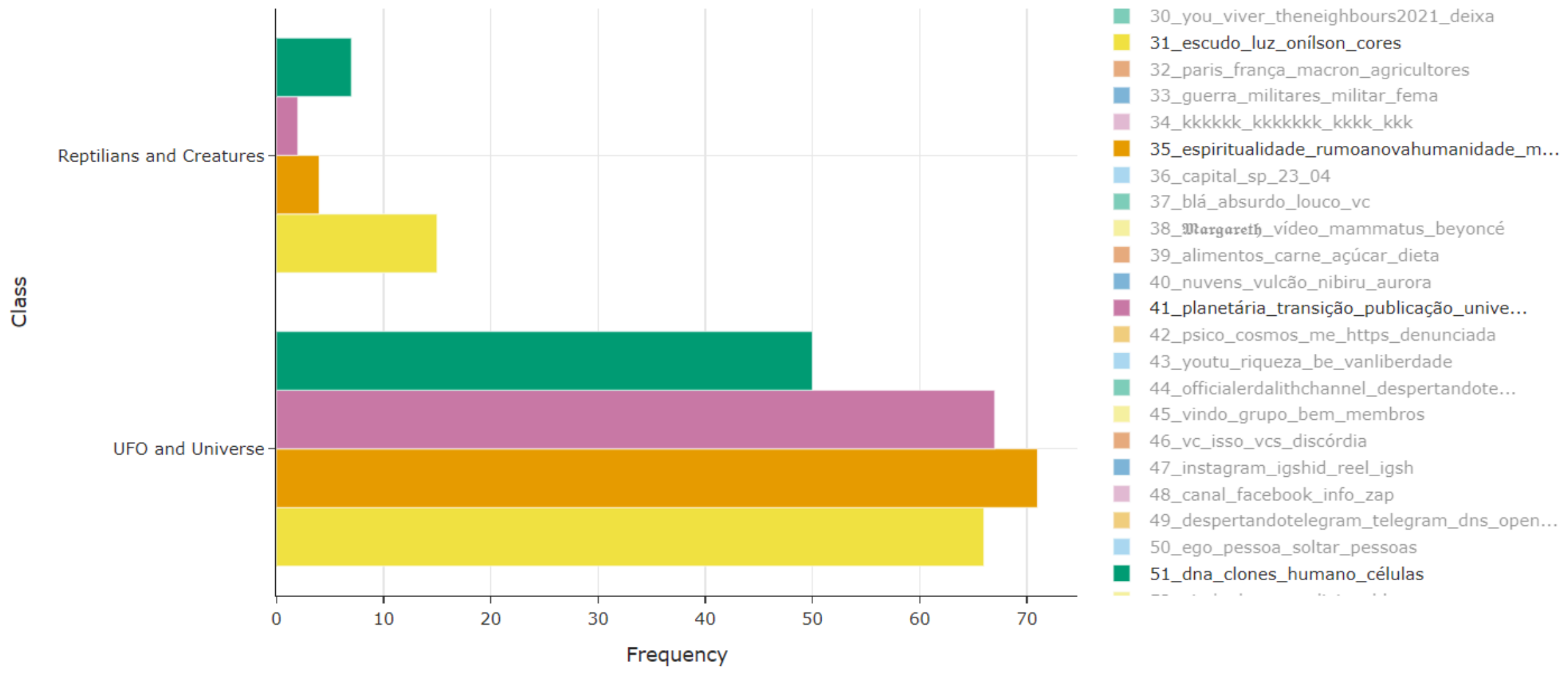

Source: Own elaboration (2024).

Figure 15 addresses how spirituality and pseudoscientific concepts, such as so-called "quantum DNA alterations", are combined with discussions about UFOs and Reptilians. Terms like "quantum" and "DNA" are frequently mentioned, suggesting that these communities see spiritual and physical evolution as directly influenced by extraterrestrial forces. This narrative creates a connection between the esoteric and the conspiratorial — something already pointed out in other investigations of the series — expanding the scope of these communities' beliefs by integrating science, spirituality, and conspiracy theories.

## 4. Reflections and future works

To answer the research question, **"how are Brazilian conspiracy theory communities on UFO, universe, reptilians and creatures topics characterized and articulated on Telegram?"**, this study adopted mirrored techniques in a series of seven publications aimed at characterizing and describing the phenomenon of conspiracy theories on Telegram, focusing on Brazil as a case study. After months of investigation, it was possible to extract a total of 66 Brazilian conspiracy theory communities on Telegram focused on UFO, universe, reptilians and creatures topics, amounting to 1,427,011 pieces of content published between May 2016 (initial publications) and August 2024 (when this study was conducted), with 141,202 users combined across the communities.

Four main approaches were adopted: **(i)** Network, which involved the creation of an algorithm to map connections between communities through invitations circulated among groups and channels; **(ii)** Time series, which used libraries like "Pandas" (McKinney, 2010) and "Plotly" (Plotly Technologies Inc., 2015) to analyze the evolution of publications and engagements over time; **(iii)** Content analysis, where textual analysis techniques were applied to identify patterns and word frequencies in the communities over the semesters; and **(iv)** Thematic agenda overlap, which utilized the BERTopic model (Grootendorst, 2020) to group and interpret large volumes of text, generating coherent topics from the analyzed publications. The main reflections are detailed below, followed by suggestions for future works.

### 4.1. Main reflections

**UFO and Universe communities act as gateways to theories about Reptilians and Creatures, connecting narratives of global control and extraterrestrials:** Communities focused on UFOs and the universe, with 1,268,407 posts, serve as a gateway to more extreme discussions about Reptilians and other creatures. These communities directed 690 invitations to Reptilian and Creature groups, suggesting that initial interest in extraterrestrial life often leads to a deeper exploration of theories that combine elements of global control with non-human beings;

**Theories about Reptilians and Creatures are strongly linked to the New World Order, reinforcing narratives of global domination:** Theories about Reptilians and Creatures have a strong connection to the New World Order (NWO), with 1,159 invitations between these communities. This intersection highlights how beliefs in Reptilian entities

merge with narratives about global control, creating a unified discourse that suggests a conspiracy of world domination by non-human forces;

**Discussions about UFOs and the Universe experienced a 4,400% growth during the pandemic, indicating renewed interest in extraterrestrial phenomena:** Between 2020 and 2021, mentions of UFOs and the universe increased from 1,000 to 45,000, driven by the release of government reports on UFO sightings. This increase reflects growing interest in extraterrestrial phenomena, often accompanied by distrust of official explanations;

**Reptilians and Creatures remain a significant subculture within conspiracy theories, with a 1,000% growth during the pandemic:** Although on a smaller scale, discussions about Reptilians and Creatures also grew significantly, rising from 1,000 to 10,000 mentions between 2020 and 2021. This growth highlights the continued interest in extreme theories involving non-human entities, often associated with narratives of control;

**UFO and Universe communities are strongly interconnected with occultism and esotericism, forming a network of alternative beliefs:** With 6,081 invitations to occult and esoteric communities, discussions about UFOs and the universe frequently intersect with esoteric beliefs. This interconnection suggests that the search for alternative explanations for unexplained phenomena extends to mystical and occult practices, creating a network of beliefs that challenge conventional science;

**The narrative that natural tragedies are planned and controlled by extraterrestrial forces has gained traction in UFO communities:** Discussions linking natural disasters, such as earthquakes and tsunamis, to extraterrestrial intervention are common in UFO communities. This belief aligns with widespread distrust of scientific explanations and suggests that natural events are actually part of a larger plan for global manipulation and domination;

**The thematic overlap between UFOs, Reptilians, and esotericism reveals a cohesive ecosystem of disinformation that is difficult to dismantle:** The analysis of thematic overlap shows that these communities not only share members and content but also mutually reinforce their beliefs. The interconnection between these themes creates a cycle of disinformation where different narratives support each other, making factual correction an even greater challenge;

**The figure of Reptilians as part of a global control plan constantly resurfaces, reinforcing the persistence of this narrative:** Despite the lack of evidence, the idea that Reptilians are part of a conspiracy persists within these communities. This narrative is repeatedly reintroduced, strengthening distrust of authorities and fueling theories of control;

**UFO communities function as amplifiers for other conspiracy theories, connecting different themes and strengthening the disinformation network:** UFO-centered communities not only focus on extraterrestrial phenomena but also act as hubs that connect their members to a variety of theories, such as occultism, NWO, and apocalypse. This amplification function makes these communities central to the spread of disinformation;

**Discussions about UFOs and Reptilians are persistently interconnected, suggesting that these narratives are rooted in a broader conspiratorial worldview:** The analysis of invitation flows between communities reveals that discussions about UFOs and Reptilians are frequently overlapping, indicating that these themes are not viewed in isolation but as part of a larger set of beliefs that challenge the conventional worldview.

### 4.2. Future works

Future studies should explore how UFO and Universe communities act as gateways to other conspiracy narratives, investigating how initial interest in extraterrestrial phenomena can lead to broader acceptance of theories related to global domination and control. Analyzing these dynamics can help understand how these beliefs spread within online communities.

It is also important to investigate the interactions between esoteric themes and extraterrestrial conspiracies, particularly in relation to occultism and alternative spirituality. With the observed interconnection between UFOs and esoteric practices, future studies could focus on how these beliefs reinforce each other and how this impacts the worldview of members within these communities.

Another relevant point is the persistence of Reptilian narratives as part of a global control plan. Future studies could investigate the mechanisms that allow the survival and constant reintroduction of this narrative, even in the face of repeated debunking. Understanding these mechanisms could be crucial for developing more effective strategies to combat disinformation.

Furthermore, future research should focus on analyzing the networks of invitations between communities, exploring how these connections facilitate the spread of disinformation and create an environment conducive to the mutual reinforcement of conspiracy beliefs. Developing tools that can map and identify these flows in real-time could be an effective way to mitigate the spread of disinformation during critical moments.

Finally, studies on the resilience of communities centered on UFOs and Reptilians in the face of global crises may offer insights into how these narratives adapt and remain relevant. Understanding how these communities respond to new information and events could help better predict and respond to outbreaks of disinformation.

## 6. Author biography

**Ergon Cugler de Moraes Silva** has a Master’s degree in Public Administration and Government (FGV), Postgraduate MBA in Data Science & Analytics (USP) and Bachelor’s degree in Public Policy Management (USP). He is associated with the Bureaucracy Studies Center (NEB FGV), collaborates with the Interdisciplinary Observatory of Public Policies (OIPP USP), with the Study Group on Technology and Innovations in Public Management (GETIP USP) with the Monitor of Political Debate in the Digital Environment (Monitor USP) and with the Working Group on Strategy, Data and Sovereignty of the Study and Research Group on International Security of the Institute of International Relations of the University of Brasília (GEPSI UnB). He is also a researcher at the Brazilian Institute of Information in Science and Technology (IBICT), where he works for the Federal Government on strategies against disinformation. Brasília, Federal District, Brazil. Web site: https://ergoncugler.com/.

[ versão português / english above ]

# **Comunidades de OVNIs, universo, reptilianos e criaturas no Telegram brasileiro:** quando o céu não é o limite e o conspiracionismo busca respostas para além da humanidade

*Ergon Cugler de Moraes Silva*

Instituto Brasileiro de Informação
em Ciência e Tecnologia (IBICT)
Brasília, Distrito Federal, Brasil

contato@ergoncugler.com
www.ergoncugler.com

## Resumo

O interesse por fenômenos extraterrestres e teorias de conspiração envolvendo OVNIs e reptilianos tem crescido no Telegram brasileiro, especialmente em tempos de incerteza global, como durante a Pandemia da COVID-19. Dessa forma, esse estudo busca responder à pergunta de pesquisa: **como são caracterizadas e articuladas as comunidades de teorias da conspiração brasileiras sobre temáticas de OVNIs, universo, reptilianos e criaturas no Telegram?** Vale ressaltar que este estudo faz parte de uma série de um total de sete estudos que possuem como objetivo principal compreender e caracterizar as comunidades brasileiras de teorias da conspiração no Telegram. Esta série de sete estudos está disponibilizada abertamente e originalmente no arXiv da Cornell University, aplicando um método espelhado nos sete estudos, mudando apenas o objeto temático de análise e provendo uma replicabilidade de investigação, inclusive com códigos próprios e autorais elaborados, somando-se à cultura de software livre e de código aberto. No que diz respeito aos principais achados deste estudo, observa-se: Comunidades de OVNIs atuam como portas de entrada para teorias sobre reptilianos, conectando narrativas de controle global com seres extraterrestres; Discussões sobre OVNIs e universo cresceram significativamente durante a Pandemia, refletindo um interesse renovado por fenômenos extraterrestres; Reptilianos permanecem como uma subcultura significativa dentro das teorias conspiratórias, com um crescimento notável durante a Pandemia; A sobreposição temática entre OVNIs, reptilianos e esoterismo revela um ecossistema coeso de desinformação, tornando a correção factual um desafio; Comunidades de OVNIs funcionam como amplificadores de outras teorias conspiratórias, conectando diferentes temas e fortalecendo a rede de desinformação.

### Principais descobertas

➔ Comunidades de OVNIs e universo atuam como portas de entrada para teorias sobre reptilianos e criaturas, conectando narrativas de controle global e extraterrestres: Com 1.268.407 publicações, essas comunidades direcionaram 690 convites para grupos de reptilianos, sugerindo que o interesse por vida extraterrestre frequentemente leva a uma exploração mais profunda de teorias que combinam controle global com seres não humanos;

➔ Teorias sobre reptilianos e criaturas estão fortemente vinculadas à Nova Ordem Mundial, reforçando narrativas de dominação global: Com 1.159 convites entre essas comunidades, a crença em entidades reptilianas se funde com narrativas sobre controle global, criando um discurso unificado que sugere uma conspiração mundial por forças não humanas;

- ➔ Discussões sobre OVNIs e universo experimentaram um crescimento de 4.400% durante a Pandemia, indicando um interesse renovado em fenômenos extraterrestres: Entre 2020 e 2021, menções a OVNIs aumentaram de 1.000 para 45.000, impulsionadas por relatórios governamentais sobre avistamentos, refletindo um crescente interesse acompanhado de desconfiança em explicações oficiais;

- ➔ Reptilianos e criaturas permanecem como uma subcultura significativa dentro das teorias de conspiração, com um crescimento de 1.000% durante a Pandemia: As menções sobre reptilianos subiram de 1.000 para 10.000 entre 2020 e 2021, destacando o interesse contínuo por teorias extremas envolvendo entidades não humanas e narrativas de controle;

- ➔ Comunidades de OVNIs e universo estão fortemente interligadas com o ocultismo e esoterismo, formando uma rede de crenças alternativas: Com 6.081 convites para comunidades de ocultismo, as discussões sobre OVNIs frequentemente se cruzam com crenças esotéricas, sugerindo que a busca por explicações alternativas para fenômenos inexplicáveis se estende a práticas místicas, desafiando a ciência convencional;

- ➔ A narrativa de que tragédias naturais são planejadas e controladas por forças extraterrestres ganhou força nas comunidades de OVNIs: Discussões ligando desastres naturais à intervenção extraterrestre são comuns, refletindo uma desconfiança nas explicações científicas e sugerindo que eventos naturais fazem parte de um plano maior de manipulação global;

- ➔ A sobreposição temática entre OVNIs, reptilianos e esoterismo revela um ecossistema coeso de desinformação que é difícil de desmantelar: Essas comunidades compartilham membros e conteúdos, criando um ciclo de desinformação onde diferentes narrativas se apoiam mutuamente, tornando a correção factual um desafio;

- ➔ A figura de reptilianos como parte de um plano de controle global ressurge constantemente, reforçando a persistência dessa narrativa: Mesmo sem evidências, a ideia de reptilianos envolvidos em conspirações globais persiste, alimentando teorias sobre dominação e reforçando a desconfiança em autoridades;

- ➔ As comunidades de OVNIs funcionam como amplificadores de outras teorias conspiratórias, conectando diferentes temas e fortalecendo a rede de desinformação: Além de focar em fenômenos extraterrestres, essas comunidades atuam como *hubs* que conectam seus membros a outras teorias conspiratórias, como ocultismo, NOM e apocalipse, centralizando a disseminação de desinformação;

- ➔ As discussões sobre OVNIs e reptilianos são persistentemente interligadas, sugerindo que essas narrativas estão enraizadas em uma visão de mundo conspiratória mais ampla: Análises mostram que esses temas são frequentemente sobrepostos, indicando que fazem parte de um conjunto maior de crenças que desafiam a visão de mundo convencional.

## 1. Introdução

Após percorrer milhares de comunidades brasileiras de teorias da conspiração no Telegram, extrair dezenas de milhões de conteúdos dessas comunidades, elaborados e/ou compartilhados por milhões de usuários que as compõem, este estudo tem o objetivo de compor uma série de um total de sete estudos que tratam sobre o fenômeno das teorias da conspiração no Telegram, adotando o Brasil como estudo de caso. Com as abordagens de identificação implementadas, foi possível alcançar um total de 66 comunidades de teorias da conspiração brasileiras no Telegram sobre temáticas de OVNIs, universo, reptilianos e criaturas, estas somando 1.427.011 de conteúdos publicados entre maio de 2016 (primeiras publicações) até agosto de 2024 (realização deste estudo), com 141.202 usuários somados dentre as comunidades. Dessa forma, este estudo tem como objetivo compreender e caracterizar as comunidades sobre temáticas de OVNIs, universo, reptilianos e criaturas presentes nessa rede brasileira de teorias da conspiração identificada no Telegram.

Para tal, será aplicado um método espelhado em todos os sete estudos, mudando apenas o objeto temático de análise e provendo uma replicabilidade de investigação. Assim, abordaremos técnicas para observar as conexões, séries temporais, conteúdos e sobreposições temáticas das comunidades de teorias da conspiração. Além desse estudo, é possível encontrar os seis demais disponibilizados abertamente e originalmente no arXiv da Cornell University. Essa série contou com a atenção redobrada para garantir a integridade dos dados e o respeito à privacidade dos usuários, conforme a legislação brasileira prevê (Lei nº 13.709/2018).

Portanto questiona-se: **como são caracterizadas e articuladas as comunidades de teorias da conspiração brasileiras sobre temáticas de OVNIs, universo, reptilianos e criaturas no Telegram?**

## 2. Materiais e métodos

A metodologia deste estudo está organizada em três subseções, sendo: **2.1. Extração de dados**, que descreve o processo e as ferramentas utilizadas para coletar as informações das comunidades no Telegram; **2.2. Tratamento de dados**, onde são abordados os critérios e métodos aplicados para classificar e anonimizar os dados coletados; e **2.3. Abordagens para análise de dados**, que detalha as técnicas utilizadas para investigar as conexões, séries temporais, conteúdos e sobreposições temáticas das comunidades de teorias da conspiração.

### 2.1. Extração de dados

Este projeto teve início em fevereiro de 2023, com a publicação da primeira versão do TelegramScrap (Silva, 2023), uma ferramenta própria e autoral, de software livre e de código aberto, que faz uso da Application Programming Interface (API) da plataforma Telegram por meio da biblioteca Telethon e organiza ciclos de extração de dados de grupos e canais abertos no Telegram. Ao longo dos meses, a base de dados pôde ser ampliada e qualificada fazendo uso de quatro abordagens de identificação de comunidades de teorias da conspiração:

**(i) Uso de palavras chave:** no início do projeto, foram elencadas palavras-chave para identificação diretamente no buscador de grupos e canais brasileiros no Telegram, tais como "apocalipse", "sobrevivencialismo", "mudanças climáticas", "terra plana", "teoria da conspiração", "globalismo", "nova ordem mundial", "ocultismo", "esoterismo", "curas alternativas", "qAnon", "reptilianos", "revisionismo", "alienígenas", dentre outras. Essa primeira abordagem forneceu algumas comunidades cujos títulos e/ou descrições dos grupos e canais contassem com os termos explícitos relacionados a teorias da conspiração. Contudo, com o tempo foi possível identificar outras diversas comunidades cujas palavras-chave elencadas não davam conta de abarcar, algumas inclusive propositalmente com caracteres trocados para dificultar a busca de quem a quisesse encontrar na rede;

**(ii) Mecanismo de recomendação de canais do Telegram:** com o tempo, foi identificado que canais do Telegram (exceto grupos) contam com uma aba de recomendação chamada de “canais similares”, onde o próprio Telegram sugere dez canais que tenham alguma similaridade com o canal que se está observando. A partir desse mecanismo de recomendação do próprio Telegram, foi possível encontrar mais comunidades de teorias da conspiração brasileiras, com estas sendo recomendadas pela própria plataforma;

**(iii) Abordagem de bola de neve para identificação de convites:** após algumas comunidades iniciais serem acumuladas para a extração, foi elaborado um algoritmo próprio autoral de identificação de urls que contivessem “t.me/”, sendo o prefixo de qualquer convite para grupos e canais do Telegram. Acumulando uma frequência de centenas de milhares de links que atendessem a esse critério, foram elencados os endereços únicos e contabilizadas as suas repetições. Dessa forma, foi possível fazer uma investigação de novos grupos e canais brasileiros mencionados nas próprias mensagens dos já investigados, ampliando a rede. Esse processo foi sendo repetido periodicamente, buscando identificar novas comunidades que tivessem identificação com as temáticas de teorias da conspiração no Telegram;

**(iv) Ampliação para tweets publicados no X que mencionassem convites:** com o objetivo de diversificar ainda mais a fonte de comunidades de teorias da conspiração brasileiras no Telegram, foi elaborada uma query de busca própria que pudesse identificar as palavras-chave de temáticas de teorias da conspiração, porém usando como fonte tweets publicados no X (antigo Twitter) e que, além de conter alguma das palavras-chave, contivesse também o “t.me/”, sendo o prefixo de qualquer convite para grupos e canais do Telegram, “https://x.com/search?q=lang%3Apt%20%22t.me%2F%22%20TERMO-DE-BUSCA”.

Com as abordagens de identificação de comunidades de teorias da conspiração implementadas ao longo de meses de investigação e aprimoramento de método, foi possível construir uma base de dados do projeto com um total de 855 comunidades de teorias da conspiração brasileiras no Telegram (considerando as demais temáticas também não incluídas nesse estudo), estas somando 27.227.525 de conteúdos publicados entre maio de 2016 (primeiras publicações) até agosto de 2024 (realização deste estudo), com 2.290.621 usuários somados dentre as comunidades brasileiras. Há de se considerar que este volume de usuários conta com dois elementos, o primeiro é que trata-se de uma variável, pois usuários podem entrar e sair diariamente, portanto este valor representa o registrado na data de extração de publicações da comunidade; além disso, é possível que um mesmo usuário esteja em mais de um grupo e, portanto, é contabilizado mais de uma vez. Nesse sentido, o volume ainda sinaliza ser expressivo, mas pode ser levemente menor quando considerado o volume de cidadãos deduplicados dentro dessas comunidades brasileiras de teorias da conspiração.

## 2.2. Tratamento de dados

Com todos os grupos e canais brasileiros de teorias da conspiração no Telegram extraídos, foi realizada uma classificação manual considerando o título e a descrição da comunidade. Caso houvesse menção explícita no título ou na descrição da comunidade a alguma temática, esta foi classificada entre: (i) "Anticiência"; (ii) "Anti-Woke e Gênero"; (iii) "Antivax"; (iv) "Apocalipse e Sobrevivencialismo"; (v) "Mudanças Climáticas"; (vi) Terra Plana; (vii) "Globalismo"; (viii) "Nova Ordem Mundial"; (ix) "Ocultismo e Esoterismo"; (x) "Off Label e Charlatanismo"; (xi) "QAnon"; (xii) "Reptilianos e Criaturas"; (xiii) "Revisionismo e Discurso de Ódio"; (xiv) "OVNI e Universo". Caso não houvesse nenhuma menção explícita relacionada às temáticas no título ou na descrição da comunidade, esta foi classificada como (xv) "Conspiração Geral". No Quadro a seguir, podemos observar as métricas relacionadas à classificação dessas comunidades de teorias da conspiração no Brasil.

**Tabela 01.** Comunidades de teorias da conspiração no Brasil (métricas até agosto de 2024)

| Categorias | Grupos | Usuários | Publicações | Comentários | Total |
|---|---|---|---|---|---|
| **Anticiência** | 22 | 58.138 | 187.585 | 784.331 | 971.916 |
| **Anti-*Woke* e Gênero** | 43 | 154.391 | 276.018 | 1.017.412 | 1.293.430 |
| **Antivacinas (*Antivax*)** | 111 | 239.309 | 1.778.587 | 1.965.381 | 3.743.968 |
| **Apocalipse e Sobrevivência** | 33 | 109.266 | 915.584 | 429.476 | 1.345.060 |
| **Mudanças Climáticas** | 14 | 20.114 | 269.203 | 46.819 | 316.022 |
| **Terraplanismo** | 33 | 38.563 | 354.200 | 1.025.039 | 1.379.239 |
| **Conspirações Gerais** | 127 | 498.190 | 2.671.440 | 3.498.492 | 6.169.932 |
| **Globalismo** | 41 | 326.596 | 768.176 | 537.087 | 1.305.263 |
| **Nova Ordem Mundial (NOM)** | 148 | 329.304 | 2.411.003 | 1.077.683 | 3.488.686 |
| **Ocultismo e Esoterismo** | 39 | 82.872 | 927.708 | 2.098.357 | 3.026.065 |
| **Medicamentos *off label*** | 84 | 201.342 | 929.156 | 733.638 | 1.662.794 |
| **QAnon** | 28 | 62.346 | 531.678 | 219.742 | 751.420 |
| **Reptilianos e Criaturas** | 19 | 82.290 | 96.262 | 62.342 | 158.604 |
| **Revisionismo e Ódio** | 66 | 34.380 | 204.453 | 142.266 | 346.719 |
| **OVNI e Universo** | 47 | 58.912 | 862.358 | 406.049 | 1.268.407 |
| **Total** | **855** | **2.296.013** | **13.183.411** | **14.044.114** | **27.227.525** |

Fonte: Elaboração própria (2024).

Com esse volume de dados extraídos, foi possível segmentar para apresentar neste estudo apenas comunidades e conteúdos referentes às temáticas de OVNIs, universo, reptilianos e criaturas. Em paralelo, as demais temáticas de comunidades brasileiras de teorias da conspiração também contaram com estudos elaborados para a caracterização da extensão e da dinâmica da rede, estes sendo disponibilizados abertamente e originalmente no arXiv da Cornell University.

Além disso, cabe citar que apenas foram extraídas comunidades abertas, isto é, não apenas identificáveis publicamente, mas também sem necessidade de solicitação para acessar ao conteúdo, estando aberto para todo e qualquer usuário com alguma conta do Telegram sem que este necessite ingressar no grupo ou canal. Além disso, em respeito à legislação brasileira e especialmente da Lei Geral de Proteção de Dados Pessoais (LGPD), ou Lei nº 13.709/2018, que trata do controle da privacidade e do uso/tratamento de dados pessoais, todos os dados extraídos foram anonimizados para a realização de análises e investigações. Dessa forma, nem mesmo a identificação das comunidades é possível por meio deste estudo, estendendo aqui a privacidade do usuário ao anonimizar os seus dados para além da própria comunidade à qual ele se submeteu ao estar em um grupo ou canal público e aberto no Telegram.

### 2.3. Abordagens para análise de dados

Totalizando 66 comunidades selecionadas nas temáticas de OVNIs, universo, reptilianos e criaturas, contendo 1.427.011 publicações e 141.202 usuários somados, quatro abordagens serão utilizadas para investigar as comunidades de teorias da conspiração selecionadas para o escopo do estudo. Tais métricas são detalhadas no Quadro a seguir:

**Tabela 02.** Comunidades selecionadas para análise (métricas até agosto de 2024)

| Categorias | Grupos | Usuários | Publicações | Comentários | Total |
|---|---|---|---|---|---|
| OVNI e Universo | 47 | 58.912 | 862.358 | 406.049 | 1.268.407 |
| Reptilianos e Criaturas | 19 | 82.290 | 96.262 | 62.342 | 158.604 |
| **Total** | **66** | **141.202** | **958.620** | **468.391** | **1.427.011** |

Fonte: Elaboração própria (2024).

**(i) Rede:** com a elaboração de um algoritmo próprio para a identificação de mensagens que contenham o termo de “t.me/” (de convite para entrarem em outras comunidades), propomos apresentar volumes e conexões observadas sobre como **(a)** as comunidades da temática de OVNIs, universo, reptilianos e criaturas circulam convites para que os seus usuários conheçam mais grupos e canais da mesma temática, reforçando os sistemas de crença que comungam; e como **(b)** essas mesmas comunidades circulam convites para que os seus usuários conheçam grupos e canais que tratem de outras teorias da conspiração, distintas de seu propósito explícito. Esta abordagem é interessante para observar se essas comunidades utilizam a si próprias como fonte de legitimação e referência e/ou se embasam-se em demais temáticas de teorias da conspiração, inclusive abrindo portas para que seus usuários conheçam outras conspirações. Além disso, cabe citar o estudo de Rocha *et al.* (2024) em que uma abordagem de identificação de rede também foi aplicada em comunidades do Telegram, porém observando conteúdos similares a partir de um ID gerado para cada mensagem única e suas similares;

**(ii) Séries temporais:** utiliza-se da biblioteca “Pandas” (McKinney, 2010) para organizar os data frames de investigação, observando **(a)** o volume de publicações ao longo dos meses; e **(b)** o volume de engajamento ao longo dos meses, considerando metadados de visualizações, reações e comentários coletados na extração; Além da volumetria, a biblioteca “Plotly” (Plotly Technologies Inc., 2015) viabilizou a representação gráfica dessa variação;

**(iii) Análise de conteúdo:** além da análise geral de palavras com identificação das frequências, são aplicadas séries temporais na variação das palavras mais frequentes ao longo dos semestres — observando entre maio de 2016 (primeiras publicações) até agosto de 2024 (realização deste estudo). E com as bibliotecas “Pandas” (McKinney, 2010) e “WordCloud” (Mueller, 2020), os resultados são apresentados tanto volumetricamente quanto graficamente;

**(iv) Sobreposição de agenda temática:** seguindo a abordagem proposta por Silva & Sátiro (2024) para identificação de sobreposição de agenda temática em comunidades do Telegram, utilizamos o modelo “BERTopic” (Grootendorst, 2020). O BERTopic é um algoritmo de modelagem de tópicos que facilita a geração de representações temáticas a partir de grandes quantidades de textos. Primeiramente, o algoritmo extrai embeddings dos documentos usando modelos transformadores de sentenças, como o “all-MiniLM-L6-v2”. Em seguida, essas embeddings têm sua dimensionalidade reduzida por técnicas como “UMAP”, facilitando o processo de agrupamento. A clusterização é realizada usando “HDBSCAN”, uma técnica baseada em densidade que identifica clusters de diferentes formas e tamanhos, além de detectar outliers. Posteriormente, os documentos são tokenizados e representados em uma estrutura de bag-of-words, que é normalizada (L1) para considerar as diferenças de tamanho entre os clusters. A representação dos tópicos é refinada usando uma versão modificada do “TF-IDF”, chamada “Class-TF-IDF”, que considera a importância das palavras dentro de cada cluster (Grootendorst, 2020). Cabe destacar que, antes de aplicar o BERTopic, realizamos a limpeza da base removendo “stopwords” em português, por meio da biblioteca “NLTK” (Loper & Bird, 2002). Para a modelagem de tópicos, utilizamos o backend “loky” para otimizar o desempenho durante o ajuste e a transformação dos dados.

Em síntese, a metodologia aplicada compreendeu desde a extração de dados com a ferramenta própria autoral TelegramScrap (Silva, 2023), até o tratamento e a análise de dados coletados, utilizando diversas abordagens para identificar e classificar comunidades de teorias da conspiração brasileiras no Telegram. Cada uma das etapas foi cuidadosamente implementada para garantir a integridade dos dados e o respeito à privacidade dos usuários, conforme a legislação brasileira prevê. A seguir, serão apresentados os resultados desses dados, com o intuito de revelar as dinâmicas e as características das comunidades estudadas.

## 3. Resultados

A seguir, os resultados são detalhados na ordem prevista na metodologia, iniciando com a caracterização da rede e suas fontes de legitimação e referência, avançando para as séries temporais, incorporando a análise de conteúdo e concluindo com a identificação de sobreposição de agenda temática dentre as comunidades de teorias da conspiração.

### 3.1. Rede

A análise das redes internas entre comunidades que discutem OVNIs, o universo, reptilianos e outras criaturas misteriosas revela uma complexa teia de interconexões que transcende as barreiras entre diferentes narrativas conspiratórias. Esses gráficos expõem como temas aparentemente distintos, como a vida extraterrestre e a existência de seres reptilianos, estão intrinsecamente ligados, formando um ecossistema conspiratório coeso. As comunidades que sustentam essas crenças atuam não apenas como veículos de disseminação de ideias, mas também como portas de entrada e saída para um vasto universo de teorias correlatas, sugerindo que o envolvimento inicial com um desses temas frequentemente leva a uma imersão mais profunda em outras teorias conspiratórias inter-relacionadas. Ao analisar as conexões e fluxos de convites entre essas comunidades, podemos observar como essas narrativas se reforçam mutuamente, criando um ambiente onde as crenças em fenômenos sobrenaturais e conspirações globais são amplificadas e perpetuadas. Essas redes não apenas sustentam as teorias individuais, mas também constroem uma visão de mundo onde o inexplicável e o oculto ganham coerência e sentido dentro de uma estrutura narrativa que desafia a lógica científica e a autoridade institucional.

**Figura 01.** Rede interna entre OVNIs, universo, reptilianos e criaturas

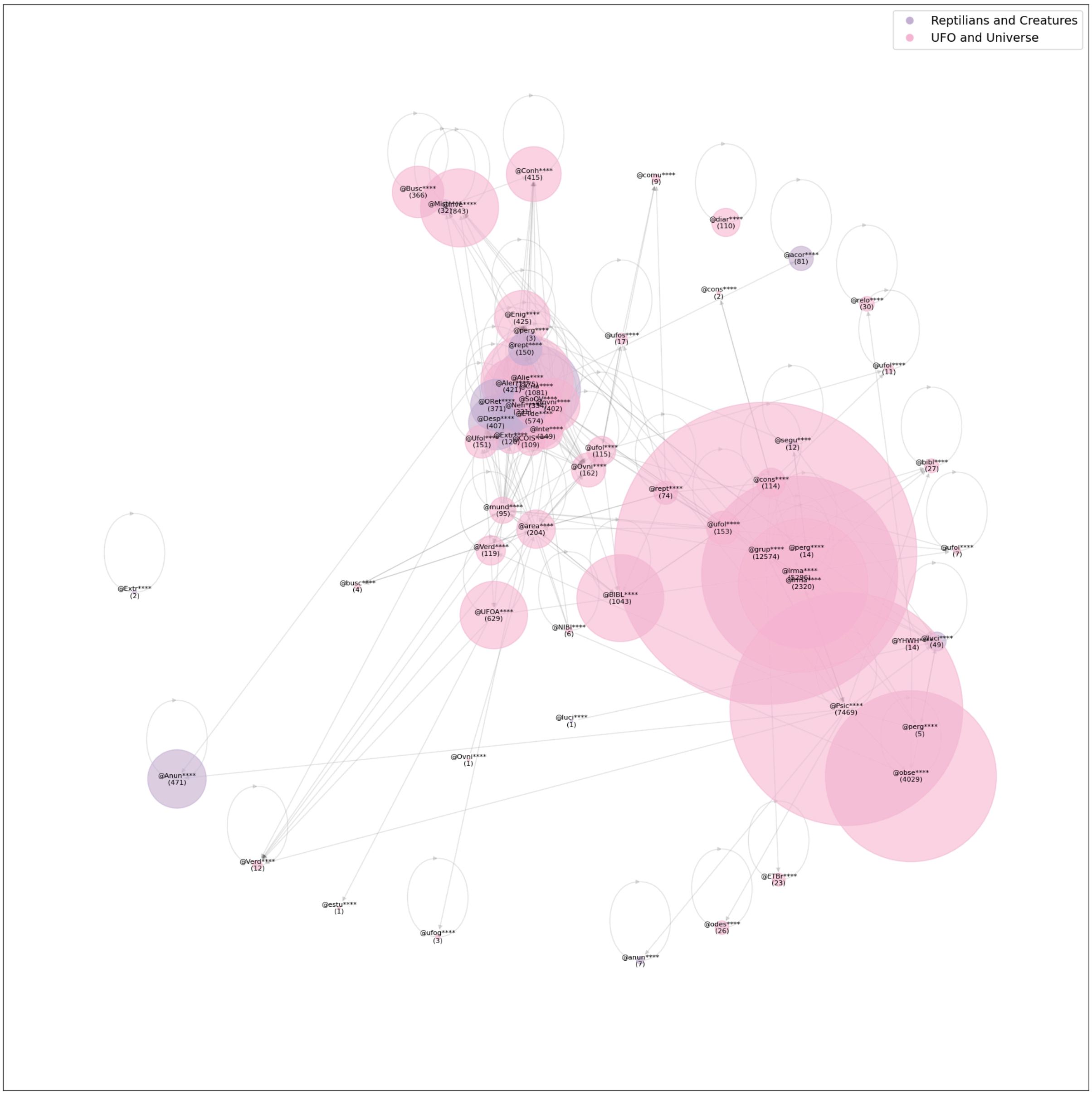

Fonte: Elaboração própria (2024).

Este gráfico retrata a rede interna entre comunidades que discutem OVNIs, o universo, reptilianos e criaturas misteriosas. A rede, composta por dois núcleos principais, revela como as teorias relacionadas à vida extraterrestre e conspirações sobre seres reptilianos estão profundamente conectadas. As interações sugerem que essas comunidades se alimentam mutuamente, com o interesse por OVNIs frequentemente levando à exploração de teorias mais radicais sobre reptilianos e criaturas desconhecidas. Os grandes nós indicam que certos grupos possuem uma influência desproporcional na propagação dessas narrativas, funcionando como centros de referência para outras comunidades menores. A rede, embora menos densa que algumas outras, ainda demonstra uma forte coesão interna, sugerindo que os seguidores são constantemente expostos a uma variedade de teorias relacionadas que reforçam e ampliam a crença na existência de conspirações extraterrestres e entidades ocultas.

**Figura 02.** Rede de comunidades que abrem portas para a temática (porta de entrada)

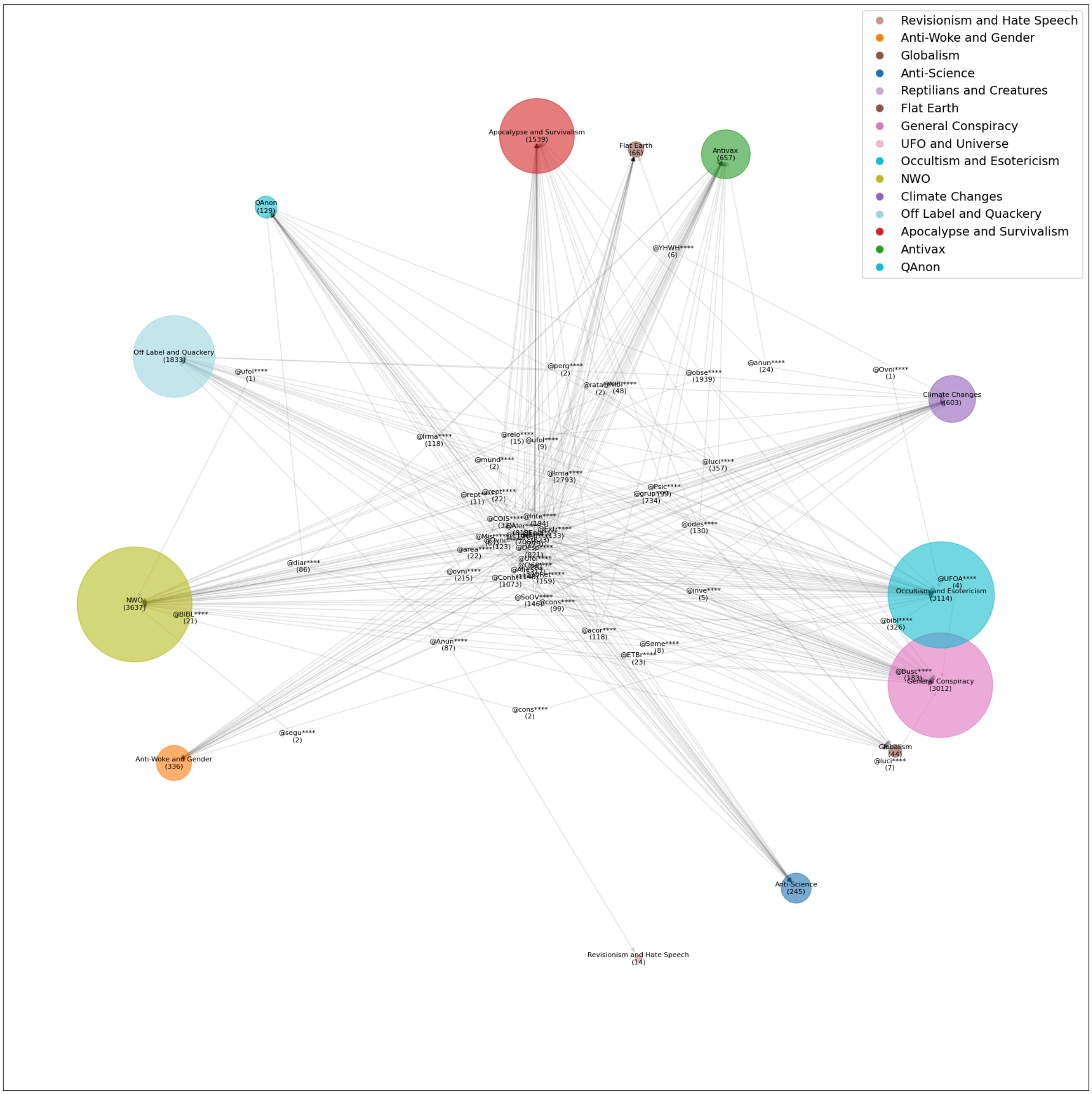

Fonte: Elaboração própria (2024).

Esta figura apresenta as comunidades que atuam como portas de entrada para as discussões sobre OVNIs, reptilianos e outras criaturas misteriosas. A rede é caracterizada por uma forte concentração em torno de algumas comunidades centrais que atraem aqueles curiosos sobre a vida extraterrestre e teorias sobre seres reptilianos. O gráfico evidencia como o interesse por OVNIs e o universo pode facilmente levar a explorações mais profundas em teorias conspiratórias sobre reptilianos e criaturas. Essas comunidades maiores não apenas servem como *hubs* de disseminação dessas narrativas, mas também conectam os seguidores a uma vasta rede de teorias correlatas, sugerindo que uma vez dentro desta rede, os indivíduos são expostos a uma ampla gama de ideias conspiratórias, reforçando suas crenças e ampliando seu envolvimento com outras comunidades semelhantes.

[ versão português / english above ]

**Figura 03. UFO** Rede de comunidades cuja temática abre portas (porta de saída)

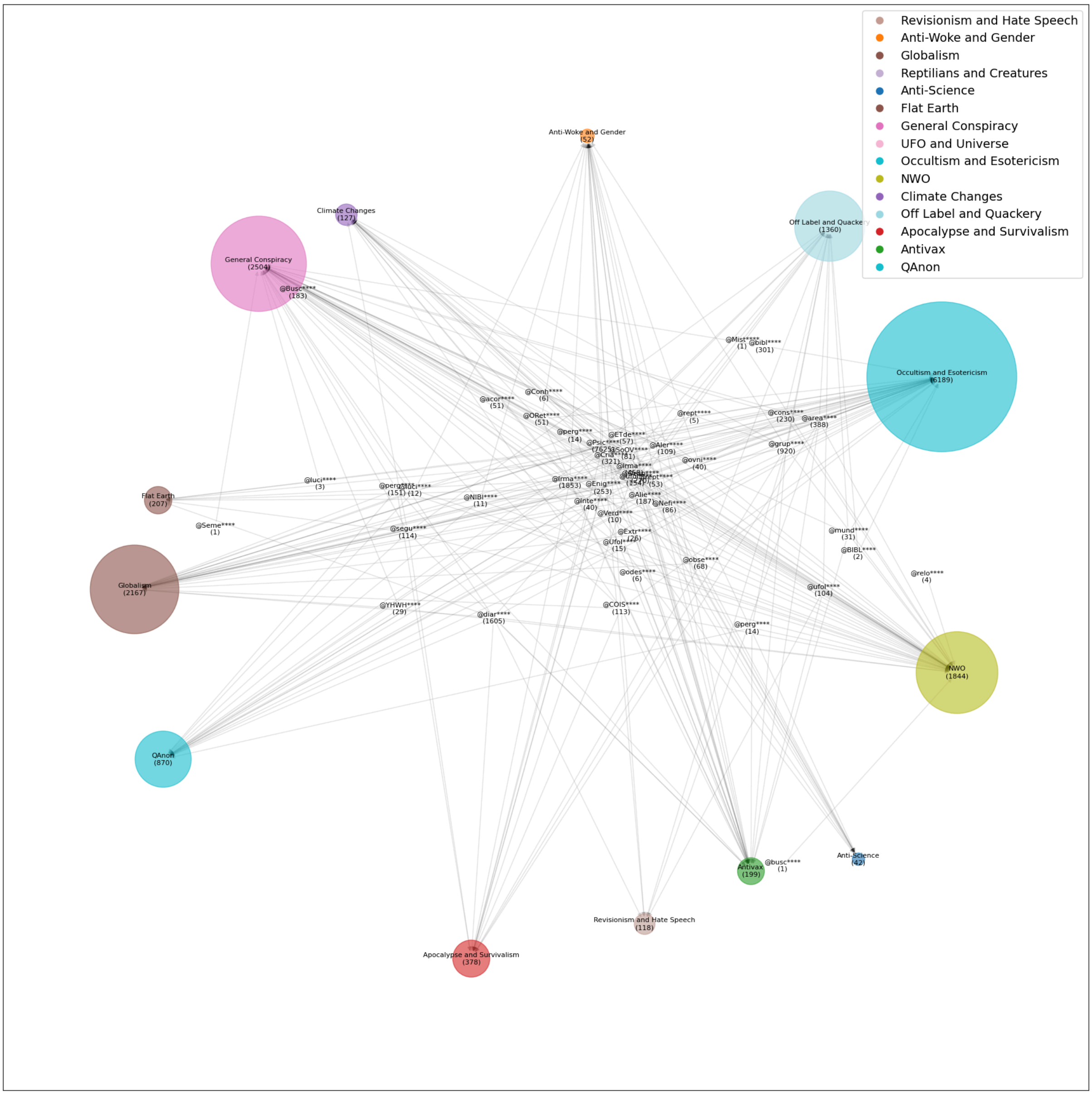

Fonte: Elaboração própria (2024).

O gráfico de redes focado em OVNIs, universo, reptilianos e criaturas destaca como essas temáticas, que podem parecer isoladas, estão profundamente interligadas com outras teorias da conspiração. As conexões entre essas comunidades e outras temáticas, como Ocultismo, Terra Plana, e Globalismo, sugerem que as discussões sobre OVNIs e vida extraterrestre frequentemente abrem caminho para a exploração de narrativas mais amplas de conspiração global e controle governamental secreto. A figura sugere que o interesse inicial por teorias de OVNIs pode levar os seguidores a uma imersão em teorias mais complexas e interconectadas, relacionando inclusive com crenças de ocultismos e esoterismos, onde as explicações sobrenaturais e a desconfiança em relação às autoridades se entrelaçam, criando um ecossistema de crenças que alimenta uma visão de mundo ancorando-se em crenças.

**Figura 04.** Fluxo de links de convites entre comunidades de OVNIs,e universo

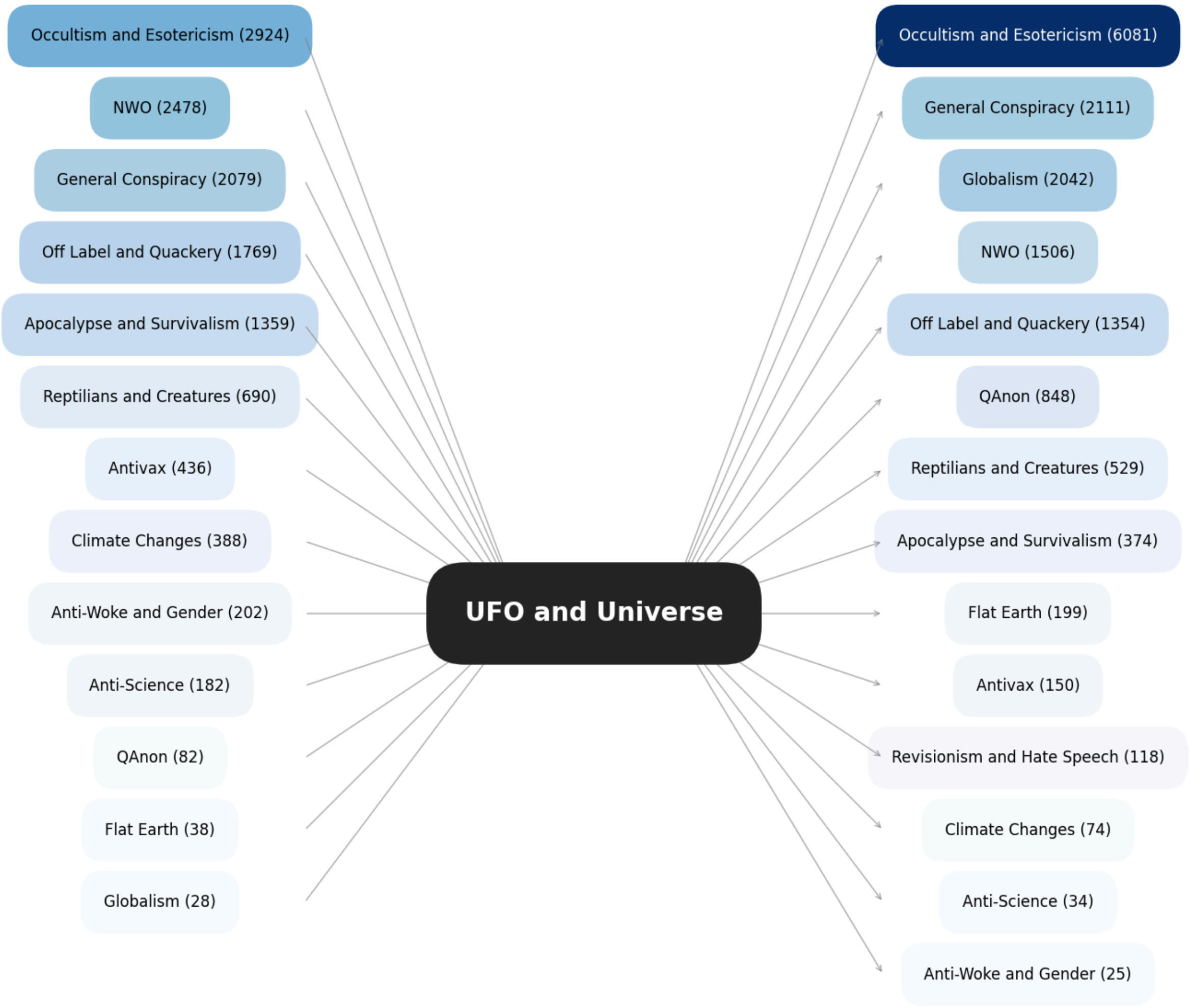

Fonte: Elaboração própria (2024).

O gráfico de OVNIs e Universo mostra que essa temática está fortemente interligada a Ocultismo e Esoterismo (2.924 links) e Conspirações Gerais (2.079 links), sugerindo que a crença em vida extraterrestre e fenômenos inexplicáveis funciona como uma entrada para uma rede maior de teorias que questionam a realidade percebida. A conexão com Medicamentos *off label* (1.769 links) e Apocalipse e Sobrevivência (1.359 links) aponta para uma tendência onde a desconfiança em relação à ciência e a preparação para eventos cataclísmicos se entrelaçam com o interesse por fenômenos não explicados pela ciência convencional. Quando OVNIs e Universo emite convites para Ocultismo e Esoterismo (6.081 links) e Conspirações Gerais (2.111 links), a análise indica que esses grupos funcionam como difusores de uma visão de mundo onde o inexplicável não só existe, mas é parte de um plano maior de manipulação e ocultação por forças poderosas. Esse ciclo de crença não apenas sustenta a narrativa conspiratória, mas também cria um ambiente onde os adeptos são incentivados a buscar constantemente por novas "verdades" que validem sua visão de mundo. Isso mostra como o tema OVNIs e Universo pode funcionar como um amplificador de teorias que desafiam a visão de mundo racional e científica, abrindo portas para uma aceitação generalizada de narrativas conspiratórias.

**Figura 05.** Fluxo de links de convites entre comunidades de reptilianos e criaturas

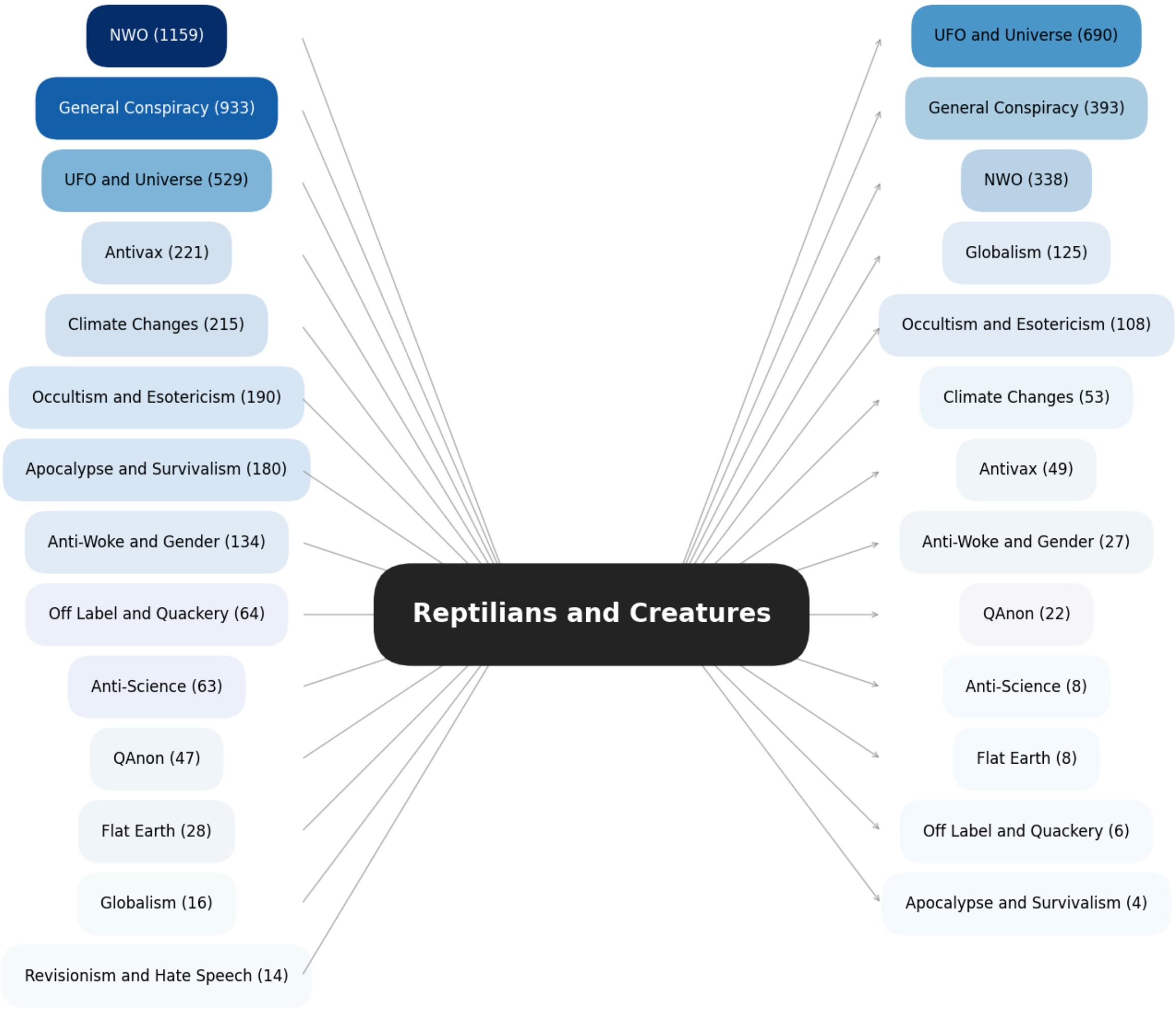

Fonte: Elaboração própria (2024).

O gráfico de Reptilianos e Criaturas destaca a NOM como uma das principais fontes de convites (1.159 links), o que sugere que as teorias sobre criaturas reptilianas estão profundamente enraizadas em narrativas de controle global e dominação. Esse vínculo reflete uma fusão de crenças onde o “outro” é demonizado e visto como parte de um plano mais amplo de subjugação da humanidade. O fato de Reptilianos e Criaturas emitir convites principalmente para OVNIs e Universo (690 links) e Conspirações Gerais (393 links) indica que essa narrativa atua como uma ponte entre a crença em entidades não-humanas e uma visão conspiratória mais abrangente. Essa dinâmica implica que a aceitação da existência de reptilianos não é apenas uma crença isolada, mas parte de um processo mais amplo de radicalização, onde cada narrativa é usada para reforçar a outra. Assim, a temática de Reptilianos e Criaturas não apenas introduz novos adeptos ao mundo conspiratório, mas também serve como um meio de perpetuar e expandir a influência de outras teorias, criando um ciclo de crença e desconfiança que é difícil de romper.

[ versão português / english above ]

### 3.2. Séries temporais

Na análise das séries temporais que se segue, veremos como as discussões sobre OVNIs e Universo, bem como Reptilianos e Criaturas, se intensificaram após 2020. O gráfico demonstra que as menções a OVNIs tiveram um crescimento impressionante, impulsionado por novos relatos governamentais e o aumento do interesse público em fenômenos extraterrestres. Paralelamente, embora em menor escala, as discussões sobre Reptilianos e outras criaturas também cresceram, refletindo um interesse contínuo em teorias mais extremas. Estes dados sugerem que, mesmo em um contexto de crescente interesse por temas extraterrestres apoiados por evidências governamentais, teorias mais radicais continuam a manter sua relevância dentro de nichos específicos.

**Figura 06.** Gráfico de linhas do período

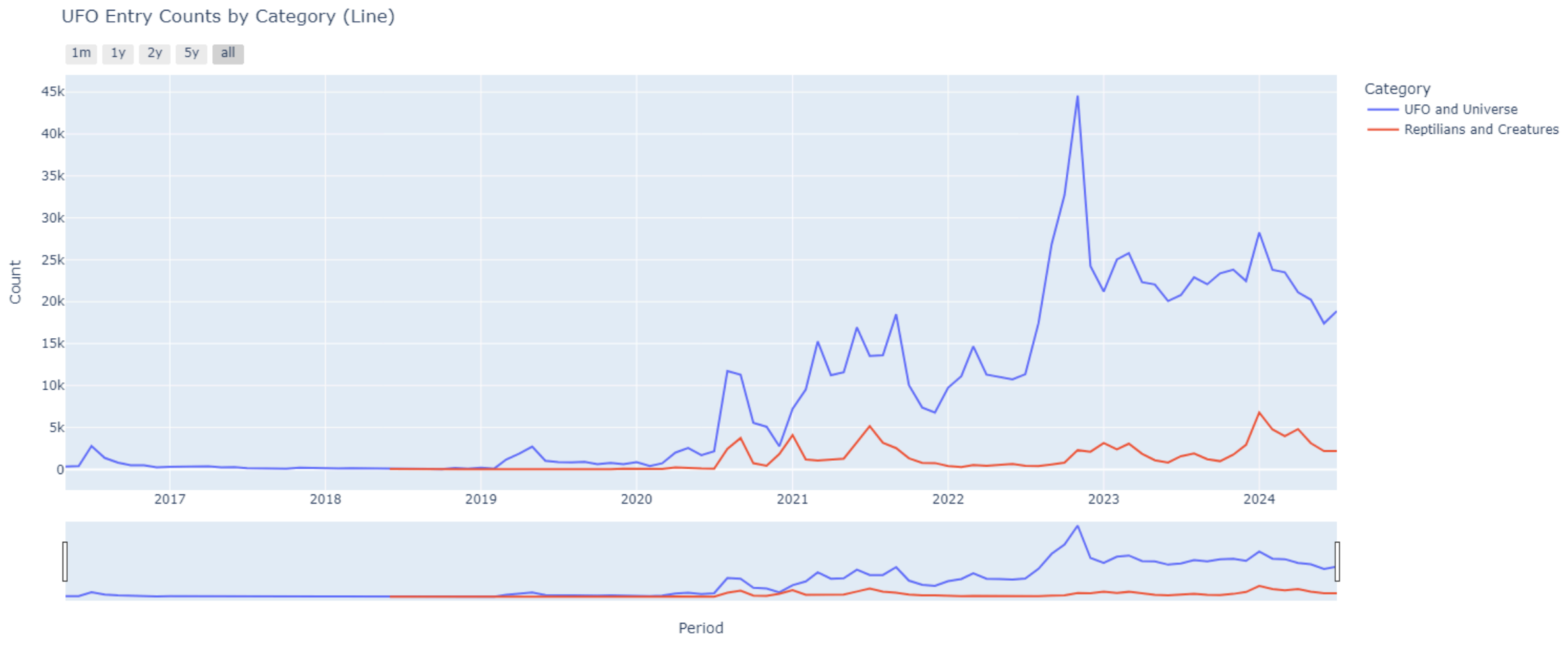

Fonte: Elaboração própria (2024).

Publicações sobre OVNIs e Universo apresentaram um aumento de 4.400% em suas menções entre 2020 e junho de 2021, subindo de 1.000 para 45.000 menções. Este aumento foi impulsionado pela divulgação de relatórios governamentais sobre avistamentos de OVNIs, que revitalizaram o interesse público. Reptilianos e Criaturas apresentaram um aumento menor em suas publicações, de cerca de 1.000% entre 2020 e 2021, subindo de 1.000 para 10.000 menções. Embora menor em volume absoluto comparado a OVNIs, o crescimento percentual é significativo, indicando um interesse renovado por teorias mais extremas em relação à presença de criaturas extraterrestres. Após os picos de 2021, ambos os temas apresentam uma estabilização em níveis mais altos do que os pré-2020, com OVNIs e Universo ainda mantendo cerca de 20.000 menções mensais, e Reptilianos e Criaturas em torno de 5.000, sugerindo que essas narrativas continuam a captar o interesse público.

[ versão português / english above ]

**Figura 07.** Gráfico de área absoluta do período

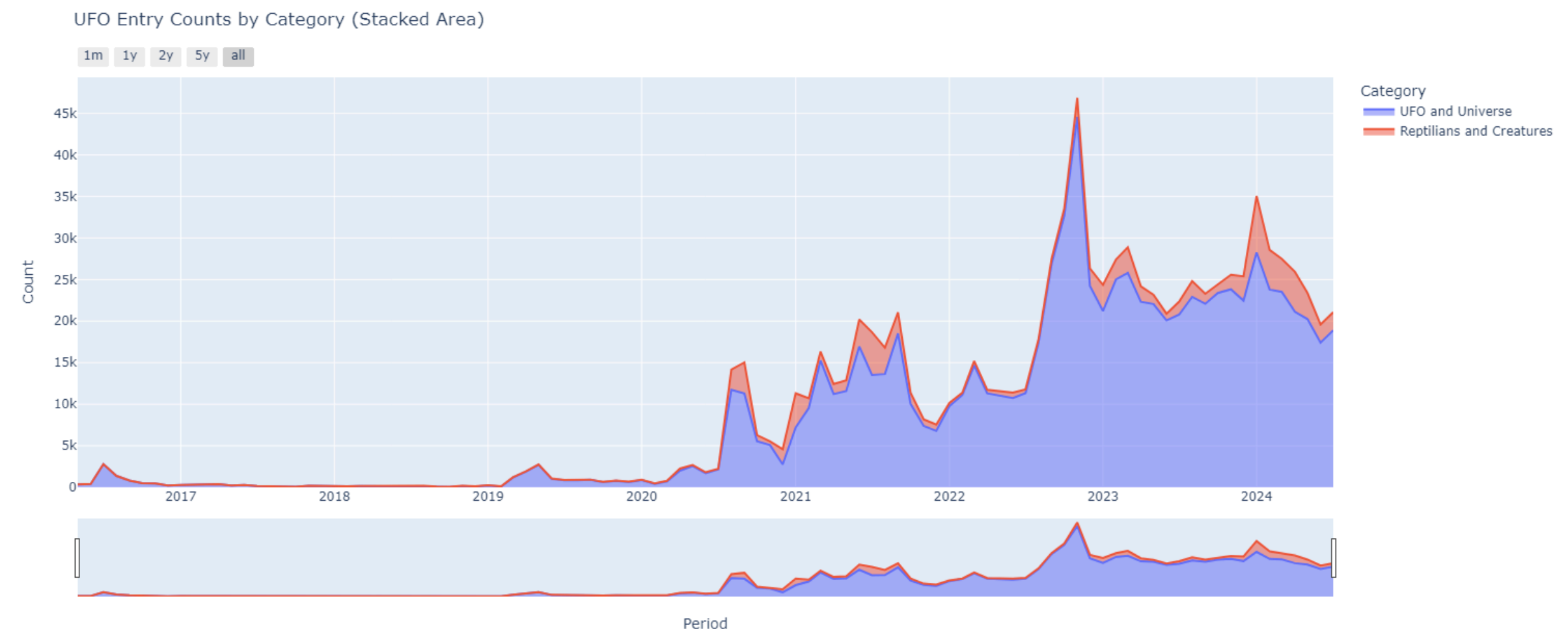

Fonte: Elaboração própria (2024).

O gráfico de área absoluta revela que as discussões em torno de OVNIs e Universo e Reptilianos e Criaturas crescem consideravelmente após 2020, com um pico acentuado em 2022. OVNIs e Universo dominam o gráfico em termos absolutos, especialmente após novos relatos governamentais sobre fenômenos aéreos não identificados e o crescente interesse da mídia em vida extraterrestre. O pico coincide com a liberação de relatórios sobre OVNIs pelo governo dos EUA, que aumentou o interesse público e revitalizou discussões sobre vida alienígena e conspirações relacionadas. Embora Reptilianos e Criaturas tenham uma presença menor, seu crescimento durante os mesmos períodos sugere que essas narrativas estão entrelaçadas, muitas vezes vinculadas a teorias de controle global por entidades não humanas.

**Figura 08.** Gráfico de área relativa do período

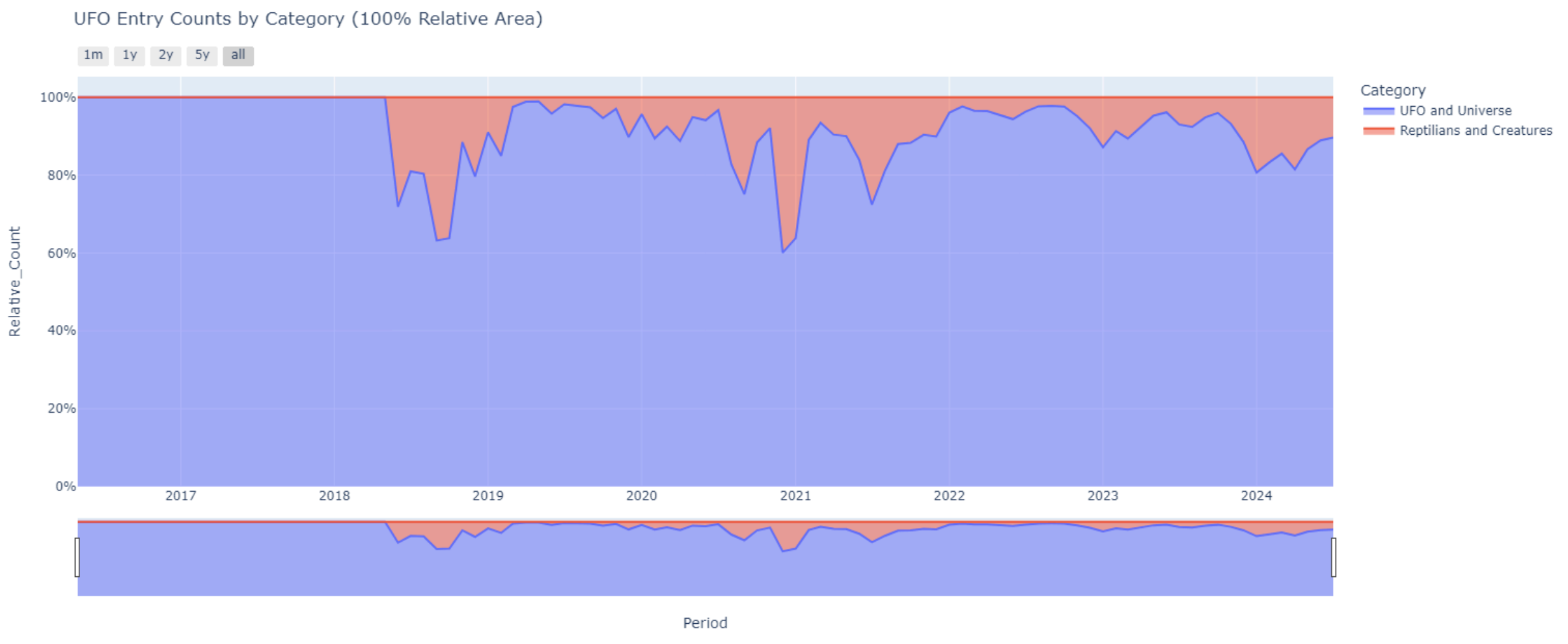

Fonte: Elaboração própria (2024).

O gráfico de área relativa do período destaca como publicações das comunidades centradas em OVNIs e Universo continuam a dominar as discussões ao longo do tempo, representando a maior parte das entradas, especialmente após 2021. Reptilianos e Criaturas, embora menos predominante, ainda mantém uma presença constante, com picos menores que indicam uma persistência dessa narrativa dentro do discurso conspiratório. Este gráfico sugere que, enquanto as discussões sobre OVNIs ganham uma relevância mais mainstream, especialmente com o apoio de relatórios governamentais, as teorias sobre reptilianos permanecem mais restritas a nichos específicos, mas continuam sendo um componente significativo dentro das discussões mais amplas sobre conspirações globais. A análise mostra como diferentes facetas das teorias de conspiração extraterrestres podem coexistir e se reforçar dentro de um mesmo ecossistema de desinformação.

### 3.3. Análise de conteúdo

A análise de conteúdo das comunidades de OVNIs, universo, reptilianos e criaturas através de nuvens de palavras permite explorar como essas temáticas interagem e se consolidam no imaginário coletivo de seus membros ao longo do tempo. As palavras destacadas nas nuvens, como “vida”, “luz”, “tempo” e “Deus”, revelam uma profunda conexão entre essas crenças e uma visão de mundo que vai além do meramente científico, envolvendo dimensões espirituais, esotéricas e existenciais. Essas narrativas se entrelaçam com um sentido de urgência e transformação, refletido nos termos “agora” e “energia”, que indicam tanto uma busca por compreensão quanto uma preparação para eventos cósmicos ou mudanças globais. A análise dessas palavras-chave oferece uma compreensão mais profunda de como essas comunidades estruturam seus discursos e mantêm suas crenças, muitas vezes em oposição à ciência convencional, e como utilizam plataformas para perpetuar essas ideias.

**Figura 09.** Nuvem de palavras consolidadas de OVNIs, universo, reptilianos e criaturas

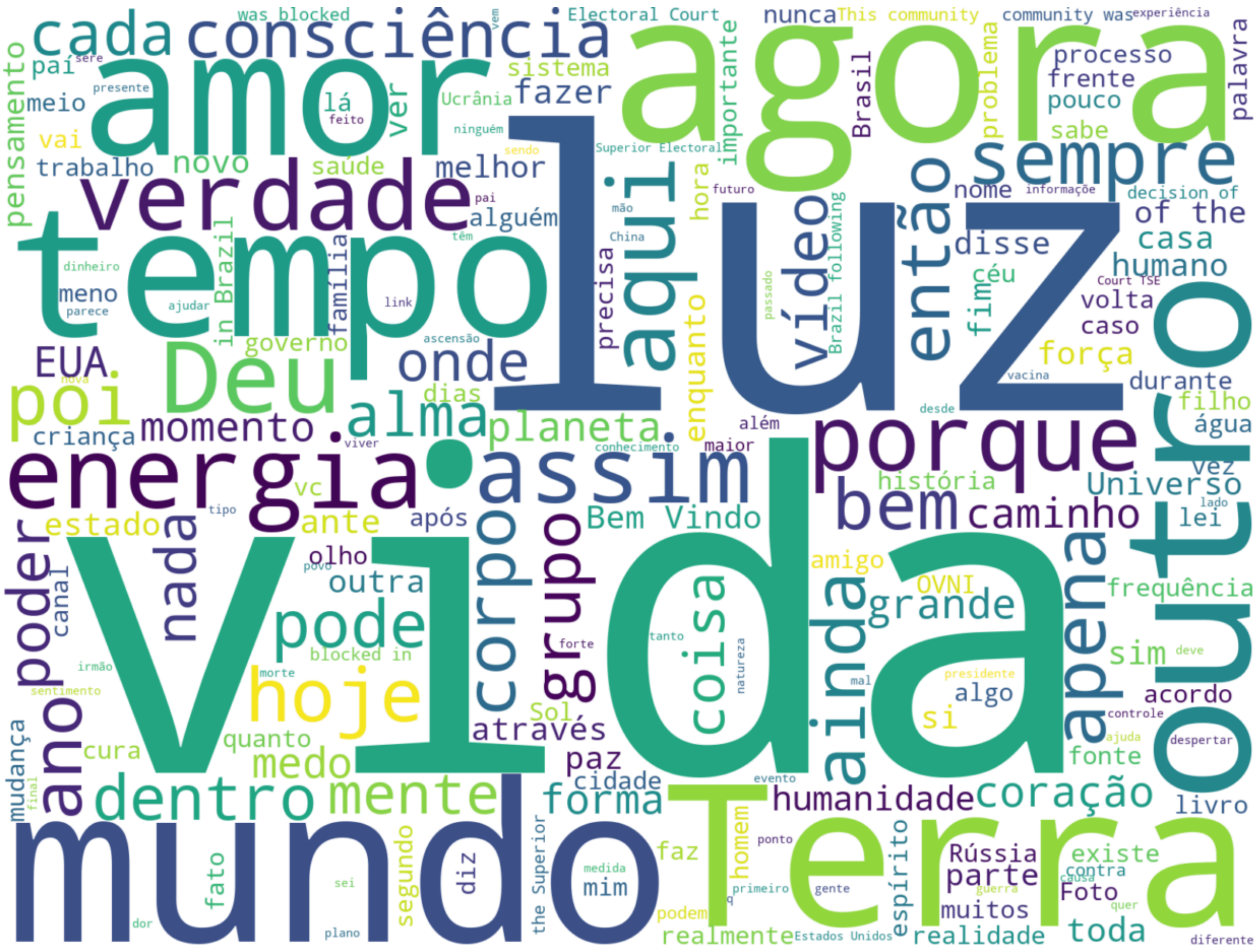

Fonte: Elaboração própria (2024).

A nuvem de palavras consolidada das temáticas de OVNIs, universo, reptilianos e criaturas evidencia uma concentração em termos como “vida”, “luz”, “tempo”, “Deus” e “energia”. A palavra “vida” aparece com destaque, sugerindo que a discussão gira em torno de conceitos que transcendem o físico e o material, envolvendo também questões espirituais e existenciais. A presença recorrente de “luz” e “energia” indica uma preocupação com forças invisíveis ou cósmicas, que são frequentemente interpretadas como componentes essenciais de uma realidade maior que a percebida. Termos como “Deus” e “verdade” reforçam a ideia de que essas discussões são moldadas por uma busca por significados profundos e verdades ocultas, muitas vezes associadas a teorias da conspiração ou visões alternativas do universo. O uso de “tempo” e “agora” sugere um senso de urgência ou iminência, possivelmente relacionado a eventos previstos ou transformações esperadas no cenário global ou cósmico.

[ versão português / english above ]

**Quadro 01.** Nuvem de palavras em série temporal de OVNIs,e universo

2016

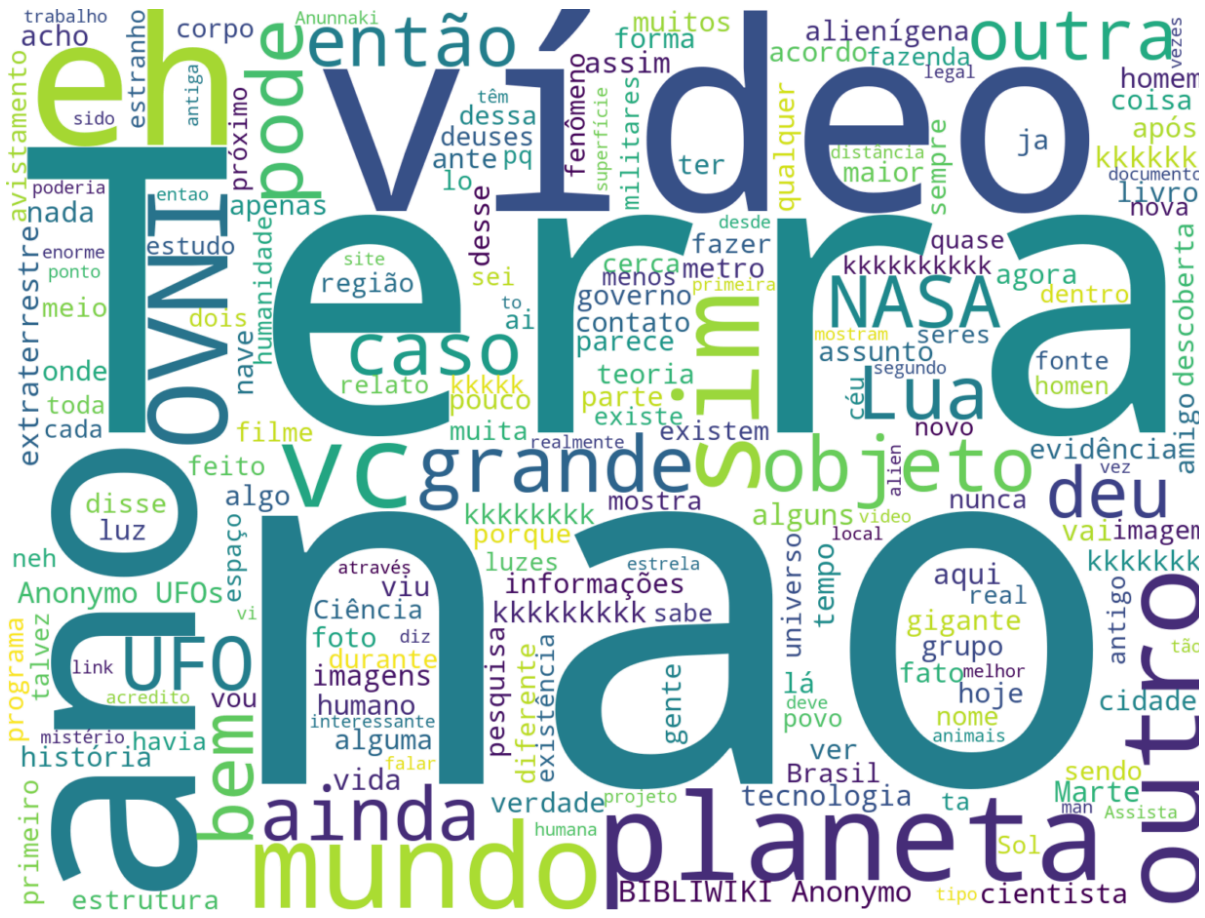

2017

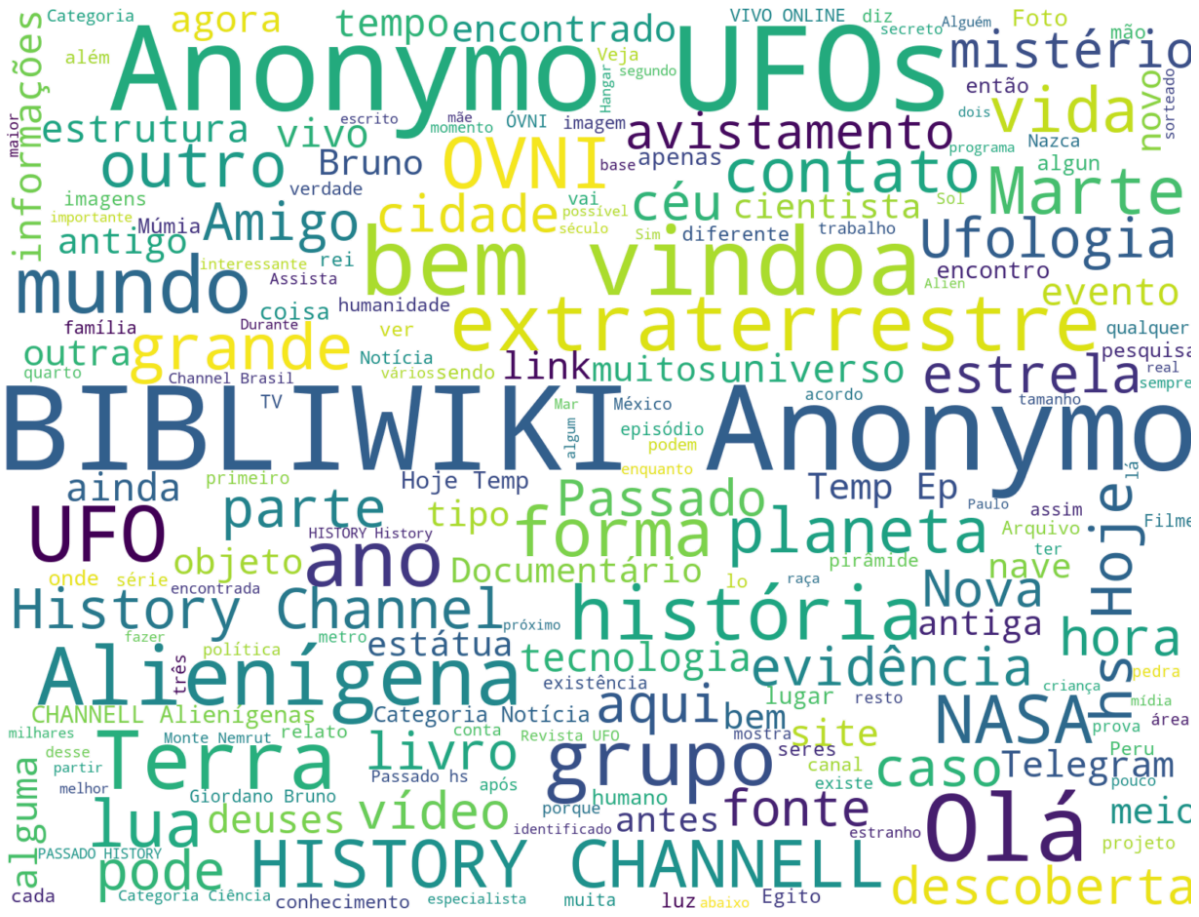

2018

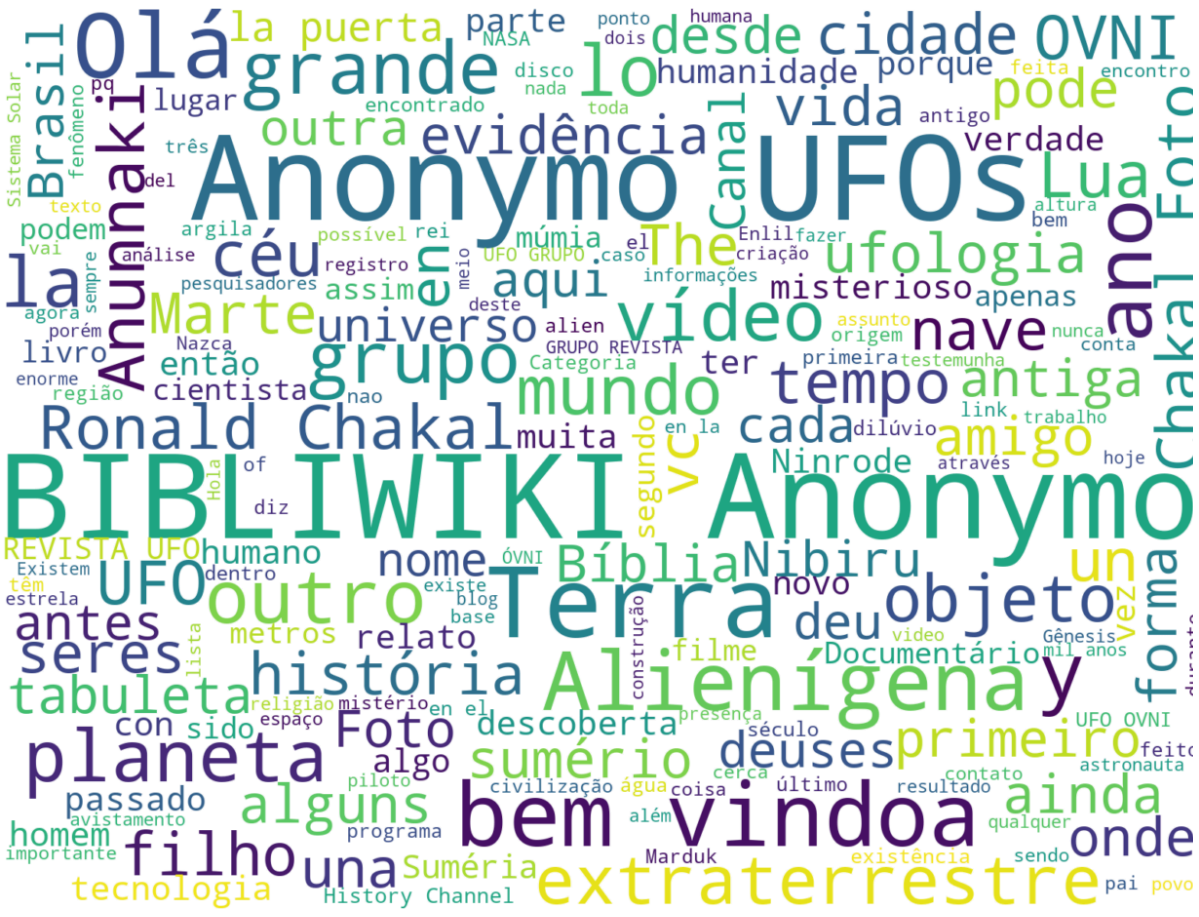

2019

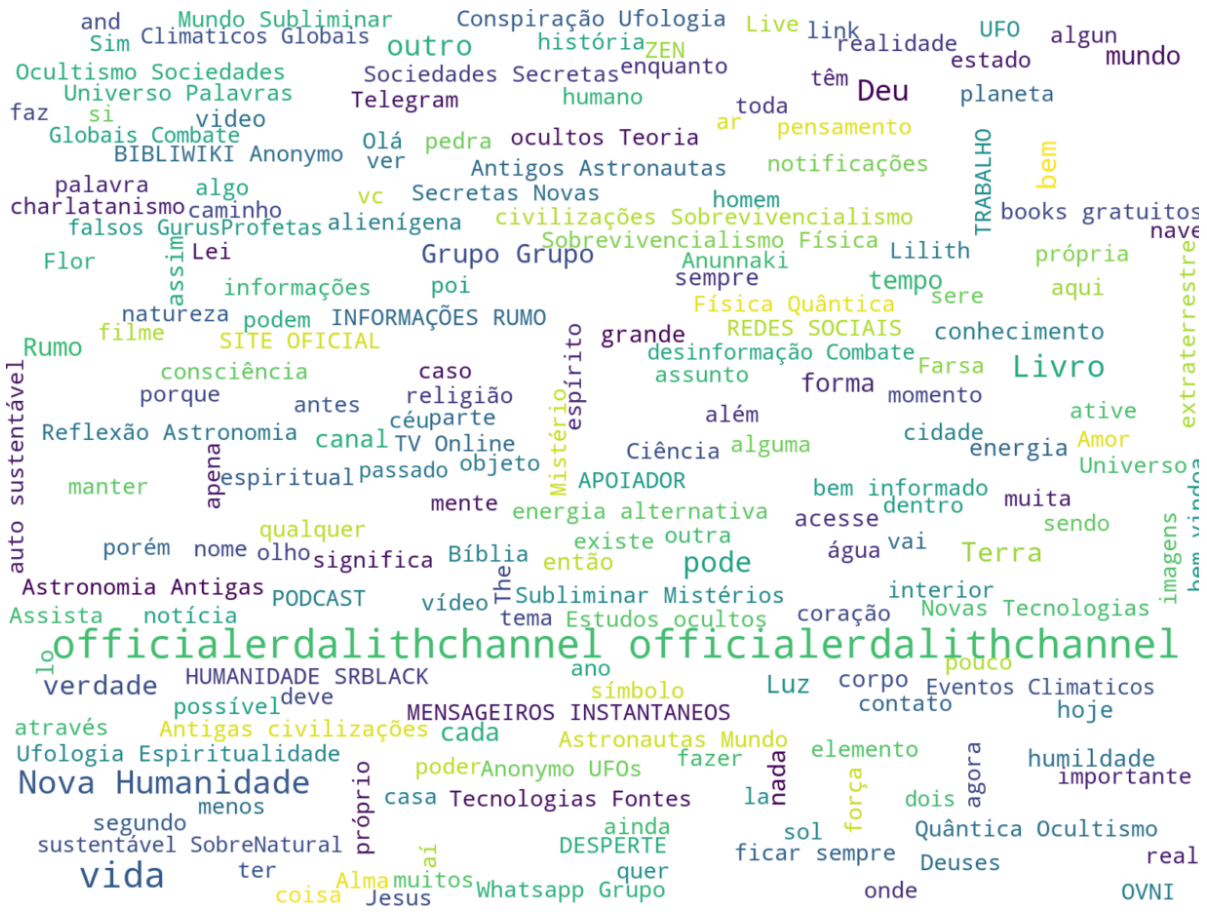

2020

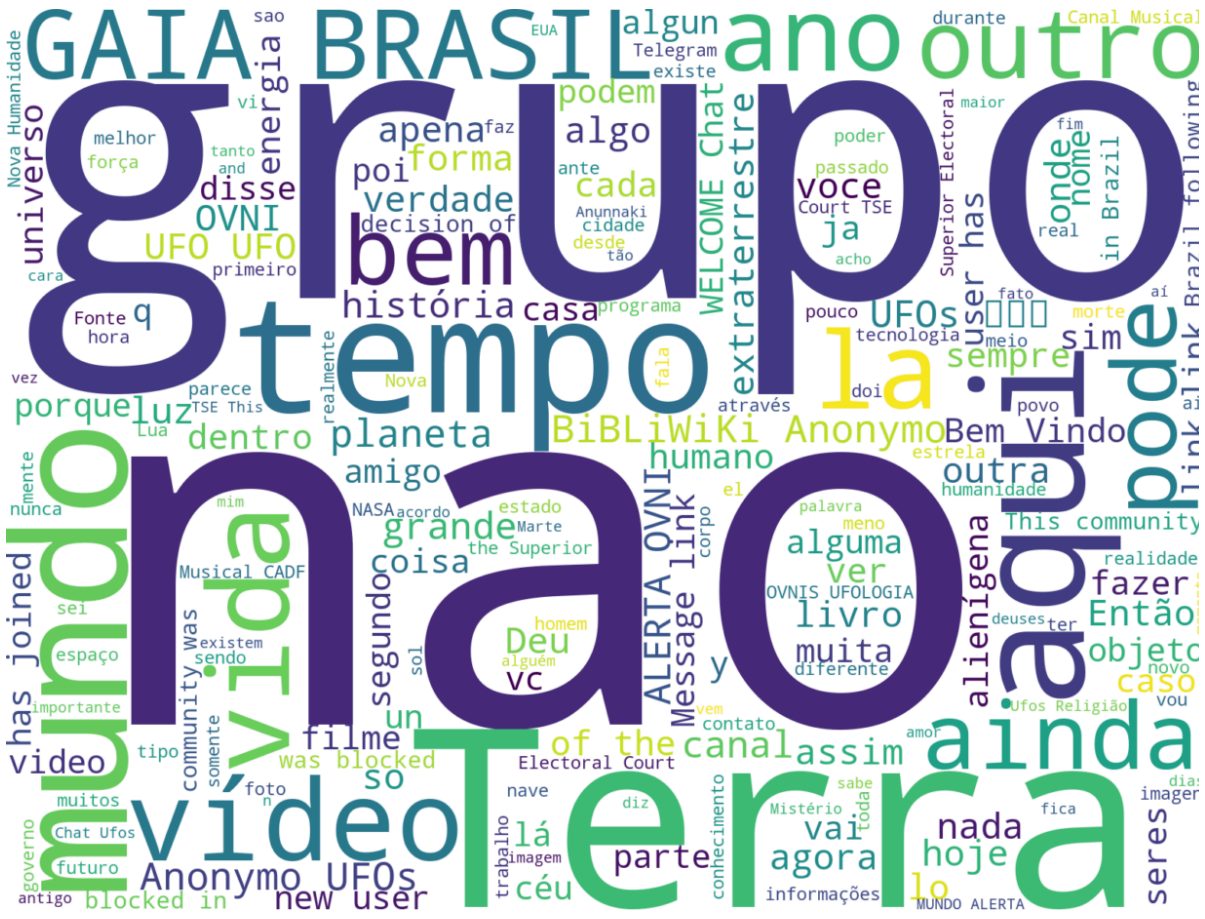

2021

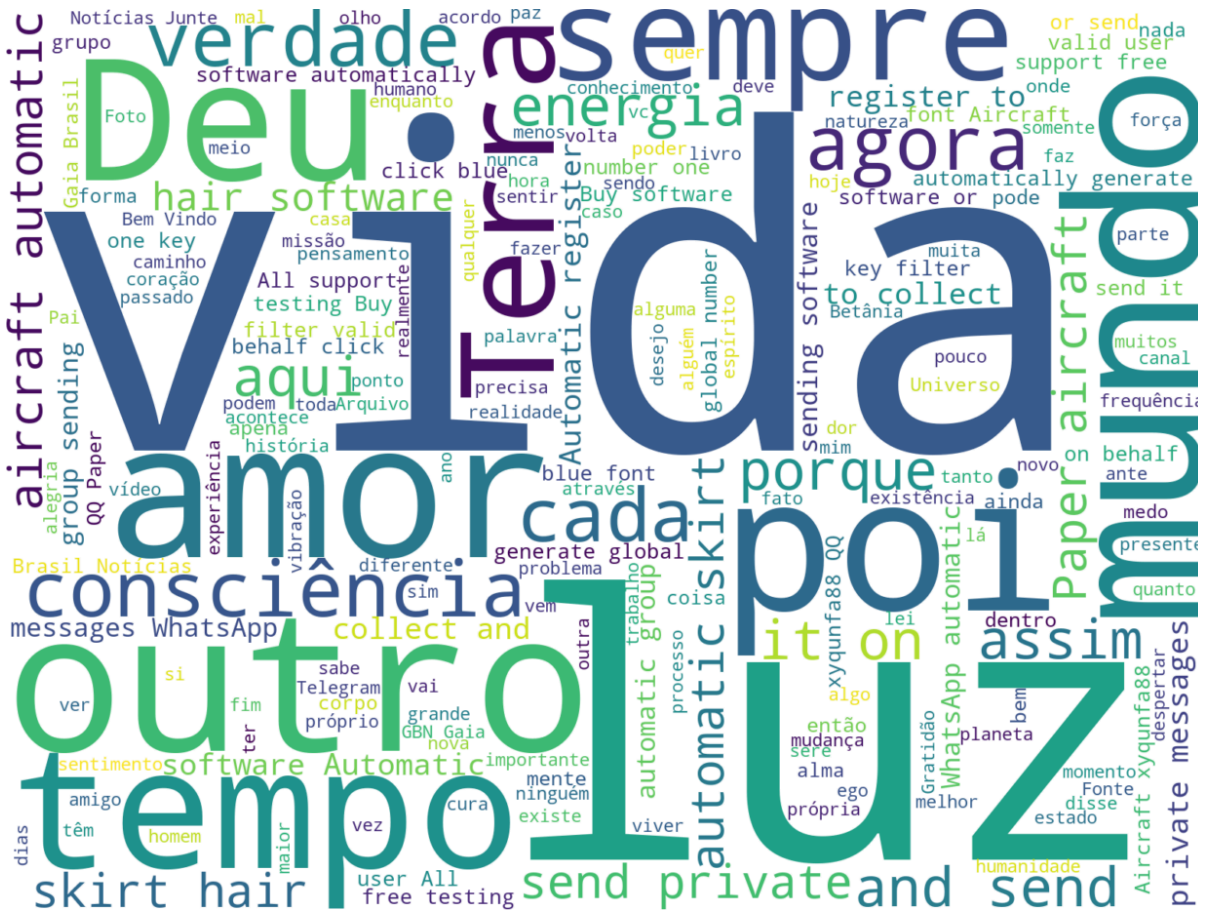

[ versão português / english above ]

2022

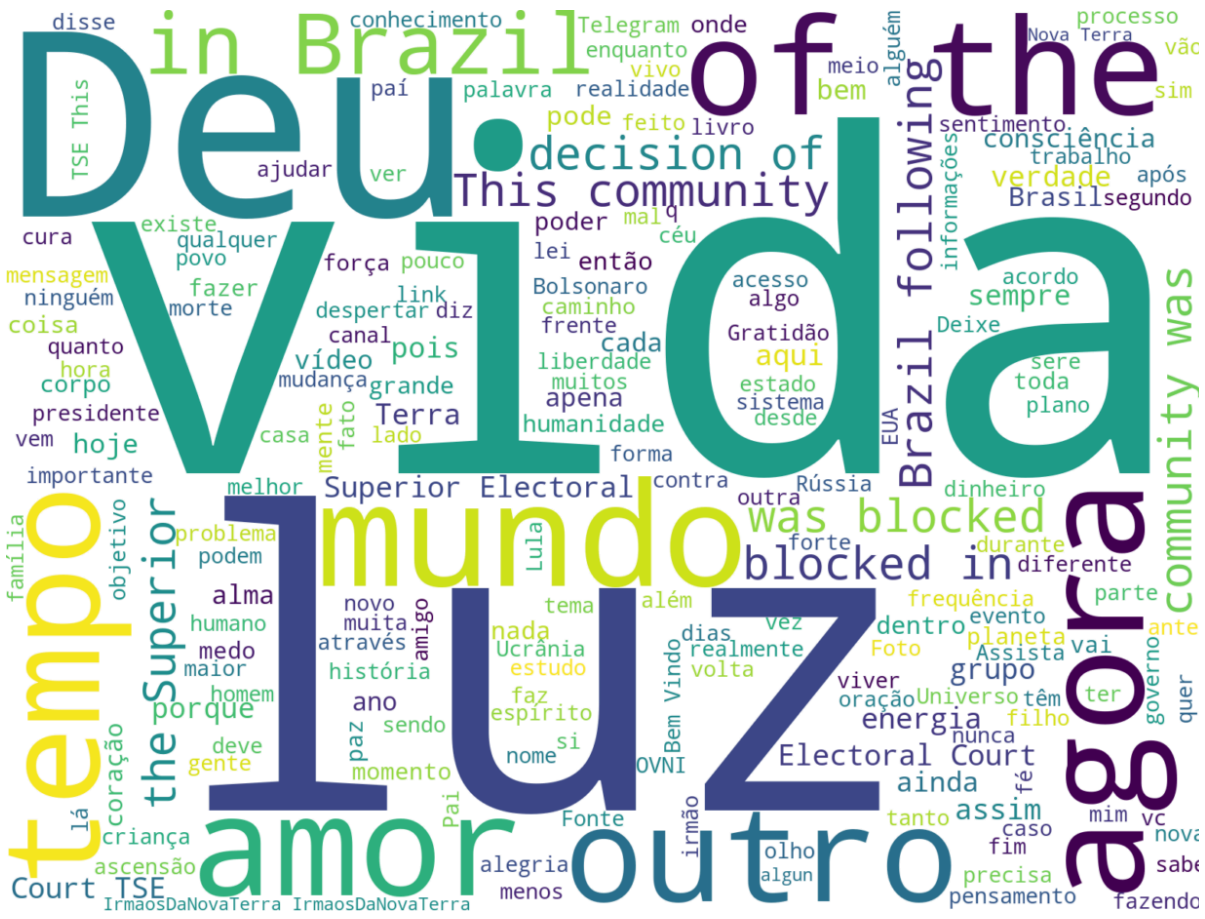

2023

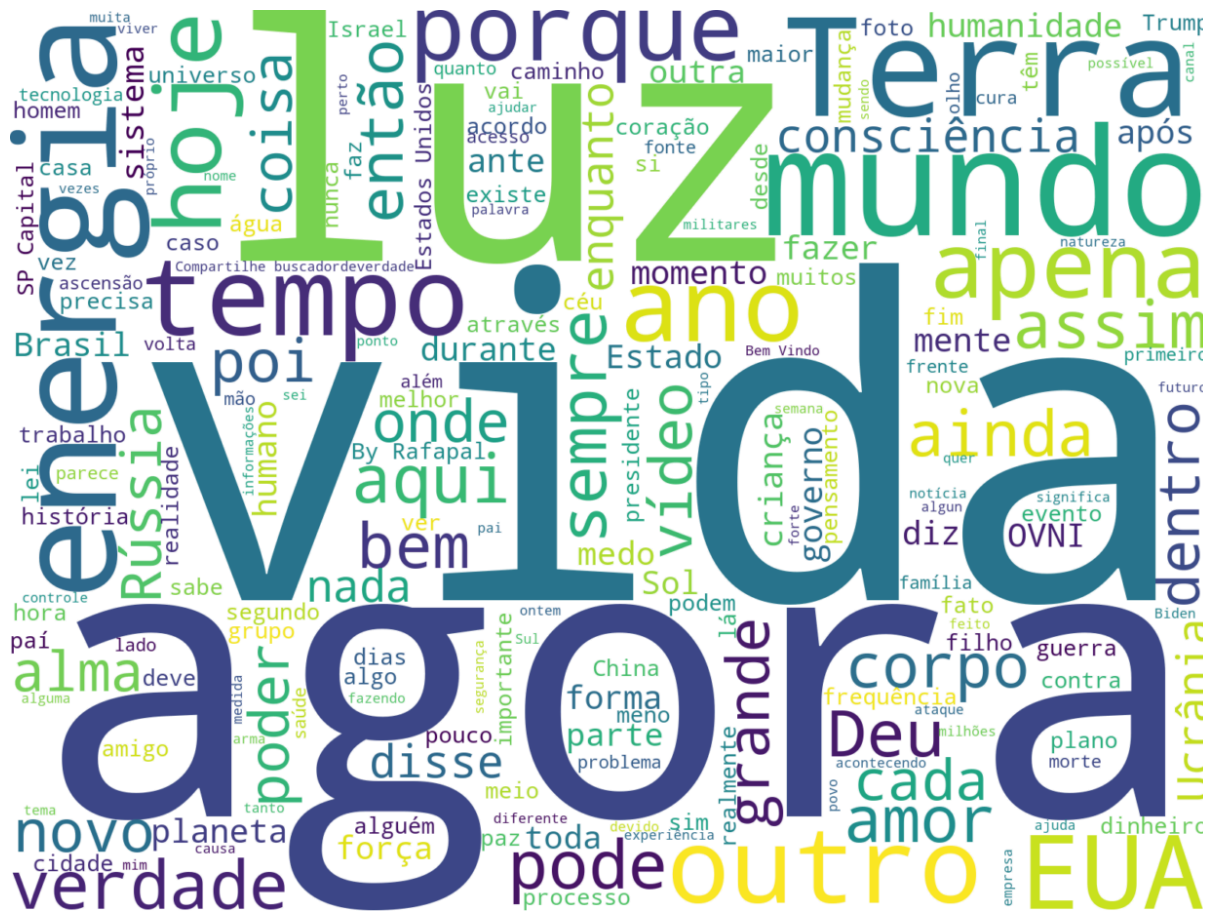

2024

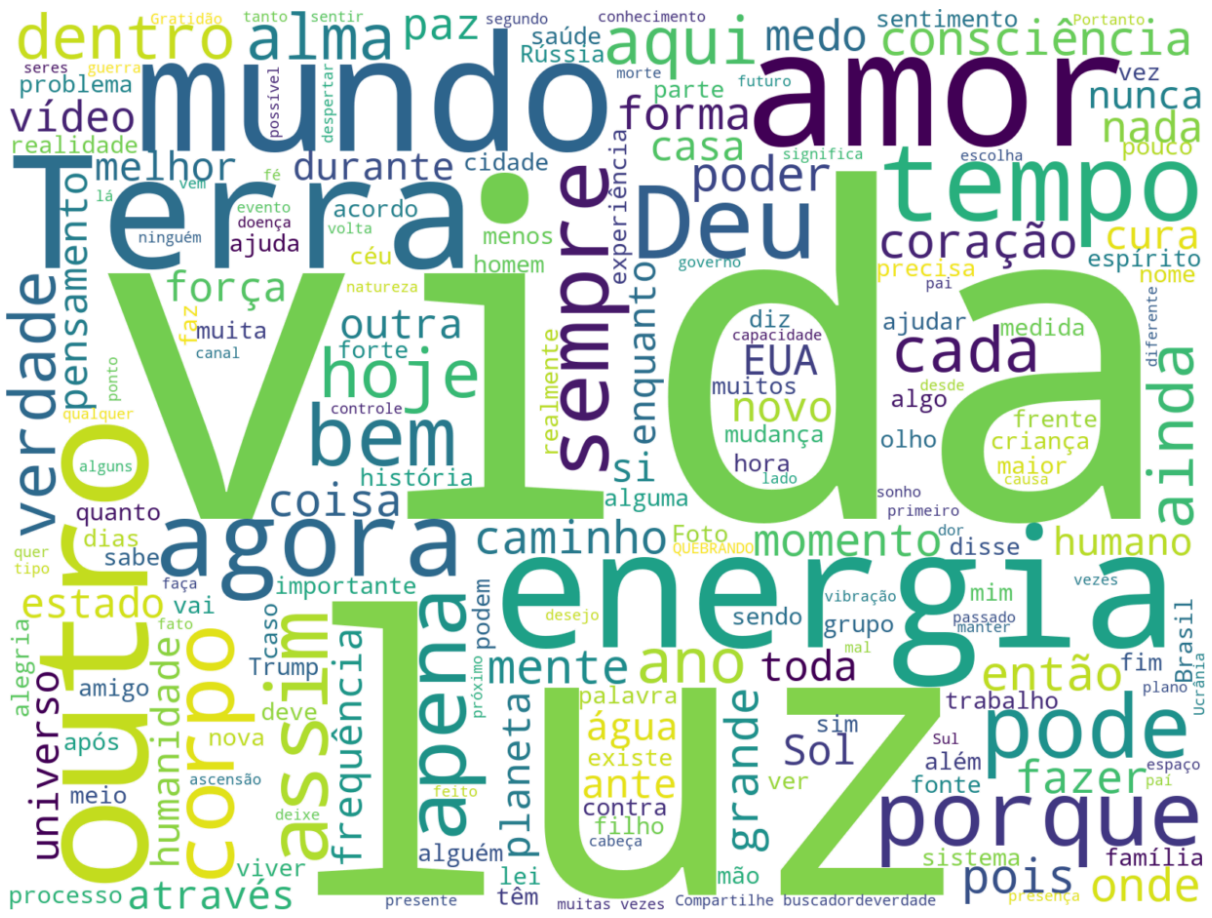

Fonte: Elaboração própria (2024).

No Quadro 01, a análise temporal das palavras revela como as discussões sobre OVNIs e universo evoluíram entre 2016 e 2024. Em 2016, termos como “vídeo” e “cara” dominam as discussões, possivelmente indicando um foco na evidência visual e em debates pessoais sobre a existência de vida extraterrestre. Ao longo dos anos, observa-se uma transição para temas mais abstratos e espirituais, com a entrada de palavras como “vida”, “luz” e “Deus” em 2020, refletindo uma mudança do concreto para o esotérico. Em 2022 e 2023, as discussões parecem se consolidar em torno da ideia de “vida” e “tempo”, sugerindo uma preocupação crescente com questões existenciais e a relação do ser humano com o universo. O ano de 2024 mantém essa tendência, com a inclusão de “energia” e “consciência”, reforçando a visão de que essas comunidades buscam um entendimento do cosmos.

[ versão português / english above ]

**Quadro 01.** Nuvem de palavras em série temporal de reptilianos e criaturas

2018

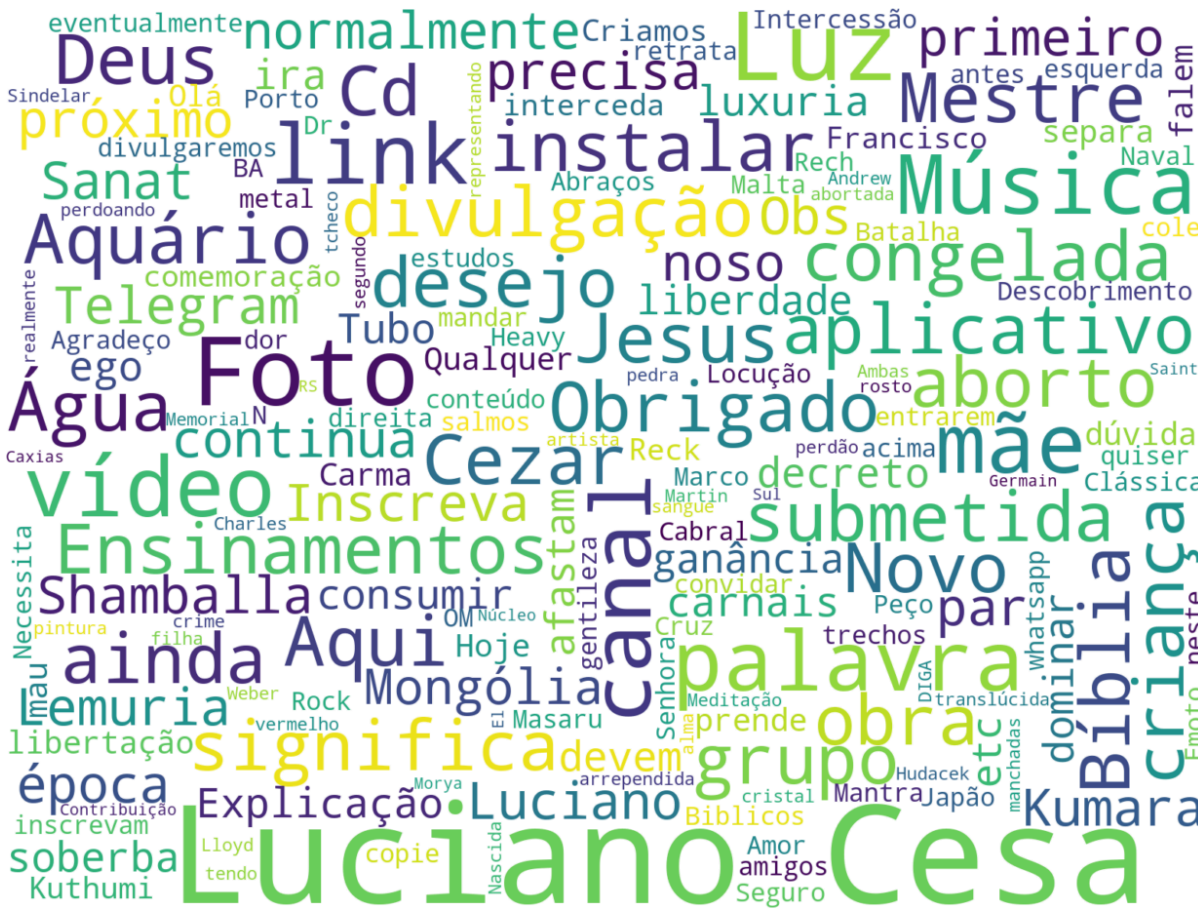

2019

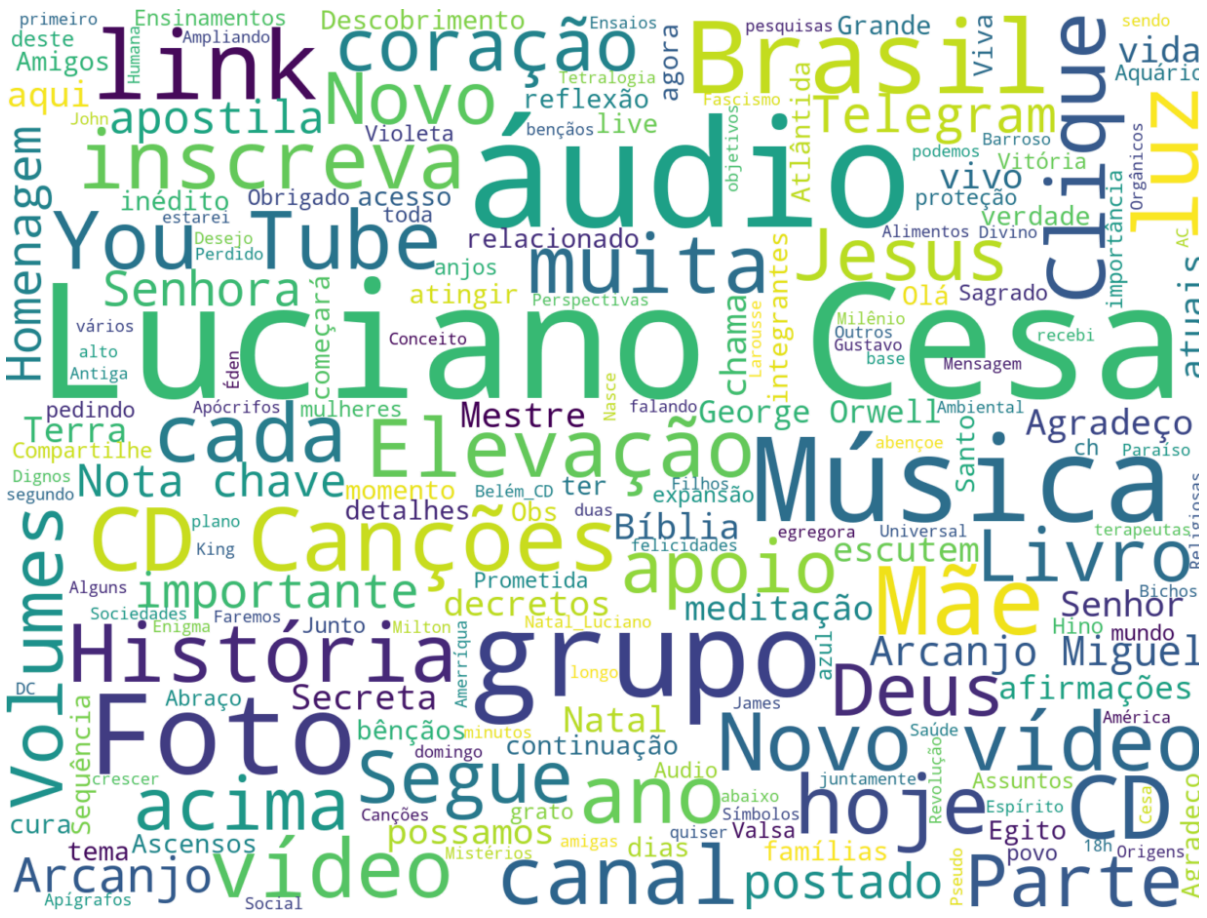

2020

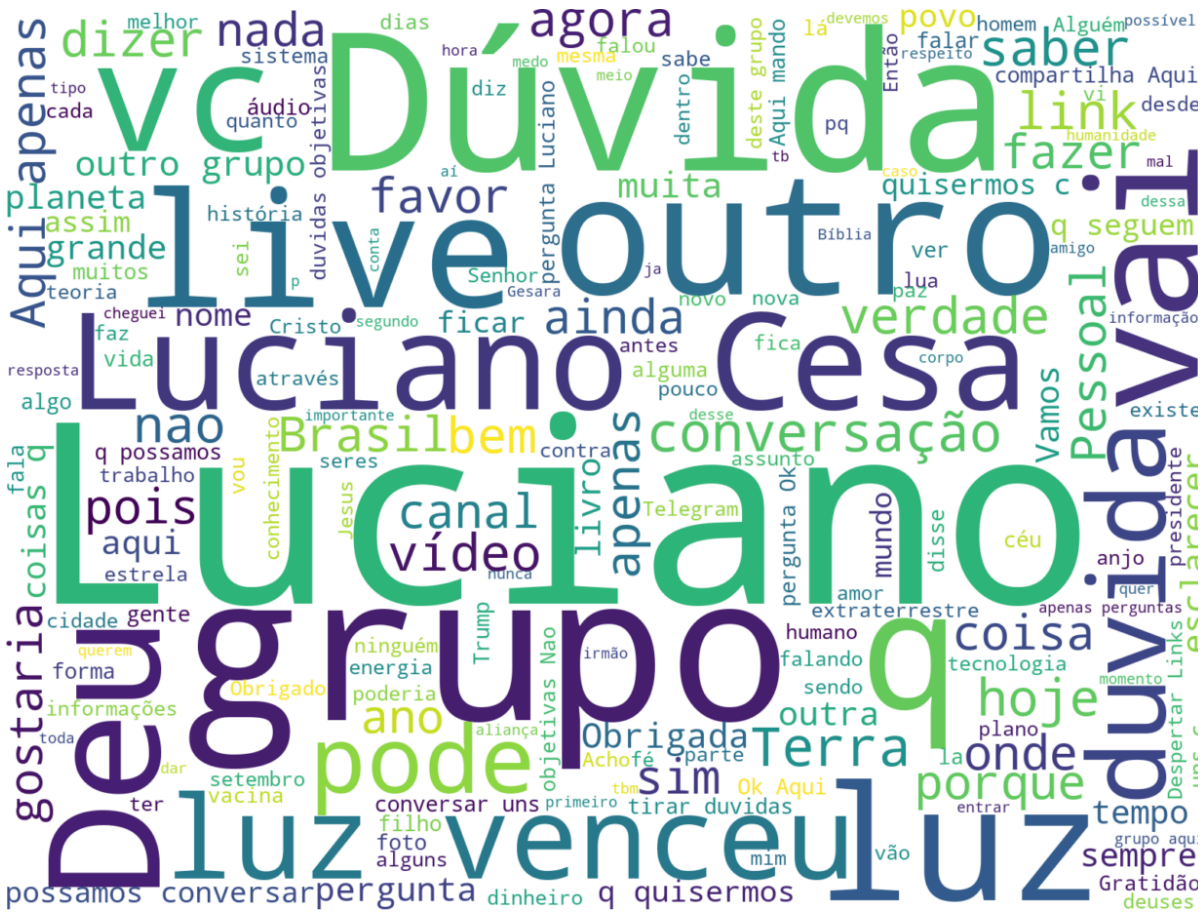

2021

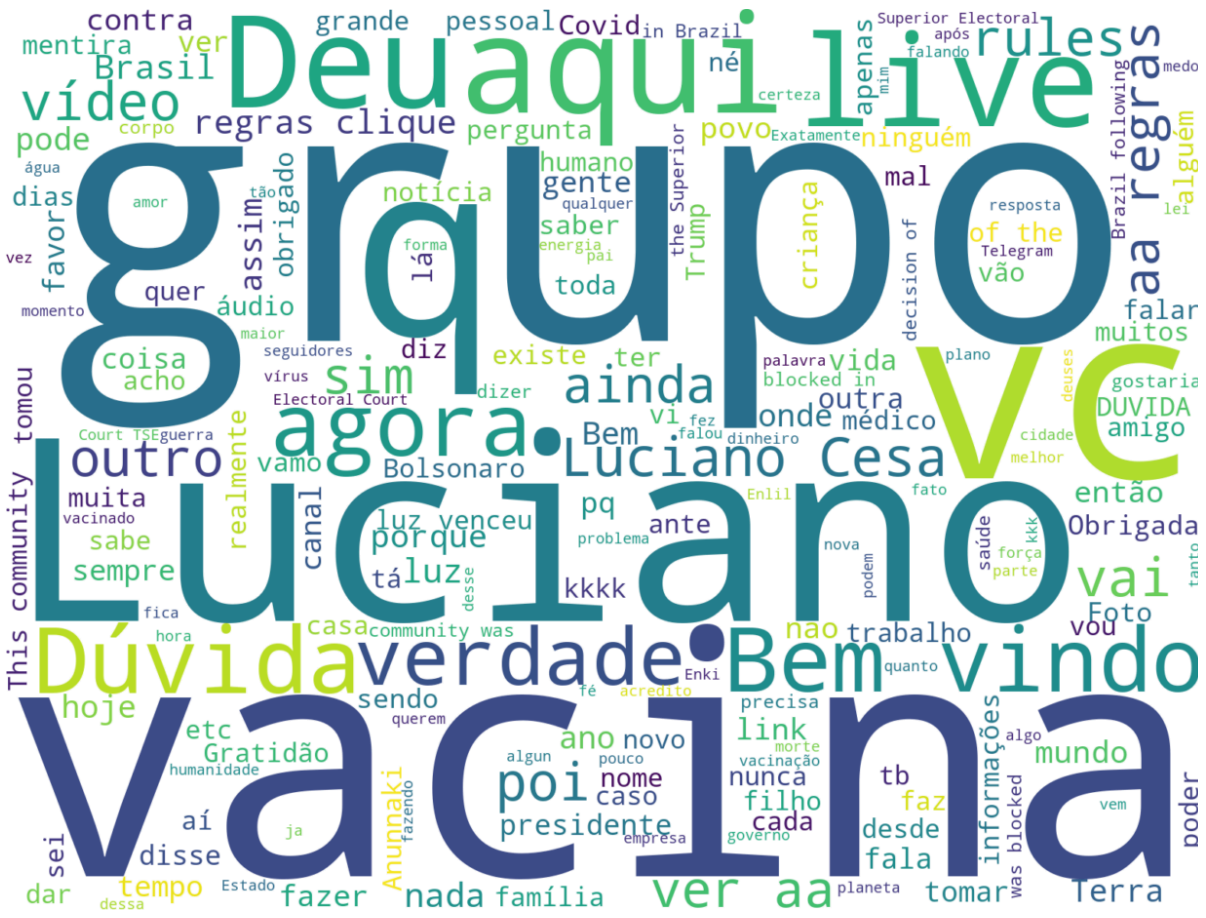

2022

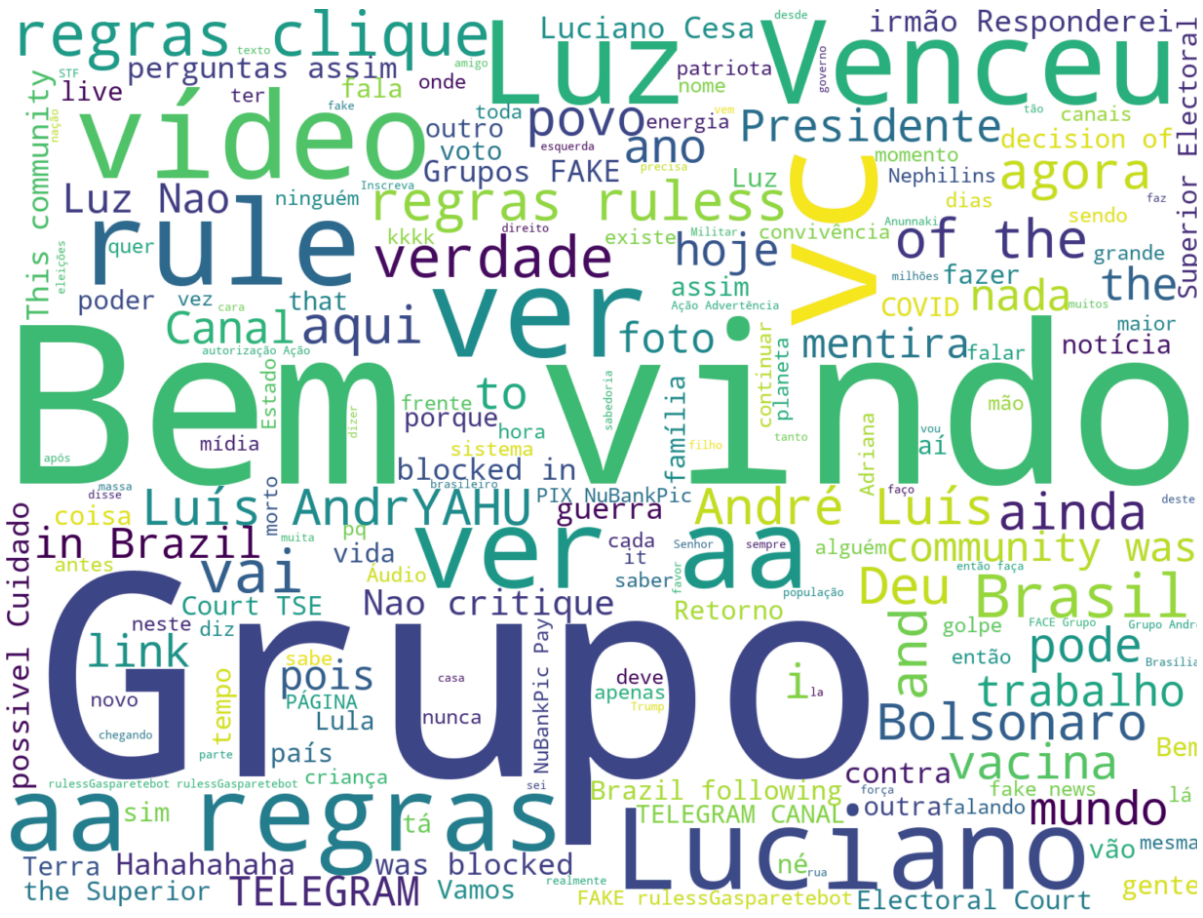

2023

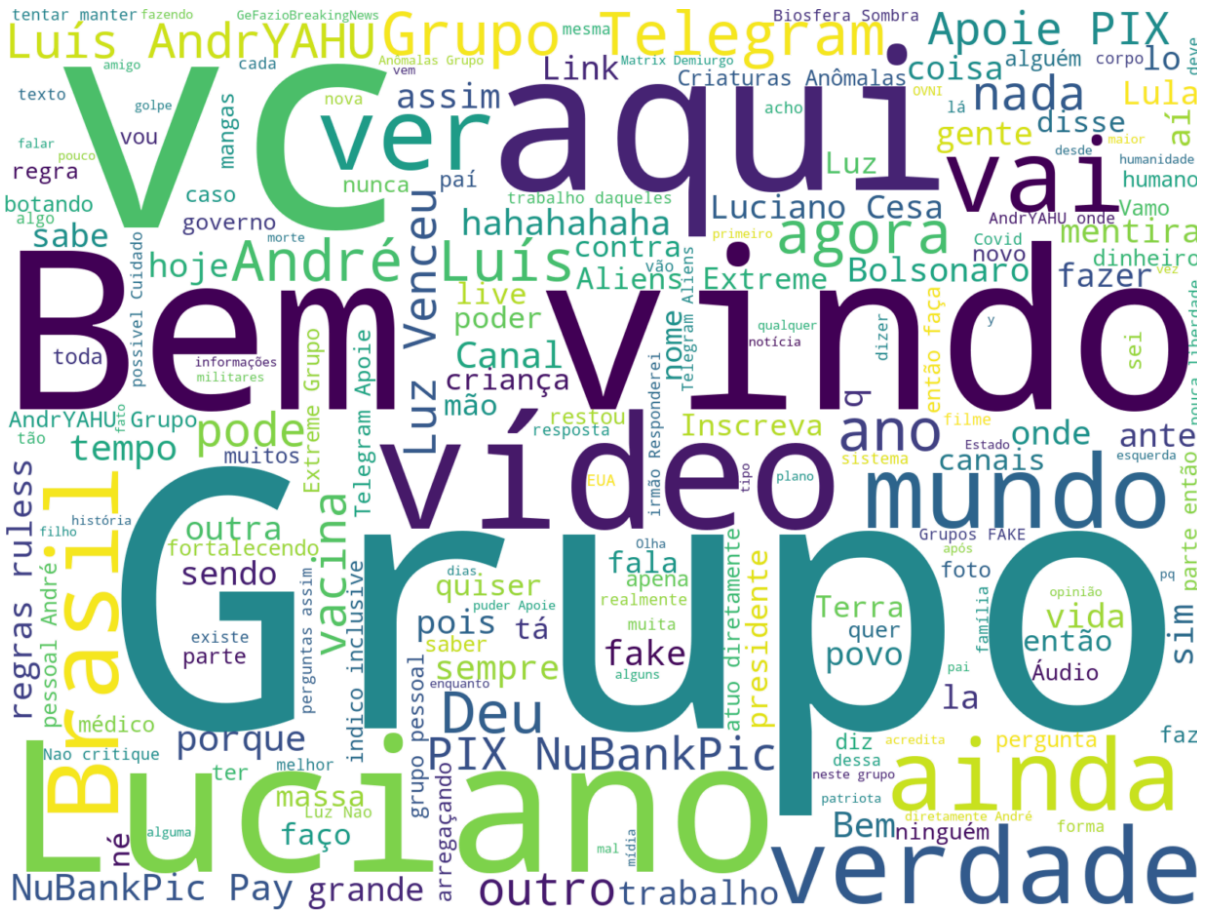

[ versão português / english above ]

2024

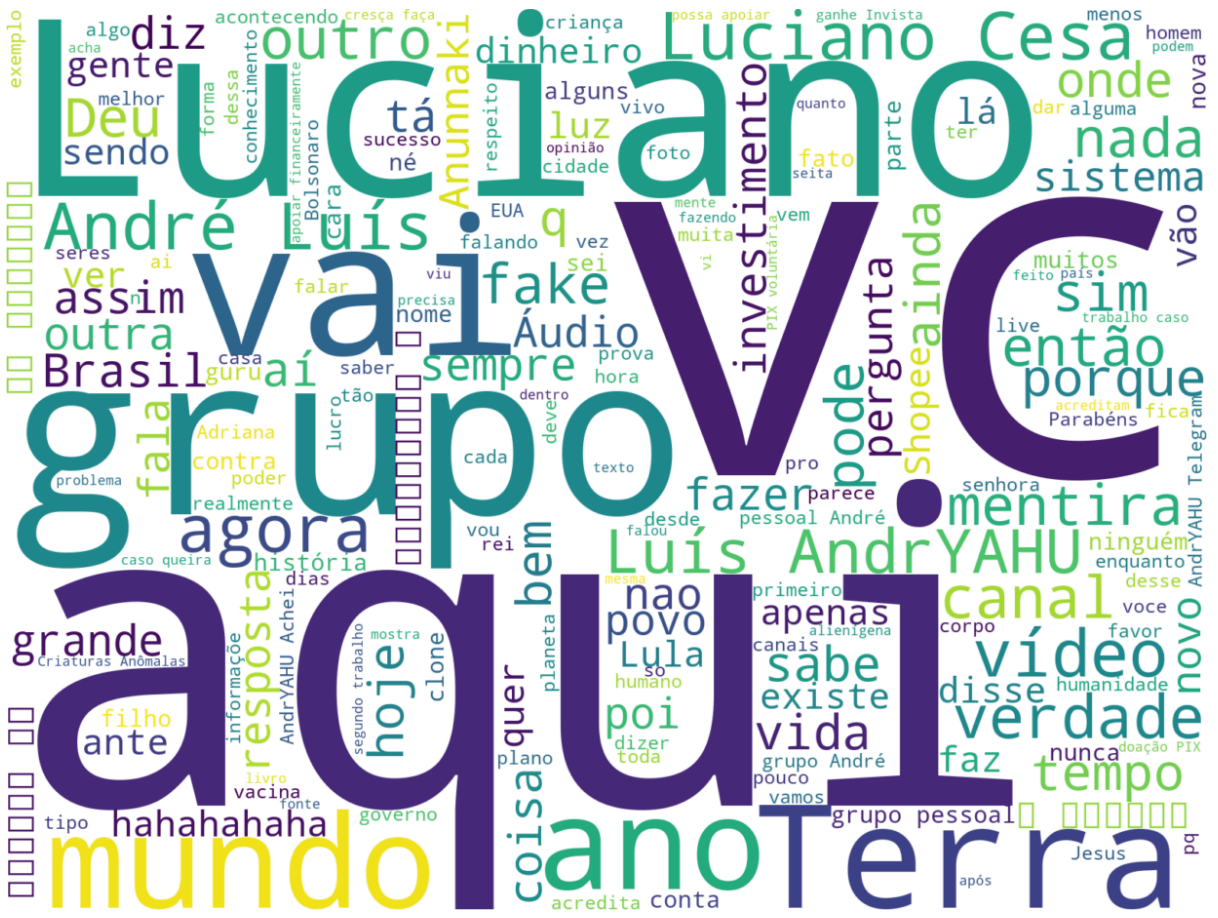

Fonte: Elaboração própria (2024).

O Quadro 02 apresenta a evolução das discussões sobre reptilianos e criaturas ao longo do tempo, de 2018 a 2024. Em 2018, a palavra "Luciano Cesa" (*coach espiritual*) destaca-se, indicando a importância de figuras específicas dentro dessas comunidades para a disseminação de narrativas. A palavra "grupo" também aparece com frequência, sugerindo que essas discussões são amplamente colaborativas e ocorrem em ambientes comunitários, como grupos online. Ao longo dos anos, vemos uma repetição e reafirmação desses termos, especialmente "grupo" e "Luciano Cesa", o que pode indicar a formação de subculturas ou grupos específicos dentro das comunidades mais amplas, que seguem líderes ou influenciadores particulares. A presença de "vacina" em 2022 e 2023 sugere uma interseção entre as narrativas de reptilianos e as discussões sobre saúde pública e Pandemia, destacando como essas teorias da conspiração podem se adaptar a eventos atuais para manter relevância e atratividade. Em 2024, a constância de termos como "vida", "verdade" e "grupo" sugere uma continuidade dessas narrativas, onde a busca por uma compreensão alternativa do mundo continua a ser mediada por figuras centrais e comunidades coesas.

### 3.4. Sobreposição de agenda temática

As figuras a seguir analisam como as temáticas relacionadas a OVNIs, Universo, Reptilianos e Criaturas se sobrepõem em discussões dentro de comunidades de teorias da conspiração. Essas temáticas não apenas atraem o interesse de indivíduos curiosos sobre o desconhecido, mas também são utilizadas para sustentar narrativas de desconfiança em relação a instituições e eventos globais. Ao integrar tópicos como extraterrestres, desaparecimentos misteriosos e tragédias naturais planejadas, essas comunidades constroem um discurso coeso que reforça crenças conspiratórias e dificulta a correção factual.

**Figura 10.** Temáticas de OVNIs

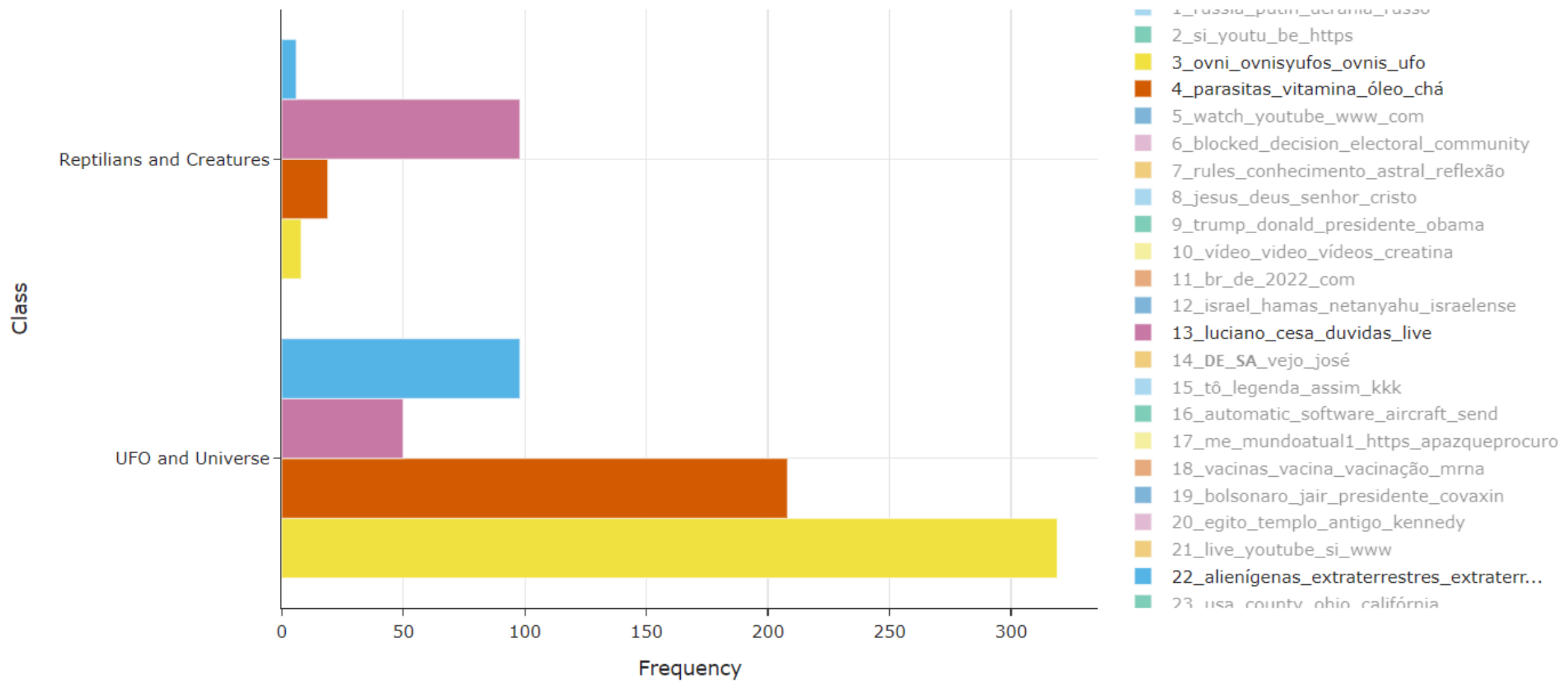

Fonte: Elaboração própria (2024).

A Figura 10 destaca como os tópicos relacionados a OVNIs e ao Universo dominam as discussões nessas comunidades, em comparação com os temas sobre Reptilianos e Criaturas. Tópicos como “alienígenas” e “extraterrestres” aparecem com alta frequência, indicando que essas comunidades estão profundamente envolvidas em narrativas que sugerem interações ou conspirações envolvendo seres de outros planetas. Essas discussões frequentemente se misturam com teorias sobre governos ocultando informações, criando uma narrativa que sustenta a desconfiança generalizada em relação a autoridades.

**Figura 11.** Temáticas de geopolítica e disputas eleitorais

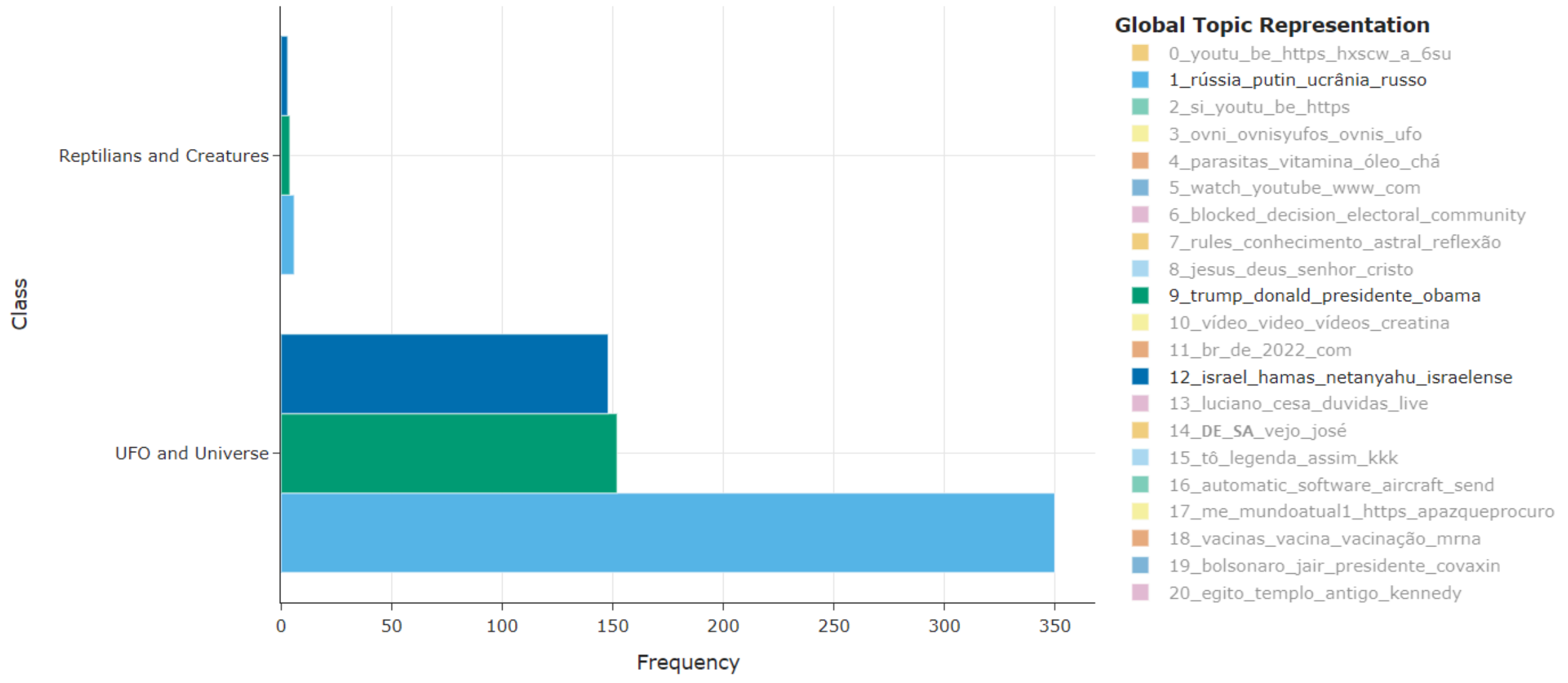

Fonte: Elaboração própria (2024).

Na Figura 11, a sobreposição de temas geopolíticos com discussões sobre OVNIs e Universo é evidente. Tópicos como “Rússia” e “Putin” são frequentemente associados a narrativas de conspirações globais, sugerindo que eventos geopolíticos são interpretados como parte de um plano maior envolvendo seres extraterrestres. A integração desses temas com disputas eleitorais reflete a politização das teorias de conspiração, onde líderes políticos são vistos como figuras centrais em conspirações envolvendo forças além do nosso planeta.

**Figura 12.** Temáticas de supostos desaparecimentos ou quedas de aviões

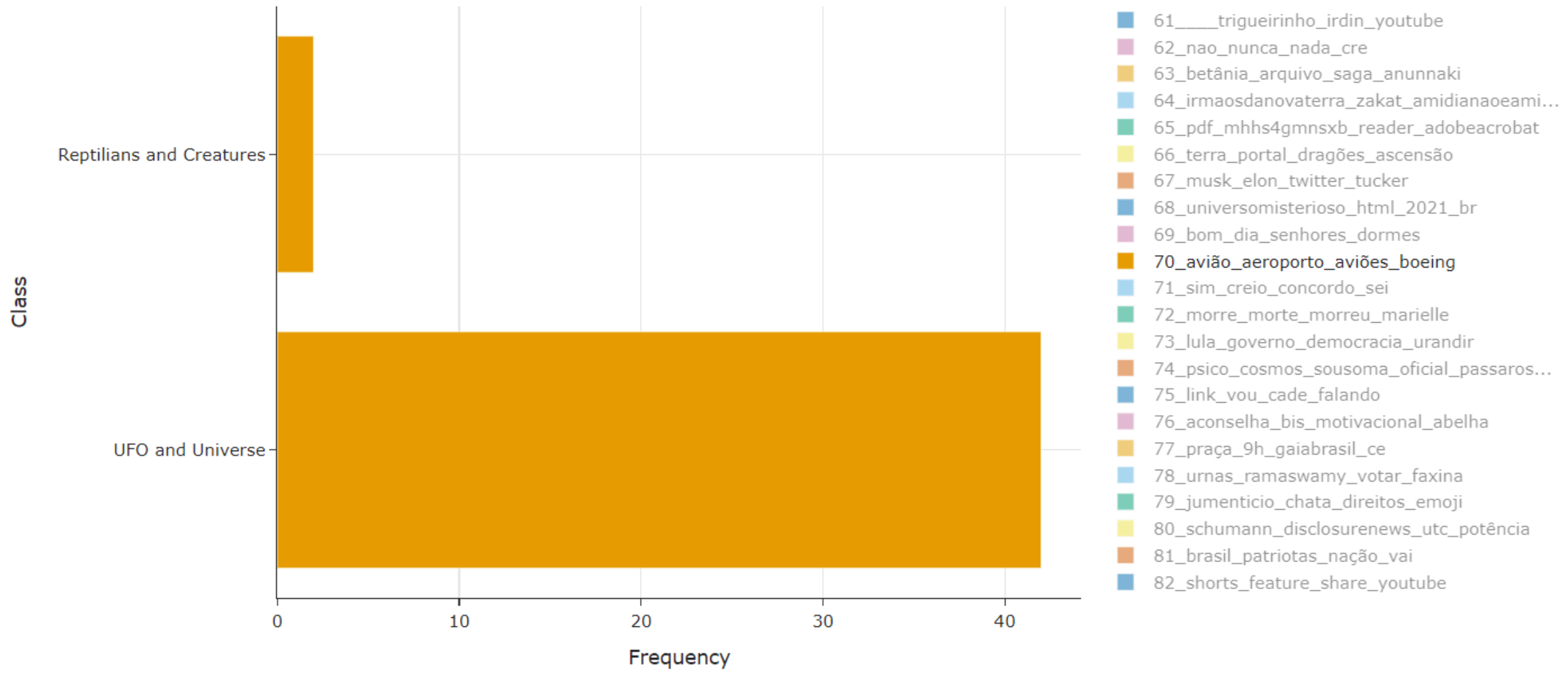

Fonte: Elaboração própria (2024).

A Figura 12 aborda como as discussões sobre OVNIs e Universo se entrelaçam com temas de desaparecimentos misteriosos e quedas de aviões. Tópicos como “desaparecimentos” e “Boeing” são destacados, sugerindo que essas comunidades associam incidentes a intervenções extraterrestres ou conspirações ocultas. A prevalência desses temas dentro das discussões sobre OVNIs reflete uma tendência dessas comunidades de interpretar tragédias e eventos inexplicáveis como evidência de atividade alienígena ou de manipulação por parte de poderes secretos.

**Figura 13.** Temáticas de tragédias naturais tratadas como planejadas

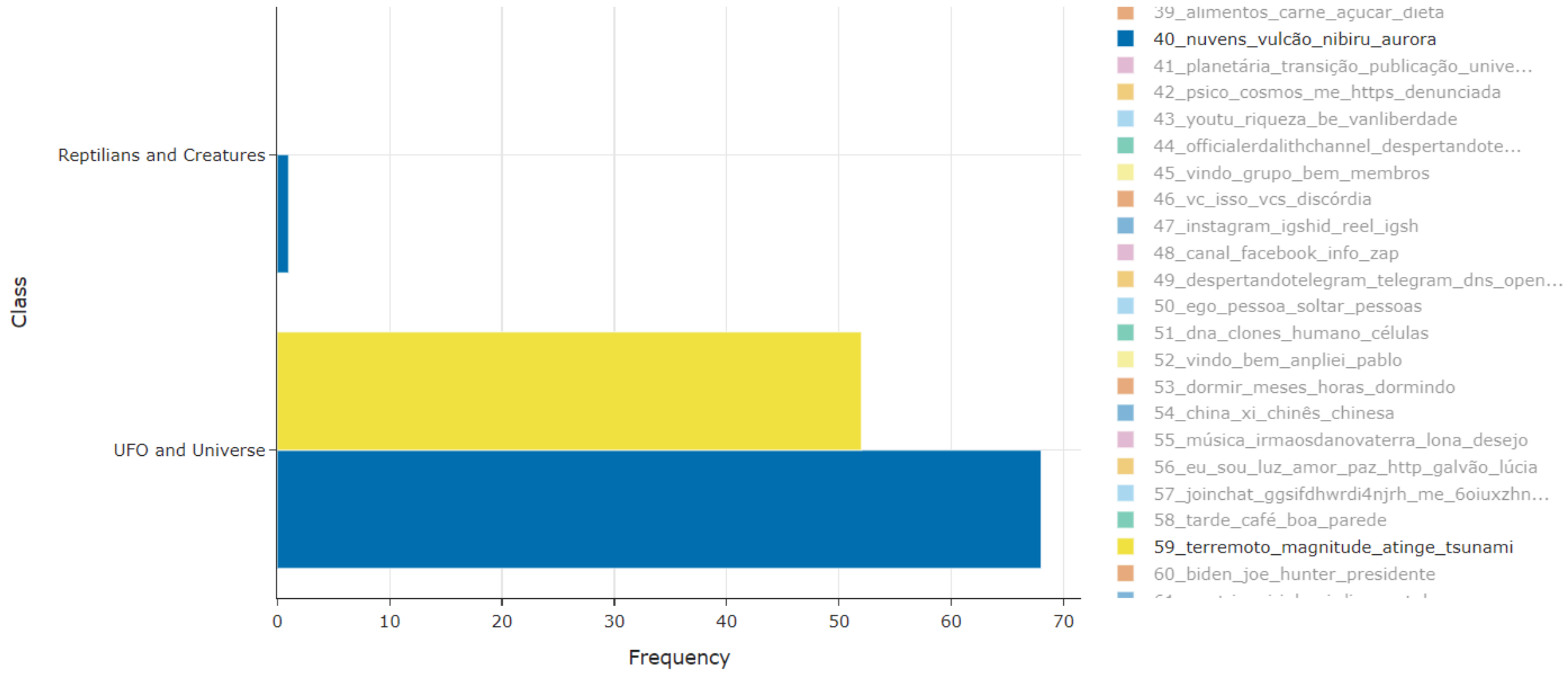

Fonte: Elaboração própria (2024).

A Figura 13 revela como as discussões sobre tragédias naturais, como terremotos e tsunamis, são integradas às narrativas sobre OVNIs e Reptilianos. A ideia de que desastres naturais são, na verdade, planejados ou controlados por forças extraterrestres ou elites ocultas é uma temática recorrente. Esses tópicos fortalecem a visão de que o mundo está sendo manipulado de maneiras que são deliberadamente ocultadas da população em geral, alimentando uma desconfiança generalizada em relação às explicações científicas.

**Figura 14.** Temáticas de religião e fé

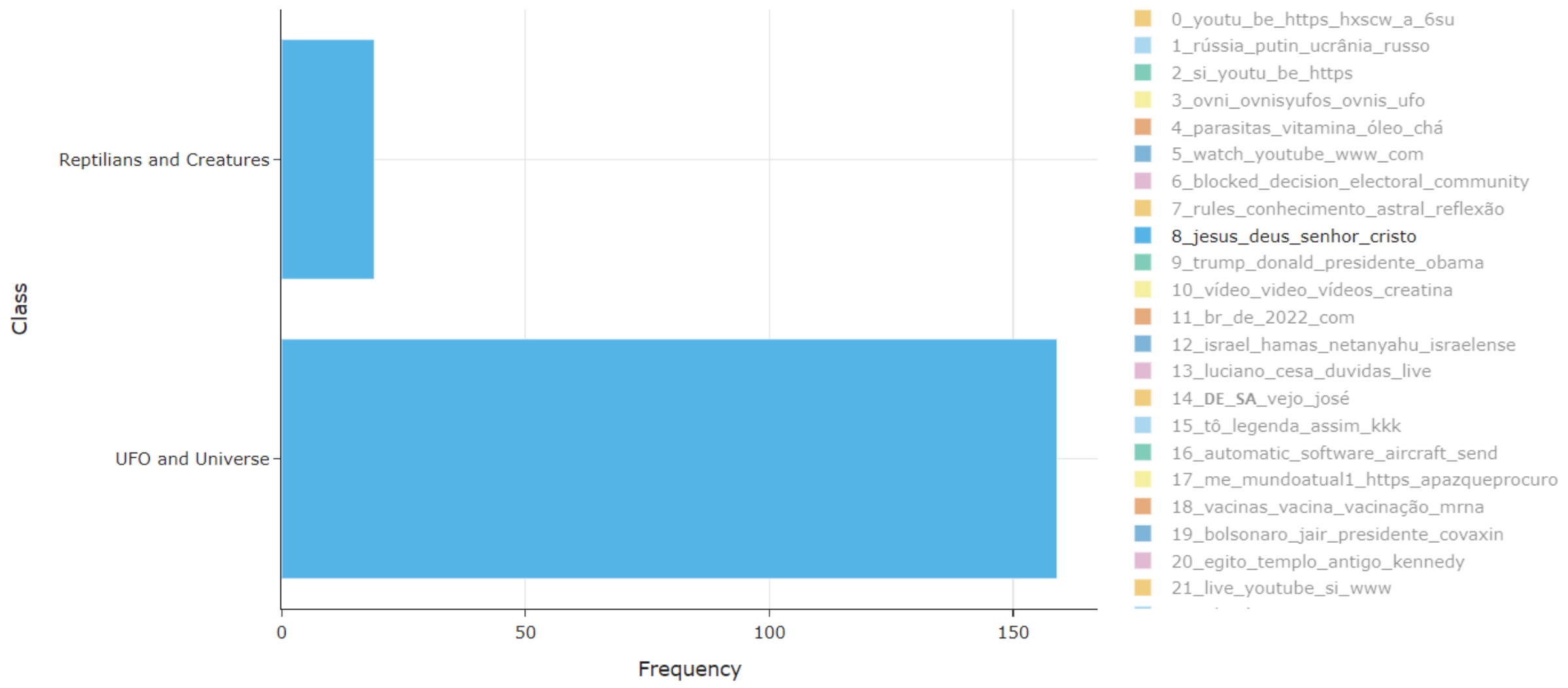

Fonte: Elaboração própria (2024).

Na Figura 14, as temáticas de OVNIs e Universo são combinadas com discussões sobre religião e fé. Tópicos como "Jesus" e "Deus" surgem em conjunto com narrativas extraterrestres, sugerindo que essas comunidades integram crenças religiosas com teorias sobre vida extraterrestre. Essa fusão reflete uma tentativa de explicar fenômenos religiosos ou espirituais como interações com seres de outros mundos, reinterpretando dogmas religiosos sob a luz de teorias conspiratórias sobre OVNIs.

**Figura 15.** Temáticas de espiritualidade e alterações chamadas de quânticas, no DNA

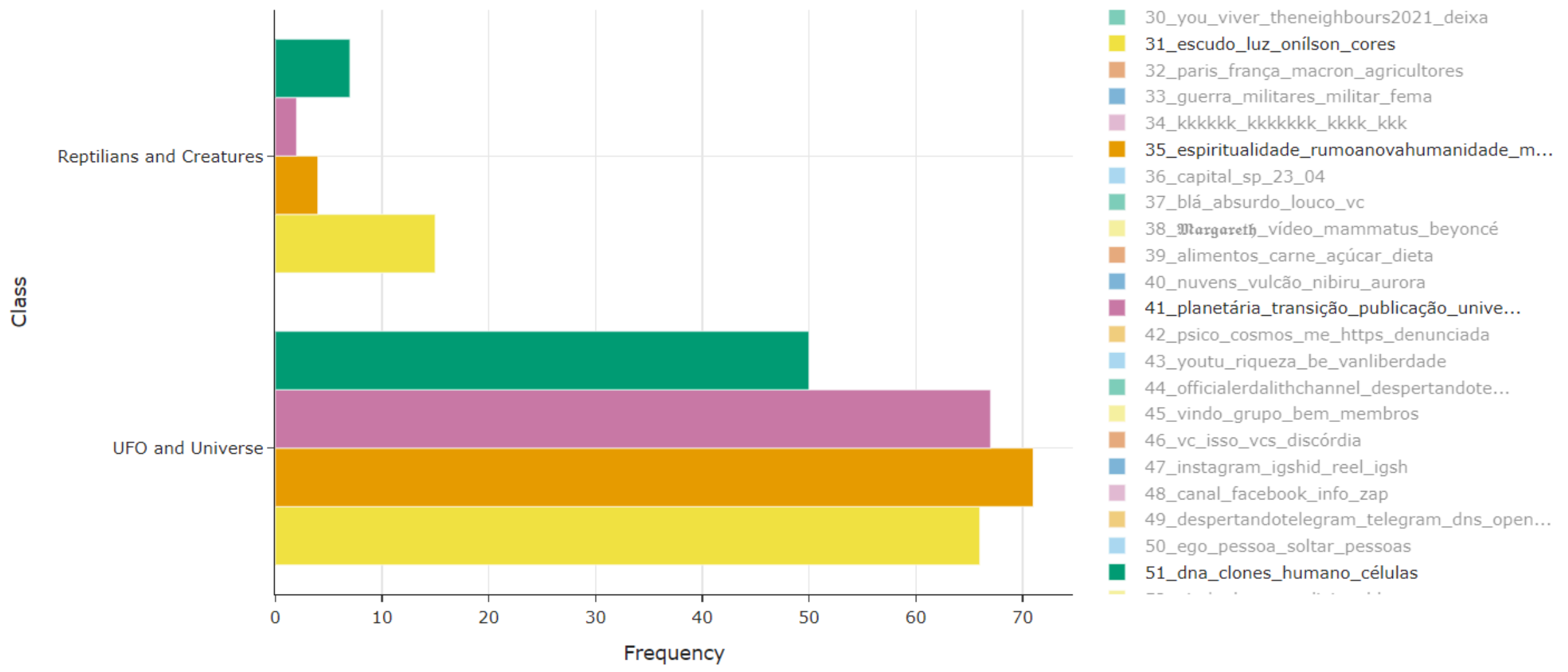

Fonte: Elaboração própria (2024).

A Figura 15 aborda como a espiritualidade e conceitos pseudocientíficos, como as chamadas “alterações quânticas no DNA”, são combinados com discussões sobre OVNIs e Reptilianos. Tópicos como “quântica” e “DNA” são frequentemente mencionados, sugerindo que essas comunidades veem a evolução espiritual e física como diretamente influenciada por forças extraterrestres. Essa narrativa cria uma conexão entre o esotérico e o conspiratório — algo já apontado em outras investigações da série —, ampliando o escopo das crenças dessas comunidades ao integrar ciência, espiritualidade e teorias de conspiração.

## 4. Reflexões e trabalhos futuros

Para responder a pergunta de pesquisa “**como são caracterizadas e articuladas as comunidades de teorias da conspiração brasileiras sobre temáticas de OVNIs, universo, reptilianos e criaturas no Telegram?**”, este estudo adotou técnicas espelhadas em uma série de sete publicações que buscam caracterizar e descrever o fenômeno das teorias da conspiração no Telegram, adotando o Brasil como estudo de caso. Após meses de investigação, foi possível extrair um total de 66 comunidades de teorias da conspiração brasileiras no Telegram sobre temáticas de OVNIs, universo, reptilianos e criaturas, estas somando 1.427.011 de conteúdos publicados entre maio de 2016 (primeiras publicações) até agosto de 2024 (realização deste estudo), com 141.202 usuários somados dentre comunidades.

Foram adotadas quatro abordagens principais: **(i)** Rede, que envolveu a criação de um algoritmo para mapear as conexões entre as comunidades por meio de convites circulados entre grupos e canais; **(ii)** Séries temporais, que utilizou bibliotecas como “Pandas” (McKinney, 2010) e “Plotly” (Plotly Technologies Inc., 2015) para analisar a evolução das publicações e engajamentos ao longo do tempo; **(iii)** Análise de conteúdo, sendo aplicadas técnicas de análise textual para identificar padrões e frequências de palavras nas comunidades ao longo dos semestres; e **(iv)** Sobreposição de agenda temática, que utilizou o modelo BERTopic (Grootendorst, 2020) para agrupar e interpretar grandes volumes de textos, gerando tópicos coerentes a partir das publicações analisadas. A seguir, as principais reflexões são detalhadas, sendo seguidas por sugestões para trabalhos futuros.

### 4.1. Principais reflexões

**Comunidades de OVNIs e universo atuam como portas de entrada para teorias sobre reptilianos e criaturas, conectando narrativas de controle global e extraterrestres:** As comunidades focadas em OVNIs e universo, com 1.268.407 publicações, servem como um gateway para discussões mais extremas sobre reptilianos e outras criaturas. Essas comunidades direcionaram 690 convites para grupos de reptilianos e criaturas, sugerindo que o interesse inicial por vida extraterrestre frequentemente leva a uma exploração mais profunda de teorias que combinam elementos de controle global com seres não humanos;

**Teorias sobre reptilianos e criaturas estão fortemente vinculadas à Nova Ordem Mundial, reforçando narrativas de dominação global:** As teorias sobre reptilianos e criaturas têm uma forte conexão com a Nova Ordem Mundial (NOM), com 1.159 convites entre essas comunidades. Essa interseção destaca como as crenças em entidades reptilianas se fundem com narrativas sobre controle global, criando um discurso unificado que sugere uma conspiração de dominação mundial por forças não humanas;

**Discussões sobre OVNIs e universo experimentaram um crescimento de 4.400% durante a Pandemia, indicando um interesse renovado em fenômenos extraterrestres:** Entre 2020 e 2021, as menções a OVNIs e universo aumentaram de 1.000 para 45.000, impulsionadas pela liberação de relatórios governamentais sobre avistamentos de OVNIs. Esse aumento reflete um interesse crescente em fenômenos extraterrestres, muitas vezes acompanhado por desconfiança em relação às explicações oficiais;

**Reptilianos e criaturas permanecem como uma subcultura significativa dentro das teorias de conspiração, com um crescimento de 1.000% durante a Pandemia:** Embora em menor escala, as discussões sobre reptilianos e criaturas também cresceram significativamente, subindo de 1.000 para 10.000 menções entre 2020 e 2021. Esse crescimento destaca o interesse contínuo por teorias extremas envolvendo entidades não humanas, muitas vezes associadas a narrativas de controle e dominação;

**Comunidades de OVNIs e universo estão fortemente interligadas com o ocultismo e esoterismo, formando uma rede de crenças alternativas:** Com 6.081 convites para comunidades de ocultismo e esoterismo, as discussões sobre OVNIs e universo frequentemente se cruzam com crenças esotéricas. Essa interconexão sugere que a busca por explicações alternativas para fenômenos inexplicáveis se estende a práticas místicas e ocultas, criando uma rede de crenças que desafia a ciência convencional;

**A narrativa de que tragédias naturais são planejadas e controladas por forças extraterrestres ganhou força nas comunidades de OVNIs:** Discussões que ligam desastres naturais, como terremotos e tsunamis, à intervenção de forças extraterrestres são comuns nas comunidades de OVNIs. Essa crença se alinha à desconfiança generalizada em relação às

explicações científicas e sugere que eventos naturais são, na verdade, parte de um plano maior de manipulação e dominação global global;

**A sobreposição temática entre OVNIs, reptilianos e esoterismo revela um ecossistema coeso de desinformação que é difícil de desmantelar:** A análise da sobreposição temática mostra que essas comunidades não apenas compartilham membros e conteúdos, mas também reforçam mutuamente suas crenças. A interligação entre esses temas cria um ciclo de desinformação onde diferentes narrativas se apoiam, tornando a correção factual um desafio ainda maior;

**A figura de reptilianos como parte de um plano de controle global ressurge constantemente, reforçando a persistência dessa narrativa:** Mesmo com a falta de evidências, a ideia de que reptilianos são parte de uma conspiração global persiste dentro dessas comunidades. Essa narrativa é reintroduzida repetidamente, fortalecendo a desconfiança em relação às autoridades e alimentando teorias mais amplas sobre dominação;

**As comunidades de OVNIs funcionam como amplificadores de outras teorias conspiratórias, conectando diferentes temas e fortalecendo a rede de desinformação:** As comunidades centradas em OVNIs não apenas se focam em fenômenos extraterrestres, mas também atuam como *hubs* que conectam seus membros a uma variedade de outras teorias conspiratórias, como ocultismo, NOM e apocalipse. Essa função de amplificação torna essas comunidades centrais na disseminação de desinformação;

**As discussões sobre OVNIs e reptilianos são persistentemente interligadas, sugerindo que essas narrativas estão enraizadas em uma visão de mundo conspiratória mais ampla:** A análise dos fluxos de convites entre comunidades revela que discussões sobre OVNIs e reptilianos são frequentemente sobrepostas, indicando que esses temas não são vistos isoladamente, mas como parte de um conjunto maior de crenças que desafiam a visão de mundo convencional.

### 4.2. Trabalhos futuros

Futuros estudos devem explorar como as comunidades de OVNIs e universo atuam como portas de entrada para outras narrativas conspiratórias, investigando como o interesse inicial por fenômenos extraterrestres pode levar a uma aceitação mais ampla de teorias de dominação global e controle. A análise dessas dinâmicas pode ajudar a entender como essas crenças se propagam e se solidificam dentro de comunidades online.

Também é importante investigar as interações entre temas esotéricos e conspirações extraterrestres, especialmente em relação ao ocultismo e à espiritualidade alternativa. Com a interconexão observada entre OVNIs e práticas esotéricas, futuros estudos poderiam se concentrar em como essas crenças se reforçam mutuamente e como isso impacta a visão de mundo dos membros dessas comunidades.

Outro ponto relevante é a persistência das narrativas de reptilianos como parte de um plano de controle global. Futuros estudos poderiam investigar os mecanismos que permitem a sobrevivência e a reintrodução constante dessa narrativa, mesmo diante de repetidos desmentidos. Compreender esses mecanismos pode ser crucial para desenvolver estratégias mais eficazes de combate à desinformação.

Além disso, pesquisas futuras devem se concentrar na análise das redes de convites entre comunidades, explorando como essas conexões facilitam a disseminação de desinformação e criando um ambiente propício para o reforço mútuo de crenças conspiratórias. Desenvolver ferramentas que possam mapear e identificar esses fluxos em tempo real pode ser uma maneira eficaz de mitigar a propagação de desinformação em momentos críticos. Por fim, estudos sobre a resiliência das comunidades centradas em OVNIs e reptilianos em face de crises globais podem oferecer insights sobre como essas narrativas se adaptam e se mantêm relevantes. Entender como essas comunidades respondem a novas informações e eventos pode ajudar a prever e responder melhor a surtos de desinformação.

## 5. Referências

## 6. Biografia do autor

**Ergon Cugler de Moraes Silva** possui mestrado em Administração Pública e Governo (FGV), MBA pós-graduação em Ciência de Dados e Análise (USP) e bacharelado em Gestão de Políticas Públicas (USP). Ele está associado ao Núcleo de Estudos da Burocracia (NEB FGV), colabora com o Observatório Interdisciplinar de Políticas Públicas (OIPP USP), com o Grupo de Estudos em Tecnologia e Inovações na Gestão Pública (GETIP USP), com o Monitor de Debate Político no Meio Digital (Monitor USP) e com o Grupo de Trabalho sobre Estratégia, Dados e Soberania do Grupo de Estudo e Pesquisa sobre Segurança Internacional do Instituto de Relações Internacionais da Universidade de Brasília (GEPSI UnB). É também pesquisador no Instituto Brasileiro de Informação em Ciência e Tecnologia (IBICT), onde trabalha para o Governo Federal em estratégias contra a desinformação. Brasília, Distrito Federal, Brasil. Site: https://ergoncugler.com/.